\documentclass[epj,nopacs]{svjour}
\usepackage{subfiles}
\usepackage{amsmath}
\usepackage{amssymb}
\usepackage{cite}

\usepackage{mathrsfs}
\usepackage{graphicx}
\usepackage{epstopdf}
\usepackage{fancyhdr}
\usepackage{array}
\usepackage[all]{xy}
\usepackage{eufrak}
\usepackage{euscript}
\usepackage{enumerate}
\usepackage{slashed}
\usepackage{physics}
\usepackage{hyperref}
\usepackage{subfigure} % NEEDED IN ORDER TO USE SUBFIG
\usepackage{epstopdf} % COMPILE EPS (AND PDF) FIGURES USING PDFLATEX!!!
\usepackage[normalem]{ulem}
\usepackage{float}
\usepackage{graphics}
\usepackage{microtype}
\begin{document}
\title{Bjorken Flow of Holographic R-Charged Plasmas}

\author{Gustavo de Oliveira\inst{1}\thanks{\emph{e-mail:} gustav.o.liveira@discente.ufg.br} \and 
        Willians Barreto\inst{2}\thanks{\emph{e-mail:} wbarreto@ivic.gob.ve} \and 
        Romulo Rougemont\inst{1}\thanks{\emph{e-mail:} rougemont@ufg.br}
}                      % Do not remove                    % Do not remove
\institute{Instituto de F\'{i}sica, Universidade Federal de Goi\'{a}s, Avenida Esperan\c{c}a - Campus Samambaia, CEP 74690-900, Goi\^{a}nia, Goi\'{a}s, Brazil \and Centro de F\'{i}sica, Instituto Venezolano de Investigaciones Cient\'{i}ficas, Apartado Postal 20632, Caracas 1020-A, Venezuela}
\date{Received: date / Revised version: date}
\abstract{
We numerically investigate the time evolution of several physical observables for the so-called 2 R-Charge Black Hole (2RCBH) model undergoing Bjorken flow. The 2RCBH model corresponds to a top-down holographic construction describing a strongly interacting conformal fluid defined at finite temperature and R-charge density. Taken together with previous findings for the purely thermal $\mathcal{N}=4$ Supersymmetric Yang-Mills (SYM) plasma, and the 1 R-Charge Black Hole (1RCBH) model, our results for the 2RCBH model provide strong numerical evidence for the existence of far-from-equilibrium correlations between the non-equilibrium holographic entropy defined through the area of the apparent horizon of dynamical bulk black holes, and the expectation value of the energy-momentum tensor of the dual boundary quantum field theory. Such correlations are relevant in the pre-hydrodynamic stages of some initial data evolved in time, and seem to hold at least for strongly interacting conformal fluids, be they charged or neutral.
%
%\PACS{
%      {PACS-key}{discribing text of that key}   \and
%      {PACS-key}{discribing text of that key}
%     } % end of PACS codes
} %end of abstract
\maketitle

\section{Introduction}
\label{sec:intro}

Out-of-equilibrium properties of dynamical systems, like entropy production and the time evolution of different physical observables, may be used to describe how these systems evolve from detailed initial out-of-equilibrium states toward later stationary states, generally characterized by a set with just a few state variables. Therefore, the investigation of far-from-equilibrium dynamics constitutes an endeavor of great importance in different areas of physics~\cite{Rigol:2007juv,Rigol:2009zz,Herrera:2011kd,Muller:2011ra,Florkowski:2017olj,Romatschke:2017ejr,Muller:2020ziz}. Of particular relevance to the field of heavy-ion collisions, there is the so-called Bjorken flow~\cite{Bjorken:1982qr}, which is a kind of out-of-equilibrium and boost-invariant dynamics relevant to the modeling of the longitudinal expansion of the fluid produced in ultrarelativistic nuclear collisions.\footnote{In the Bjorken flow, one neglects the dynamics related to the fluid expansion in the plane transverse to the colliding beam axis, so that it should work better as an approximation for the fluid dynamics closer to the beam axis and at high energies.} Given the outstanding technical difficulties regarding first principles real-time calculations in QCD with a running coupling ranging from weakly to strongly-coupled physics through the time evolution of the system, the Bjorken flow has been widely studied in the literature for different models of relativistic fluids --- see e.g.~\cite{Chesler:2009cy,Heller:2011ju,Heller:2012je,Jankowski:2014lna,Pedraza:2014moa,Bellantuono:2015hxa,Heller:2016rtz,Romatschke:2017vte,DiNunno:2017obv,Spalinski:2017mel,Strickland:2017kux,Casalderrey-Solana:2017zyh,Critelli:2018osu,Kurkela:2019set,Almaalol:2020rnu,Dore:2020jye,Rougemont:2021qyk,Rougemont:2021gjm,Cartwright:2021maz,Rougemont:2022piu,Cartwright:2022hlg,Dore:2022qyz}. By analyzing models of different nature, one general idea is that one could obtain insights about possible similarities and also fundamental differences between e.g.~weakly and strongly interacting dynamics, both of which are expected to contribute to different stages of ultrarelativistic nuclear collisions~\cite{McLerran:2001sr,Iancu:2003xm,Weigert:2005us,Gelis:2010nm,Gelis:2012ri,Heinz:2013th,Shuryak:2014zxa,MUSES:2023hyz}.

In what concerns strongly interacting models, one of the most prominent theoretical frameworks of the last decades is the holographic gauge-gravity duality~\cite{Maldacena:1997re,Gubser:1998bc,Witten:1998qj,Witten:1998zw}. Such a framework posits a detailed mathematical dictionary relating physical observables of some strongly interacting quantum field theories (QFTs) defined in flat spacetime dimensions, with calculations involving (semi-)classical gravity models in asymptotically Anti-de Sitter (AdS) spacetimes. Since the QFT lives at the (conformally) flat boundary of the bulk asymptotically AdS spacetime, the bulk geometry has (at least) one extra spatial dimension relatively to the spacetime where the dual QFT is defined. Therefore, concerning applications to four-dimensional QFTs, holographic models are usually defined in terms of five-dimensional gravity actions.

A very interesting feature of gauge-gravity models is that they allow for full far-from-equilibrium numerical simulations of real-time dynamics of some strongly interacting QFTs, including systems defined at finite temperature and chemical potential. Therefore, one can study how late time hydrodynamic behavior emerges in the time evolution of holographic media subjected to different kinds of far-from-equilibrium dynamics. Although at early times the behavior of the system is highly dependent on the details of its initial state, several different initial states can typically converge at late times to the same hydrodynamic attractor characterized by just a few variables, like the values of temperature and chemical potential of the medium close to thermodynamic equilibrium. This process of hydrodynamization of the medium effectively happens at late times, when the system loses memory of details regarding the initial conditions. This effective erasing of information is accompanied by entropy production in the medium.

Previous works on the holographic Bjorken flow of the purely thermal $\mathcal{N}=4$ Supersymmetric Yang-Mills (SYM) plasma~\cite{Rougemont:2021qyk,Rougemont:2021gjm}, and the 1-R Charge Black Hole (1RCBH) model~\cite{Rougemont:2022piu} at finite R-charge density, have unveiled some important differences in the pre-hydrodynamic stages of such strongly interacting quantum fluids and classical weakly-coupled plasmas described by kinetic theory~\cite{degroot}. In fact, while the latter dictates time evolutions always satisfying classical energy conditions (such as the dominant, strong, and weak energy conditions)~\cite{HawkingEllisBook,Wald:1984rg}, holographic models can transiently violate such energy conditions in pre-hydrodynamic stages. Moreover, due to fundamentally different spectra of quasinormal modes, the approach towards equilibrium in holographic and kinetic theory systems is also generally different~\cite{Noronha:2011fi}.

Another important feature unveiled in Refs.~\cite{Rougemont:2021qyk,Rougemont:2021gjm,Rougemont:2022piu} was the existence of transient plateaus in the time evolution of the holographic non-equilibrium entropy associated to the apparent horizon of some dynamical bulk black hole solutions. These plateaus correspond to far-from-equilibrium transient stages with zero entropy production, which is also something that does not happen in kinetic theory models. Furthermore, such transient plateau structures in the apparent horizon entropy were found to be correlated with the expectation value of the energy momentum-tensor of the boundary QFT. More specifically, when such entropy plateaus begin to be formed, they anticipate the later formation of local minima in the pressure anisotropy normalized by the energy density, which may be related to some violations of the dominant energy condition~\cite{Rougemont:2021qyk,Rougemont:2021gjm,Rougemont:2022piu}.

The purpose of the present work is to numerically develop the Bjorken flow dynamics of the holographic 2-R Charge Black Hole (2RCBH) model~\cite{DeWolfe:2011ts,DeWolfe:2012uv}, which as well as the 1RCBH model, is also a conformal theory defined at finite temperature and R-charge density, but with a very different phase diagram from the latter. We confirm here, also for the 2RCBH model, the existence of the aforementioned far-from-equilibrium correlations between apparent horizon entropy plateaus and the later formation of local minima in the pressure anisotropy normalized by the energy density. This provides compeling numerical evidence for the generality of such pre-hydrodynamic correlations, at least in what concerns strongly interacting conformal fluids, be them charged or neutral. We further investigate in the present work the behavior of the non-equilibrium entropy defined through the area of the dynamical event horizon, which in contrast to the apparent horizon, is never observed to produce transient plateaus. This indicates that the apparent horizon holographically encodes some far-from-equilibrium correlations with the energy momentum-tensor of the dual boundary QFT, which are not present in the event horizon.

We also analyze the time evolution of the normalized scalar condensate of the medium and its thermalization, besides the equilibration of the normalized dynamical entropy associated to the apparent horizon, and the hydrodynamization process of the normalized pressure anisotropy of the fluid. We find that the different relaxation time scales associated to these different physical observables, with the exception of the entropy density, are delayed as the R-charge density of the fluid is increased. That is, overall, the denser the fluid, the more time it takes to hydrodynamize, and later thermalize.

It is important to remark that these two R-charged models, the 2RCBH and 1RCBH plasmas, should be viewed as toy models for two different strongly interacting media. Although physically different, as we shall see, both R-charged models display some qualitative features in common, like the possibility of transient violations of energy conditions when the system is still far from equilibrium, besides correlations between transient time windows with zero entropy production in the far-from-equilibrium regime, and the later formation of minima for the pressure anisotropy of the medium violating the dominant (or even the weak) energy condition. Since these possibilities are allowed for two different R-charged models, one may speculate or conjecture that such possibilities are generally allowed by strongly interacting quantum systems, even though such possibilities are not considered in current hybrid phenomenological models of heavy ion collisions, where hydrodinamization is modeled by classical kinetic approaches.

The present work is organized as follows. In Section~\ref{sec:sec1}, we review the definition of the 2RCBH and 1RCBH models and their equilibrium thermodynamics. In Section~\ref{sec:sec2}, we begin by reviewing the basics of Bjorken flow hydrodynamics, and then discuss the holographic algorithm employed here to dynamically evolve the bulk fields implementing the far-from-equilibrium Bjorken flow dynamics. In Section~\ref{sec:3}, we derive the relevant holographic formulas to numerically simulate the evolution of different dynamical observables at the dual boundary QFT, with our numerical results and the corresponding physical analysis being presented in Section~\ref{sec:results}. Finally, we summarize our main findings in Section~\ref{sec:conc}.

In this work, we use natural units with $\hbar = c = k_B = 1$ and a mostly plus metric signature.

\section{The Holographic R-Charged Black Hole Models}
\label{sec:sec1}

We are interested here in investigating the far-from-equilibrium dynamics for different plasma states of the four-dimensional $\mathcal{N}=4$ SYM theory at finite temperature and R-charge density, as it evolves under Bjorken flow~\cite{Bjorken:1982qr}: a boost-invariant longitudinal expansion which is homogeneous and isotropic with respect to the transverse plane. Via the gauge-gravity duality, the four-dimensional SYM plasma states at finite R-charge density are dual to charged black hole solutions\footnote{More precisely, black brane solutions with spatially extended and translationally invariant event horizons.} of the five-dimensional STU model~\cite{Behrndt:1998jd,Cvetic:1999ne}. In general, the STU black hole solutions are charged under the Abelian $U(1)\times U(1)\times U(1)$ group, which corresponds to the Cartan subgroup of the original global $SU(4)$ R-symmetry group of SYM theory in vacuum. In the grand canonical ensemble, the three independent conserved $U(1)$ charges $(Q_a, Q_b, Q_c)$ are coupled to three distinct chemical potentials $(\mu_a, \mu_b, \mu_c)$, which are holographically interpreted as the R-charge chemical potentials of a given plasma state of the dual SYM theory at finite temperature and R-charge density.

Here we consider two particular realizations of the STU model~\cite{DeWolfe:2011ts,DeWolfe:2012uv,deOliveira:2024bgh}. One case corresponds to setting two of the three conserved R-charges to zero, allowing the remaining $U(1)$ R-charge to take nonzero values, which is the so-called 1RCBH model and features a single R-charge chemical potential. The second case, which is the main focus of the present work, corresponds to setting one of the three conserved R-charges to zero, while taking the other two R-charges to be equal and allowing them to take nonzero values, which is the so-called 2RCBH model and also features a single independent chemical potential. The Bjorken flow of the 1RCBH model has been analyzed in~\cite{Critelli:2018osu,Rougemont:2022piu}, and here we shall compare it to the new results obtained for the Bjorken flow of the 2RCBH model, with the purpose of seeking for possible general properties of different strongly interacting R-charged fluids.

Both holographic models are described by the class of five-dimensional Einstein-Maxwell-Dilaton (EMD) actions of the general form,
\begin{equation}
\label{eq:EMDaction}
\begin{split}
    S = \frac{1}{2\kappa_5^2} \int_{\mathcal{M}_5} d^5x\sqrt{-g} &\bigg[ R - \frac{f(\phi)}{4} F_{\mu\nu}F^{\mu\nu}- \frac{1}{2}(\partial_\mu\phi)^2 \\
    &\quad - V(\phi) \bigg] + S_\textrm{GHY} + S_\textrm{CT},
\end{split}
\end{equation}
where $\kappa_5^2=8\pi G_5$, with $G_5$ being the five-dimensional Newton's constant, $S_\textrm{GHY}$ is the Gibbons-Hawking-York (GHY) boundary action~\cite{York:1972sj,Gibbons:1976ue}, needed for the consistency of the gravitational variational problem in asymptotically AdS spacetimes (and, more generally, in any spacetime with a boundary~\cite{Poisson:2009pwt}), and $S_\textrm{CT}$ is the holographic counterterm boundary action~\cite{Critelli:2017euk}, required to consistently removing the divergences of the total on-shell boundary action according to the holographic renormalization program~\cite{deHaro:2000vlm,Bianchi:2001kw,Skenderis:2002wp,Papadimitriou:2011qb,Lindgren:2015lia,Elvang:2016tzz}.

The most general EMD field equations are found by extremizing the bulk action in~\eqref{eq:EMDaction} with respect to the bulk fields, 
\begin{subequations}
\label{eq:EMDequations}
\begin{align}
    &R_{\mu\nu} - \frac{g_{\mu\nu}}{3} \left( V(\phi) - \frac{f(\phi)}{4} F^2_{\alpha\beta} \right) - \frac{1}{2} \partial_{\mu}\phi \partial_{\nu}\phi \nonumber\\
    &\hspace{4cm}- \frac{f(\phi)}{2} F_{\mu\rho}F_{\nu}^{\,\,\rho} = 0,\\
    &\nabla_{\mu}(f(\phi)F^{\mu\nu}) = \frac{1}{\sqrt{-g}}\partial_{\mu}(\sqrt{-g}f(\phi)F^{\mu\nu}) = 0, \\
    &\frac{1}{\sqrt{-g}}\partial_{\mu}(\sqrt{-g}g^{\mu\nu}\partial_{\nu}\phi) - \partial_{\phi}V(\phi) - \frac{\partial_{\phi}f(\phi)}{4}F^2_{\mu\nu} = 0.
\end{align}
\end{subequations}

Within the general class of EMD actions, the 1RCBH and 2RCBH models are defined by the following specifications of the dilaton potential $V(\phi)$ (the same for both models) and the Maxwell-dilaton coupling function $f(\phi)$,
\begin{subequations}
\label{eq:Vfphi1R}
\begin{align}
    &\text{1RCBH model:} &\nonumber\\
    &V(\phi) = -\frac{1}{L^2} \left( 8e^{\phi/\sqrt{6}} + 4e^{-\sqrt{\frac{2}{3}}\phi} \right), \label{eq:Vphi1R}\\
    &f(\phi) = e^{-2\sqrt{\frac{2}{3}}\phi},\label{eq:fphi1R}
\end{align}
\end{subequations}
\begin{subequations}
\label{eq:Vfphi2R}
\begin{align}
    &\text{2RCBH model:}&\nonumber\\
    &V(\phi) =- \frac{1}{L^2} \left( 8e^{\phi/\sqrt{6}} + 4e^{-\sqrt{\frac{2}{3}}\phi} \right), \label{eq:Vphi2R} \\
     &f(\phi) = e^{\sqrt{\frac{2}{3}}\phi},\label{eq:fphi2R}
\end{align}
\end{subequations}
where $L$ is the asymptotic AdS$_5$ radius.

Although both the 1RCBH and 2RCBH models descend from the same parent gravity model, the STU model, their holographic duals, corresponding to strongly interacting fluids at finite temperature and R-charge density, exhibit starkly different thermodynamic phase structures. Different thermal states in thermodynamic equilibrium at the boundary QFT are holographically dual to different static, spatially homogeneous and isotropic charged EMD black hole solutions in the bulk. Therefore, in what regards thermodynamics, one considers the following ansatze for the bulk fields, written below in modified Eddington-Finkelstein (EF) coordinates denoted with a tilde~\cite{Critelli:2017euk}, which will be convenient for comparisons with out-of-equilibrium solutions to be considered later in this work,
\begin{subequations}
\begin{align}
    ds^2&=g_{\mu\nu}dx^\mu dx^\nu\nonumber\\
    &=d\tau\left[2e^{ a(\tilde{r})+b(\tilde{r})}d\tilde{r}-e^{ 2a(\tilde{r})}h(\tilde{r})d\tau\right]+e^{2 a(\tilde{r})}d\mathbf{x}^2,\\
A_\mu &= \Phi(\tilde{r})\delta_\mu^t,\\
\phi&=\phi(\tilde{r}),
    \end{align}
\end{subequations}
\label{eq:AnsatzEqs}
where the boundary lies at $\tilde{r}\to\infty$, and the bulk EF time coordinate $\tau$ is related to the time coordinate $t$ at the boundary QFT as follows,
\begin{equation}
d\tau = dt +\sqrt{-\frac{g_{\tilde{r}\tilde{r}}}{g_{tt}}}d\tilde{r}.
\label{eq:EFtime}
\end{equation}
In thermodynamic equilibrium, one substitutes the above ansatze for the bulk fields into the general EMD field equations~\eqref{eq:EMDequations}, from which a set of coupled ordinary differential equations (ODEs) is derived. Analytical solutions for these ODEs can be found in terms of two black hole parameters corresponding to its mass $M$ and charge $Q$ as follows (from now on we consider length units of $L=1$)~\cite{DeWolfe:2011ts,DeWolfe:2012uv,Finazzo:2016psx,Critelli:2017euk,deOliveira:2024bgh},
\begin{subequations}
\label{eq:AnsatzAll}
\begin{align}
    &\text{1RCBH model:}\qquad &&\nonumber \\
   &a(\tilde{r})=\ln \tilde{r} +\frac{1}{6}\ln\left(1+\frac{Q^2}{\tilde{r}^2}\right),\qquad & &\\ 
   &b(\tilde{r})=-\ln \tilde{r} -\frac{1}{3}\ln\left(1+\frac{Q^2}{\tilde{r}^2}\right),\qquad & &\\
   &h(\tilde{r})=1-\frac{M^2}{\tilde{r}^2(\tilde{r}^2+Q^2)}, \qquad & &\\
   &\phi(\tilde{r})=-\sqrt{\frac{2}{3}}\ln\left(1+\frac{Q^2}{\tilde{r}^2}\right),& &\\
   &\Phi(\tilde{r}) = -\frac{MQ}{\tilde{r}^2+Q^2}+\frac{MQ}{\tilde{r}_{H}^2+Q^2} ,&&
   \end{align}
\end{subequations}
\begin{subequations}
\label{eq:AnsatzAll2R}
    \begin{align}
   &\text{2RCBH model:}&&\nonumber\\
   &a(\tilde{r})=\ln \tilde{r} +\frac{1}{3}\ln\left(1+\frac{Q^2}{\tilde{r}^2}\right),&&\\
   &b(\tilde{r})=-\ln \tilde{r} -\frac{2}{3}\ln\left(1+\frac{Q^2}{\tilde{r}^2}\right),&&\\
   &h(\tilde{r})=1-\frac{M^2}{(\tilde{r}^2+Q^2)^2}, &&\\
   &\phi(\tilde{r})=\sqrt{\frac{2}{3}}\ln\left(1+\frac{Q^2}{\tilde{r}^2}\right),&&\\
   &\Phi(\tilde{r}) = -\frac{\sqrt{2}MQ}{\tilde{r}^2+Q^2}+\frac{\sqrt{2}MQ}{\tilde{r}_{H}^2+Q^2}.
\end{align}
\end{subequations}
Here, the radial position of the black hole event horizon, $\tilde{r}_H$, is specified by the largest real zero of the blackening function, $h(\tilde{r})$,
\begin{align}
    &\text{1RCBH model:}&&\text{2RCBH model:}\nonumber\\
    &\tilde{r}_{H}=\sqrt{\frac{1}{2}\left(\sqrt{Q^4+4M^2}-Q^2\right)},  
    &&\tilde{r}_{H}=\sqrt{M-Q^2}.\label{eq:bbrp}
\end{align}
One can describe the different black hole solutions in equilibrium by using either the pair of parameters $(M,Q)$ or $(\tilde{r}_H,Q)$.

The physical interpretation of the black hole parameters $M$ and $Q$ is determined by the behavior of the solutions at asymptotic radial distances, $\tilde{r}\to \infty$. For both holographic models, the blackening function has the leading-order term $h(\tilde{r})\approx 1-M^2/\tilde{r}^4$. By comparing this deviation from the pure $AdS_5$ background with the standard ADM mass formalism, the parameter $M$ is identified as the black hole mass. On the other hand, the charge can be identified using Gauss Law. The bulk radial electric field, $E_{\tilde{r}}=-\partial_{\tilde{r}}\Phi(\tilde{r})$, behaves asymptotically as $E_{\tilde{r}}\propto Q/\tilde{r}^3$. Integrating the electric flux over a large sphere yields a total enclosed charge proportional to $Q$, confirming, therefore, its interpretation as the black hole charge.

From the holographic dictionary, the temperature $T$ and the $U(1)$ R-charge chemical potential $\mu$ of the dual strongly interacting fluid can be calculated, respectively, from the expressions for the Hawking temperature of the black hole horizon, and the value of the bulk Maxwell field at the boundary~\cite{DeWolfe:2011ts,DeWolfe:2012uv,Finazzo:2016psx,Critelli:2017euk,deOliveira:2024bgh},

\begin{subequations}
\label{eq:Tmu}
\begin{align}
&\text{1RCBH model:} &&\nonumber\\
&T=\frac{\sqrt{-(g_{tt})'(g^{\tilde{r}\tilde{r}})'}}{4\pi}\Bigg|_{\tilde{r}=\tilde{r}_{H}}=\frac{Q^2+2\tilde{r}_{H}^2}{2\pi\sqrt{Q^2+\tilde{r}_{H}^2}}, &&\label{eq:TThermDef1R}\\
&\mu=\lim_{\tilde{r}\to\infty}\Phi(\tilde{r})=\frac{\tilde{r}_{H}Q}{\sqrt{Q^2+\tilde{r}_{H}^2}}, &&\\
&&&\nonumber\\
&\text{2RCBH model:}&&\nonumber\\
&T=\frac{\sqrt{-(g_{tt})'(g^{\tilde{r}\tilde{r}})'}}{4\pi}\Bigg|_{\tilde{r}=\tilde{r}_{H}}=\frac{\tilde{r}_{H}}{\pi},&&\label{eq:TThermDef2R}\\
&\mu=\lim_{\tilde{r}\to\infty}\Phi(\tilde{r})=\sqrt{2}Q.&&
\end{align}
\end{subequations}

\begin{figure*}
\centering  
\subfigure[Entropy density]{\includegraphics[width=0.475\linewidth]{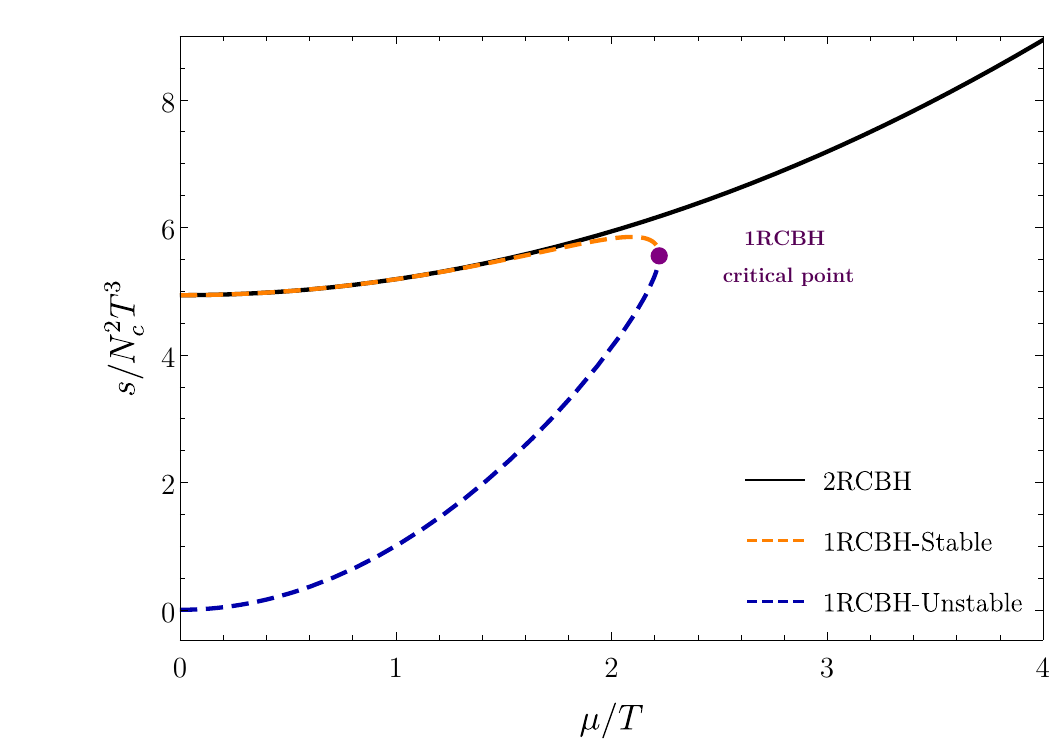}\label{fig:sThermo}}
\subfigure[R-charge density]{\includegraphics[width=0.475\linewidth]{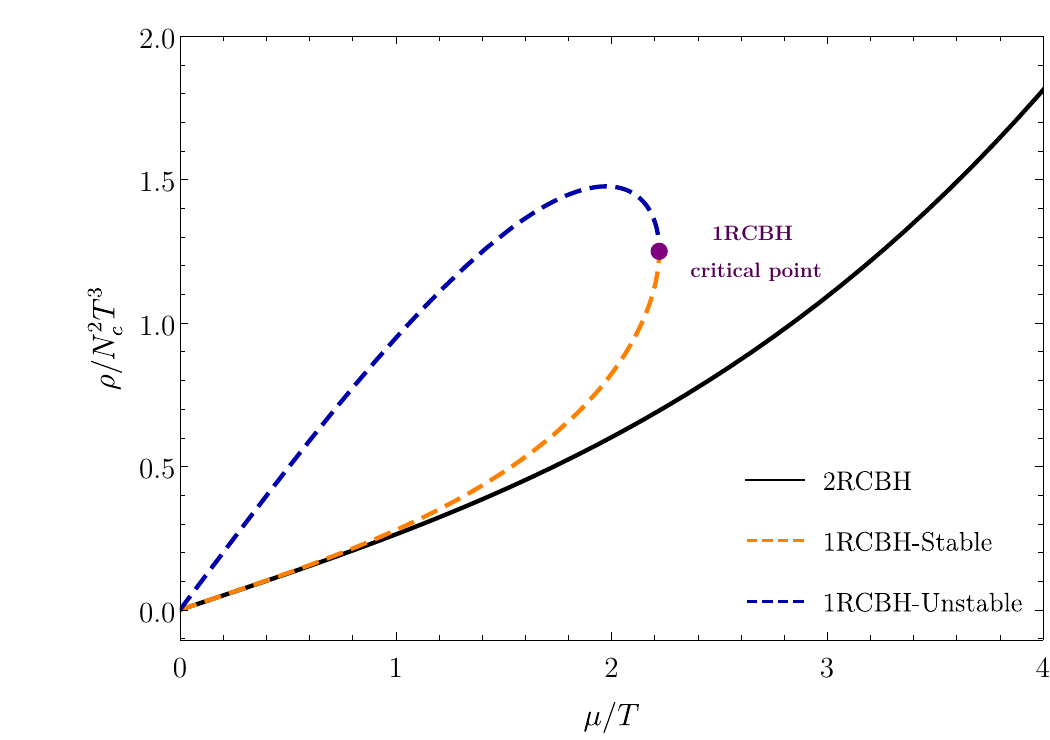}\label{fig:rhoThermo}}
\subfigure[Pressure]{\includegraphics[width=0.475\linewidth]{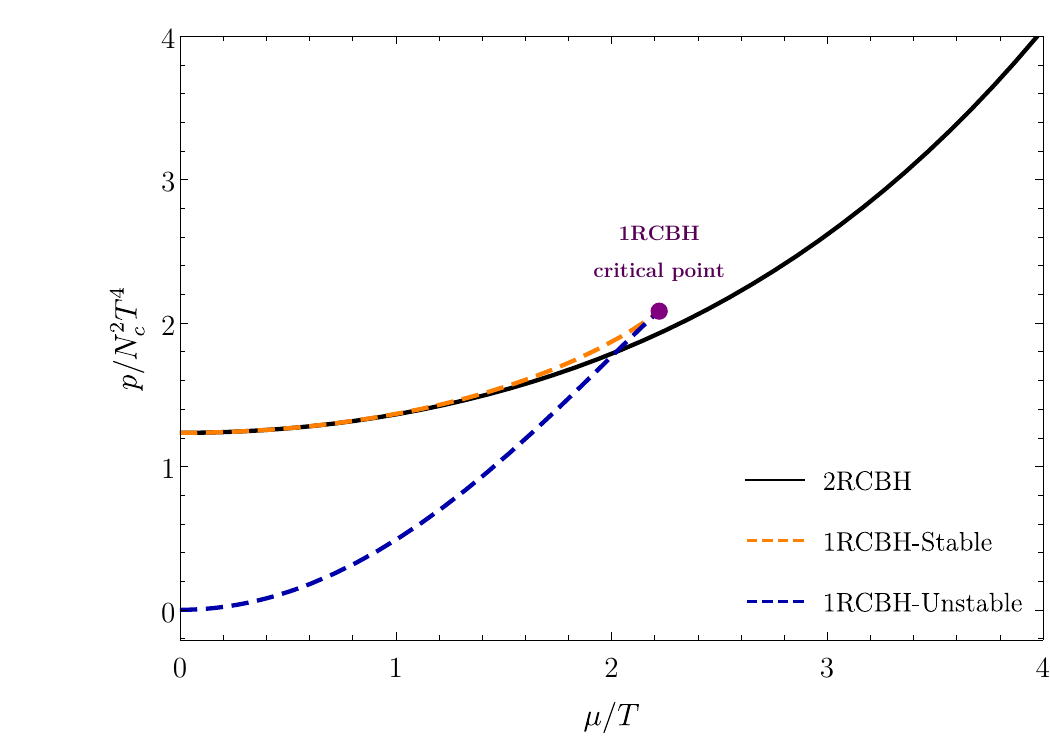}\label{fig:pThermo}}
% \subfigure[Scalar condensate]{\includegraphics[width=0.475\linewidth]{Thermo-O.pdf}\label{fig:OThermo}}
% \subfigure[Specific heat]{\includegraphics[width=0.475\linewidth]{Thermo-C.pdf}\label{fig:CThermo}}
\subfigure[R-charge susceptibility]{\includegraphics[width=0.475\linewidth]{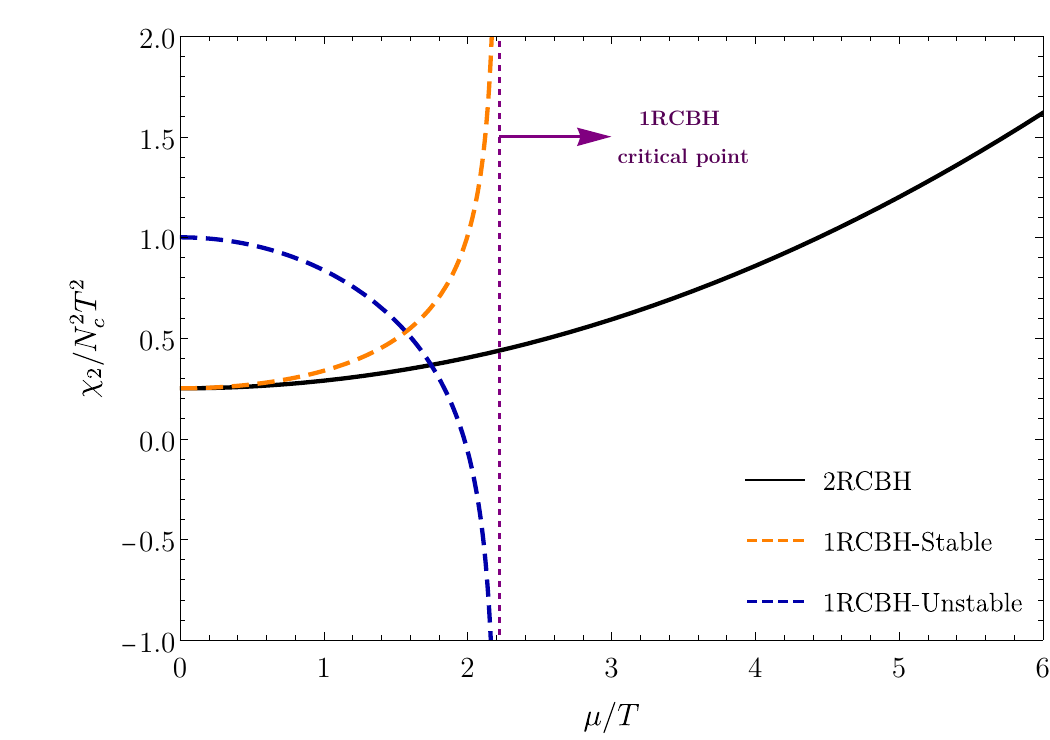}\label{fig:chi2Thermo}}
\caption{Entropy density (a), R-charge density (b), pressure (c), and R-charge susceptibility (d) for the 2RCBH and 1RCBH models in thermodynamic equilibrium.}
\label{fig:Thermo}
\end{figure*}

Due to the conformal symmetry of the holographic 2RCBH and 1RCBH models, the corresponding boundary gauge theory plasmas in thermodynamic equilibrium depend solely on the dimensionless ratio $\mu /T$, which by virtue of Eqs.~\eqref{eq:Tmu} can be written in terms of the bulk dimensionless ratio $Q/\tilde{r}_H$ as follows,
\begin{align}
&\text{1RCBH model:} &&\text{2RCBH model:}\nonumber\\
&\frac{Q}{\tilde{r}_H}=\sqrt{2}\left(\frac{1\pm\sqrt{1-\left(\frac{\mu/T}{\pi/\sqrt{2}}\right)^2}}{\frac{\mu/T}{\pi/\sqrt{2}}}\right), &&\frac{Q}{\tilde{r}_H}=\frac{1}{\pi\sqrt{2}}\frac{\mu}{T }\label{eq:Qoverr}.
\end{align}
From Eq.~\eqref{eq:Qoverr}, the non-negativity of $Q/\tilde{r}_H$ implies a finite interval $\mu/T\in [0,\pi/\sqrt2]$ for the phase diagram of the 1RCBH model, whereas for the 2RCBH model one has $\mu/T\in[0,\infty)$. In the case of the 1RCBH model, the $\pm$ signs in Eq.~\eqref{eq:Qoverr} denote the existence of two distinct values of $Q/\tilde{r}_H$ for each value of $\mu/T$. The lower sign corresponds to the thermodynamically stable branch of black hole solutions with $Q/\tilde{r}_{H} \in [0,\sqrt{2}]$, while the upper sign is associated with the thermodynamically unstable branch of black hole solutions with $Q/\tilde{r}_{H} \in [\sqrt{2},\infty)$. Both branches of solutions for the 1RCBH model meet at the critical point $Q/\tilde{r}_H=\sqrt{2}$, corresponding to the end point of its phase diagram at $\mu/T=\pi/\sqrt{2}$. In contrast, the 2RCBH model features a single branch of equilibrium black hole solutions and no critical point. Furthermore, for $Q/\tilde{r}_H\to 0\Rightarrow\mu/T \to 0$, the AdS$_5$-Schwarzschild solution, holographically dual to the purely thermal SYM plasma with zero R-charge density, is recovered as a particular limit in both the 2RCBH model and within the stable branch of solutions of the 1RCBH model~\cite{DeWolfe:2011ts,DeWolfe:2012uv,Finazzo:2016psx,Critelli:2017euk,deOliveira:2024bgh}.\footnote{In the 1RCBH model, the limit $\mu/T\to 0$ can be also achieved within the unstable branch of solutions by taking $Q/\tilde{r}_H\to\infty$: this particular solution is not a black hole, corresponding instead to a charged supersymmetric BPS solution called `superstar'~\cite{Myers:2001aq}.}

According to the holographic dictionary, the thermodynamic equilibrium entropy of the dual QFT is identified with the Bekenstein-Hawking entropy of the bulk black hole event horizon~\cite{Bekenstein:1973ur,Hawking:1975vcx},\footnote{We make use of the following top-down holographic relation, which is valid for SYM plasmas~\cite{Natsuume:2014sfa,Gubser:1996de}: $1/\kappa_5^2\equiv 1/8\pi G_5=N_c^2/4\pi^2L^3=N_c^2/4\pi^2$ (in units of $L=1$), where $N_c$ is the number of color charges at the boundary QFT.}
\begin{align}
S_H&=\frac{A_H}{4G_5}= \frac{N_c^2}{2\pi}\int_{\mathcal{M}_H}d^3x \sqrt{|\gamma_H|} \nonumber\\
&= \frac{N_c^2}{2\pi} g_{xx}^{3/2}(\tilde{r}_H)\int_{\mathcal{M}_H}d^3x = \frac{N_c^2}{2\pi} g_{xx}^{3/2}(\tilde{r}_H) V_H,
\label{eq:sBH}
\end{align}
where $A_H$ is the hyperarea of the black brane event horizon in the bulk, being $\gamma_H$ the determinant of the induced metric at the horizon and $V_H$ its spatial volume. Consequently, the entropy density, $s\equiv S_H/V_H$, for the 2RCBH and 1RCBH models in thermodynamic equilibrium is given by~\cite{DeWolfe:2011ts,DeWolfe:2012uv,Finazzo:2016psx,Critelli:2017euk,deOliveira:2024bgh},
\begin{subequations}
\label{eq:seq}
\begin{align}
&\text{1RCBH model:} \nonumber\\
&\frac{s_{\text{eq}}}{N_c^2 T^3}=\frac{\pi^2}{16}\left[3\pm \sqrt{1-\left(\frac{\mu/T}{\pi/\sqrt{2}}\right)^2}\right]^2\nonumber\\
&\hspace{1.7cm}\times\left[1\mp \sqrt{1-\left(\frac{\mu/T}{\pi/\sqrt{2}}\right)^2}\right],\\
& \nonumber\\
&\text{2RCBH model:}\nonumber\\
&\frac{s_{\text{eq}}}{N_c^2 T^3}=\frac{\pi^2}{2}\left[1+\frac{(\mu/T)^2}{2\pi^2}\right].
\end{align}
\end{subequations}
The $U(1)$ R-charge density of the dual QFT plasma can be obtained in terms of the on-shell action by the formula, $\rho_c =\lim_{\tilde{r}\to \infty}\delta S/\delta \Phi'$, which in thermodynamic equilibrium gives~\cite{DeWolfe:2011ts,DeWolfe:2012uv,Finazzo:2016psx,Critelli:2017euk,deOliveira:2024bgh},
\begin{subequations}
\label{eq:rho12}
\begin{align}
&\text{1RCBH model:}\nonumber\\
&\frac{\rho_{c,{\text{eq}}}}{N_c^2 T^3}=\frac{\mu/T}{16}\left[3\pm \sqrt{1-\left(\frac{\mu/T}{\pi/\sqrt{2}}\right)^2}\right]^2,\\
&\nonumber\\
&\text{2RCBH model:}\nonumber\\
&\frac{\rho_{c,{\text{eq}}}}{N_c^2 T^3}=\frac{\mu/T}{4}\left[1+\frac{(\mu/T)^2}{2\pi^2}\right].
\end{align}
\end{subequations}

The thermodynamic equilibrium pressure of both R-charged plasmas can be obtained by substituting~\eqref{eq:seq} or~\eqref{eq:rho12} into the Gibbs-Duhem thermodynamic relation, $dp=s dT+\rho d\mu$. As a result, one obtains~\cite{DeWolfe:2011ts}\cite{DeWolfe:2012uv}\cite{Finazzo:2016psx}\cite{Critelli:2017euk}\cite{deOliveira:2024bgh},  
\begin{subequations}
\label{eq:p12}
\begin{align}
&\text{1RCBH model:}\nonumber\\
& \frac{p_{\text{eq}}}{N_c^2 T^4}=\frac{\pi^2}{128}\left[3\pm \sqrt{1-\left(\frac{\mu/T}{\pi/\sqrt{2}}\right)^2}\right]^3\nonumber\\
&\hspace{2cm}\times\left[1\mp \sqrt{1-\left(\frac{\mu/T}{\pi/\sqrt{2}}\right)^2}\right],\\
&\nonumber\\
&\text{2RCBH model:}\nonumber\\
&\frac{p_{\text{eq}}}{N_c^2 T^4}=\frac{\pi^2}{8}\left[1+\frac{(\mu/T)^2}{2\pi^2}\right]^2.
\end{align}
\end{subequations}
Finally, using Eq.~\eqref{eq:p12}, one can calculate the specific heat at fixed chemical potential~\cite{DeWolfe:2010he}, $C_\mu=T(\partial^2p/\partial T^2)_\mu=T(\partial s/\partial T)_\mu$, and the $n$th order R-charge susceptibility, $\chi_n=(\partial^n p/\partial \mu^n)_T=(\partial^{n-1}\rho/\partial \mu^{n-1})_T$. Plots for the normalized equilibrium entropy density, $U(1)$ R-charge density, pressure, and the second order R-charge susceptibility are displayed in Fig.~\ref{fig:Thermo}. There, solid black lines represent the 2RCBH thermodynamic observables, while blue and orange dashed lines depict, respectively, the thermodynamic stable and unstable branches of solutions of the 1RCBH model. The divergence of the R-charge susceptibility, a second order derivative of the pressure, confirms that the critical point of the 1RCBH phase diagram is a point of second order phase transition.

\section{Holographic Bjorken Flow}
\label{sec:sec2}

\subsection{Viscous Relativistic Hydrodynamics with Bjorken Flow Symmetry}

In this section we discuss the relativistic hydrodynamic description for the 1RCBH and the 2RCBH fluids when constrained by the Bjorken flow symmetry: a system invariant under longitudinal boosts along some direction, which we take to be the $z$-axis, and $O(2)$ rotations around the transverse $xy$-planes. The former symmetry implies that the fluid looks the same to all observers moving at different constant velocities along the $z$-axis. This description is compatible with a \textit{uniform expansion}, where any two fluid elements move away from one another with a relative velocity directly proportional to the distance separating them. In the case of the Bjorken flow, this fluid expansion is governed by a relative velocity between fluid elements of magnitude $z/t$.\footnote{We can demonstrate this by considering an observer $\mathcal{O}'$ moving with velocity $V$ relative to the initial frame. The velocity of a fluid element in this new frame, $v_z'$, is given by the relativistic velocity-addition formula. As shown below, this formula is naturally satisfied if we define velocity as position over time in the new frame using the Lorentz-transformed coordinates ($z'=\gamma(z-Vt)$ and $t'=\gamma(t-Vz)$),
\begin{equation}
    v_z'=\frac{v_z-V_z}{1-v_zV_z}=\frac{z/t-V_z}{1-zV_z/t}=\frac{\gamma(z-Vt)}{\gamma(t-zV_z)}=\frac{z'}{t'}.
\end{equation} 
This confirms that an observer in any boosted frame finds the same simple rule governing the expansion: a fluid element's velocity is its position divided by time in that observer's own reference frame ($v_z' = z'/t'$). This boost-invariance is the defining characteristic of Bjorken flow, implying that no single inertial frame is preferred in its description.}

Since (i) the fluid velocity $z/t$ is invariant under longitudinal boosts in the $z$-direction; and (ii) boosts in the $z$-direction amount to shifts in the rapidity $\xi=\operatorname{arctanh}(z/t)$, one concludes that the fluid's properties cannot depend on $\xi$. Furthermore, due to time dilation, the evolution of the system is best described by the proper time $\tau\equiv \sqrt{t^2-z^2}$ as the ``internal clock'' for all fluid elements. Consequently, in the case of the Bjorken flow symmetry, it is more natural to adopt the Milne coordinates $(\tau,x,y,\xi)$, where we eliminate the dependence on the $z$-coordinate in the dynamical description of the system, whose evolution in these coordinates depends only on the proper time $\tau$. In Milne coordinates, the Minkowski metric takes the form,
\begin{equation}
\label{eq:metricbjorken}
    ds_{\text{Mink}_4}^2=\eta_{\mu\nu} dx^\mu dx^\nu=-d\tau^2+dx^2 + dy^2+\tau^2 d\xi^2.
\end{equation}

As an effective theory, relativistic hydrodynamics is built upon conservation laws, constitutive relations, and low energy expansions~\cite{csernai1994,Romatschke:2017ejr}. Since both the 1RCBH and 2RCBH holographic models represent, in the dual QFT, conformal fluids with finite $U(1)$ R-charge density, we will now consider the relativistic hydrodynamic description originated from the local conservation of energy-momentum and R-charge density,
\begin{equation}
\label{eq:RelHydroConserved}
\nabla^\mu T_{\mu\nu}=0, \qquad \text{and} \qquad  
\nabla_\mu J^{\mu}=0,    
\end{equation}
where $T_{\mu\nu}$ is the energy-momentum tensor, $J^{\mu}$ represent the conserved R-charge current, and $\nabla_\mu$ denotes the geometric covariant derivative. As the Bjorken expanding system is longitudinally boost-invariant, and homogeneous in the transverse plane, physical observables in such a setup, when written in terms of the Milne coordinates, depend only on the proper time $\tau$.

The Bjorken symmetry significantly simplifies the equations of motion of the system. By evaluating the covariant derivatives using the nonzero Christoffel symbols of the metric~\eqref{eq:metricbjorken}, $\Gamma^\tau_{\xi\xi} = \tau$ and $\Gamma^\xi_{\tau\xi} = 1/\tau$, the conservation laws in Eqs.~\eqref{eq:RelHydroConserved} reduce to the following set of coupled differential equations,
\begin{subequations}
\label{eq:conservation_explicit}
\begin{align}
    \nabla_\mu T^{\mu\tau} &= \partial_\mu T^{\mu\tau} + \Gamma^\mu_{\mu\lambda} T^{\lambda\tau} + \Gamma^\tau_{\mu\lambda} T^{\mu\lambda}\nonumber\\ 
    &= \partial_\tau T^{\tau\tau} + \frac{1}{\tau}T^{\tau\tau} + \tau T^{\xi\xi} = 0,\label{eq:RelHydroT} \\
    \label{eq:RelHydroJ}
    \nabla_\mu J^{\mu} &= \partial_{\mu} J^{\mu} + \Gamma^{\mu}_{\mu\lambda}J^{\lambda} = \partial_{\tau} J^{\tau} + \frac{1}{\tau}J^{\tau} = 0.
\end{align}
\end{subequations}

To proceed, we must specify the constitutive relations for $T^{\mu\nu}$ and $J^\mu$. For the charge current at leading order, one considers the ideal fluid description, where $J^{\mu}=\rho_c u^{\mu}$, with $\rho_c(\tau)$ being the local charge density and $u^\mu$ denoting the local fluid's four-velocity. In its local rest frame, where $u^\mu \equiv (1,0,0,0)$, the fluid charge current is thus written as $J^{\mu}=\left(\rho_c(\tau),0,0,0\right)$. Substituting this into Eq.~\eqref{eq:RelHydroJ} immediately yields the evolution of the charge density,
\begin{equation}
\label{eq:rhoconser}
    \partial_\tau \rho_c + \frac{\rho_c}{\tau} = 0 \quad \implies \quad \rho_c(\tau) = \frac{\rho_0}{\tau},
\end{equation}
where $\rho_0$ is an integration constant which must be chosen as part of the set of initial conditions of the Bjorken flow dynamics of the fluid.

For the energy-momentum tensor, we consider first order relativistic viscous hydrodynamics, where we have the relativistic Navier-Stokes constitutive relation,
\begin{equation}
\label{eq:constitutive-relation}
    T^{\mu\nu}=(\varepsilon+p)u^\mu u^\nu+pg^{\mu\nu}+\pi^{\mu\nu}+\Delta^{\mu\nu}\Pi,
\end{equation}
with $\varepsilon$ and $p$ being the local equilibrium density and pressure for the fluid in its local rest frame. Moreover, $\pi^{\mu\nu}=-2\eta \sigma^{\mu\nu}$ is the shear stress tensor written in terms of the fluid's shear viscosity $\eta$ and the shear tensor $\sigma^{\mu\nu}$,
\begin{equation}
    \sigma^{\mu\nu}=\frac{1}{2}\Delta^{\mu\alpha}\Delta^{\nu\beta}\left(\nabla_\alpha u_{\beta}+\nabla_\beta u_{\alpha}\right)-\frac{1}{3}\Delta^{\mu\nu}\nabla_\alpha u^{\alpha},
\end{equation}
whereas $\Pi=-\zeta  \Delta_{\lambda \sigma}\nabla^{\sigma} u^{\lambda}$ is the bulk viscous pressure written in terms of the fluid's bulk viscosity $\zeta$. Here $\Delta^{\mu\nu}=g^{\mu\nu}+u^{\mu}u^{\nu}$ is a space-like projection operator.

We note that since,
\begin{align}
u^\beta u_\beta = -1 \Rightarrow \nabla_\alpha (u^\beta u_\beta) = 0 \Rightarrow u^\beta \nabla_\alpha u_\beta = 0,    
\end{align}
it follows that,
\begin{align}
\Delta^{\alpha\beta} \nabla_\alpha u_\beta = \nabla_\alpha u^\alpha + u^\alpha u^\beta \nabla_\alpha u_\beta = \nabla_\alpha u^\alpha.  
\end{align}
Therefore, the trace of the shear tensor vanishes,
\begin{align}
\sigma^\mu\,_\mu = g_{\mu\nu} \sigma^{\mu\nu} &= \frac{1}{2} \underbrace{\Delta^{\mu\alpha}\Delta_\mu\,^\beta}_{=\Delta^{\alpha\beta}} (\nabla_\alpha u_\beta + \nabla_\beta u_\alpha)\nonumber\\
&- \frac{1}{3} \underbrace{g_{\mu\nu}\Delta^{\mu\nu}}_{=\delta^\mu_\mu+u_\mu u^\mu=4-1}\Delta^{\alpha\beta} \nabla_\alpha u_\beta\nonumber\\
&= \Delta^{\alpha\beta} \nabla_\alpha u_\beta - \Delta^{\alpha\beta} \nabla_\alpha u_\beta\nonumber\\
&= 0,    
\end{align}
where in passing from the 1st to the 2nd line of the above equation, we used the fact that, since the projector $\Delta^{\alpha\beta}$ is a symmetric tensor, we may write, $\Delta^{\alpha\beta} (\nabla_\alpha u_\beta + \nabla_\beta u_\alpha) = \Delta^{\alpha\beta} \nabla_\alpha u_\beta + \Delta^{\alpha\beta} \nabla_\beta u_\alpha = (\Delta^{\alpha\beta} + \Delta^{\beta\alpha}) \nabla_\alpha u_\beta = 2\Delta^{\alpha\beta} \nabla_\alpha u_\beta$.

The components of the energy-momentum tensor for the Bjorken flow are identified with the longitudinal pressure $p_L$ along the expansion axis of the fluid, and the transverse pressure $p_T$. Considering the specific expressions for the case of the Bjorken flow expansion, where $\nabla_{\alpha}u^{\alpha}=1/\tau$, $\sigma^{xx}=\sigma^{yy}=-1/(3\tau)$, and $\sigma^{\xi\xi}=2/\tau^3$, one can calculate the pressures through the spatial components of $T^{\mu\nu}$,
\begin{align}
    p_T(\tau) &\equiv T^x_x = \eta_{xx} T^{xx} = T^{xx} = p(\tau)+\frac{2\eta}{3\tau}+\Pi(\tau), \label{eq:pT-hydro} \\
    p_L(\tau) &\equiv T_\xi^\xi = \eta_{\xi\xi} T^{\xi\xi} = \tau^2 T^{\xi\xi} = p(\tau) - \frac{4\eta}{3\tau}+\Pi(\tau), \label{eq:pL-hydro}
\end{align}
where $\Pi(\tau)=-\zeta/\tau$. By plugging into Eq.~\eqref{eq:RelHydroT} the energy-momentum components $T^{\tau\tau}=\varepsilon -\eta \sigma^{\tau\tau}$ and $T^{\xi\xi}=p_L/\tau^2 $, which were calculated from Eq.~\eqref{eq:constitutive-relation}, we obtain,
\begin{equation}
\label{eq:Hydrodynamic}
    \partial_\tau \varepsilon(\tau) + \frac{\varepsilon(\tau)+p_L(\tau)}{\tau} = 0,
\end{equation}
where $\sigma^{\tau\tau}$ was set to zero since, by definition, the shear tensor is orthogonal to the fluid's four-velocity, i.e., $u_\nu \sigma^{\mu\nu}=0$.

Since both the 1RCBH and 2RCBH models are conformal field theories, the trace of the energy-momentum tensor must vanish, $T^\mu_{~\mu} \equiv 0$. From Eqs.~\eqref{eq:metricbjorken},~\eqref{eq:constitutive-relation},~\eqref{eq:pT-hydro}, and~\eqref{eq:pL-hydro}, one gets,
\begin{equation}
\label{eq:traceTmunu}
    T^{\mu}_{~\mu}=\eta_{\mu\nu} T^{\mu\nu}=-\varepsilon+3p+3\Pi = -\varepsilon+p_L+2p_T.
\end{equation}
Conformal symmetry also fixes the thermodynamic equation of state to satisfy $p=\varepsilon/3$. Imposing the traceless condition on the expression above therefore requires that the bulk viscous pressure $\Pi$, and consequently the bulk viscosity $\zeta$, must vanish for conformal fluids as the 1RCBH and 2RCBH models.

Using Eq.~\eqref{eq:Hydrodynamic} together with the traceless condition \eqref{eq:traceTmunu}, one can rewrite the longitudinal and transverse pressures solely in terms of the fluid's energy density,
\begin{subequations}
\label{eq:pLpT}
    \begin{align}
    p_L(\tau)&=-\varepsilon(\tau)-\tau\partial_\tau\varepsilon(\tau),\\
    p_T(\tau)&=\varepsilon(\tau)+\frac{\tau}{2}\partial_\tau\varepsilon(\tau).
\end{align}
\end{subequations}
Finally, the pressure anisotropy is defined as the difference between the transverse and longitudinal pressures, $\Delta p\equiv p_T-p_L$. Thus, from Eqs.~\eqref{eq:pLpT}, one obtains the following dimensionless ratio,
\begin{equation}
\label{eq:pressure-anisotropy-EOM}
    \frac{\Delta p}{\varepsilon}\equiv\frac{p_T-p_L}{\varepsilon}=2+\frac{3}{2}\tau\frac{\partial_\tau\varepsilon}{\varepsilon}.
\end{equation}

Returning to the discussion of the hydrodynamic equation for the energy density $\varepsilon(\tau)$, by substituting Eq.~\eqref{eq:pL-hydro} (with $\zeta=0$) into Eq.~\eqref{eq:Hydrodynamic}, we get,
\begin{equation}
\label{eq:final-eom-fluid}
    \partial_\tau \varepsilon + \frac{4}{3}\frac{\varepsilon}{\tau} = \frac{4\eta}{3\tau^2}.
\end{equation}
Since both the 1RCBH and 2RCBH plasmas are holographically described in the bulk by effective gravitational actions with at most two derivatives of the metric, and since at asymptotically late times the medium equilibrates into homogeneous and isotropic states, the famous Kovtun-Son-Starinets (KSS) result~\cite{Kovtun:2004de} for the shear viscosity over entropy density ratio guarantees that for both holographic fluids, $\eta/s=1/4\pi\Rightarrow 4\eta=s/\pi$. Moreover, using the thermodynamic relation $s=\varepsilon/T+p/T-\mu\rho_c/T$ together with the conformally symmetric equation of state $\varepsilon=3p$, we rewrite Eq.~\eqref{eq:final-eom-fluid} as,
\begin{equation}
\label{eq:eom-fluid-bjorken}
    \partial_\tau \varepsilon + \frac{4}{3}\frac{\varepsilon}{\tau} = \frac{1}{3\pi \tau^2}\left(\frac{4}{3}\frac{\varepsilon}{T}-\frac{\mu\rho_c}{T}\right).
\end{equation}

Solutions to Eq.~\eqref{eq:eom-fluid-bjorken} can be found by first assuming that $x\equiv \mu/T$ is kept fixed (as is the case for the late time equilibrium behavior), and second by considering the 1RCBH and 2RCBH thermodynamic expressions for the R-charge density $\rho_c$ and temperature $T$ obtained from Eqs.~\eqref{eq:rho12} and \eqref{eq:p12} (with the conformal result $\varepsilon=3p$),
\begin{align}
    \rho_c=\kappa_5^{-2}h(\mu/T) T^3, \quad \text{and}\quad T=\kappa_5^{1/2}f(\mu/T)^{-1/4}\varepsilon^{1/4},
\label{eq:30}
\end{align}
where,
\begin{subequations}
\begin{align}
    &\text{1RCBH model:} \nonumber \\
    &f(x)=\frac{3\pi^4}{32}\left[3-\sqrt{1-\left(\frac{x}{\pi/\sqrt{2}}\right)^2}\right]^3\nonumber\\
    &\hspace{1.5cm}\times\left[1+\sqrt{1-\left(\frac{x}{\pi/\sqrt{2}}\right)^2}\right], \label{eq:31}\\
    &h(x)=\frac{\pi^2}{4}\, x \left(3-\sqrt{1-\left(\frac{x}{\pi/\sqrt{2}}\right)^2}\right)^2, \label{eq:32}\\
    &\nonumber\\
    &\text{2RCBH model:} \nonumber \\
    & f(x)=\frac{3\pi^4}{2}\left(1 + \frac{x^2}{2\pi^2}\right)^2,\\
    & h(x)=\pi^2 x\left(1 + \frac{x^2}{2\pi^2}\right).
\end{align}
\end{subequations}
In terms of the above expressions, we rewrite~\eqref{eq:eom-fluid-bjorken} as follows,
\begin{equation}
\label{eq:final-eom-fluid-bjorken}
    \partial_\tau \varepsilon(\tau) + \frac{4}{3\tau} \varepsilon(\tau) =\frac{\mathcal{A}(x)}{\tau^2}\varepsilon^{3/4}(\tau),
\end{equation}
where
\begin{equation}
    \mathcal{A}(x)=\frac{1}{9 \pi \kappa_5^{1/2} }\frac{4f(x)-3xh(x)}{ f(x)^{3/4}}.
\end{equation}
The general and exact solution to Eq.~\eqref{eq:final-eom-fluid-bjorken}, which is a Bernoulli differential equation, is given by,
\begin{align}
    \varepsilon(\tau)&=\frac{c_0}{\tau^{4/3}}\left(1-\frac{3\mathcal{A}(x)}{8 c_0^{1/4}\tau^{2/3}}\right)^4\nonumber\\
    &=\frac{81 \mathcal{A}(x)^4}{4096 \tau ^4}-\frac{27 \mathcal{A}(x)^3 c_0^{1/4}}{128 \tau ^{10/3}}+\frac{27 \mathcal{A}(x)^2 c_0^{2/4}}{32 \tau ^{8/3}}\nonumber\\
    &-\frac{3 \mathcal{A}(x) c_0^{3/4}}{2 \tau^2}+\frac{c_0}{\tau ^{4/3}},\label{eq:solution-eom-density}
\end{align}
where $c_0$ is an integration constant. In the long time regime, we may truncate~\eqref{eq:solution-eom-density} as follows,
\begin{equation}
\label{eq:energytau}
    \varepsilon(\tau)=\frac{c_0}{\tau^{4/3}}+\frac{1}{6\pi \kappa_5^{1/2}}\frac{c_0^{3/4}}{\tau^2}\frac{3xh(x)-4f(x)}{ f(x)^{3/4}}+\mathcal{O}(\tau^{-8/3}).
\end{equation}
Finally, by fixing the value of the constant $c_0$ in such a way that the dominant term in the long time regime of~\eqref{eq:energytau} matches the corresponding $\mathcal{N}=4$ SYM result for the energy density of the plasma in the limit of zero charge density, $x\equiv\mu/T\to 0$, one gets, $c_0= 3\pi^3 \Lambda^{8/3}/2\kappa_5^2$, so that~\cite{Critelli:2018osu},
\begin{align}
    \hat{\varepsilon}(\tau)&\equiv\kappa_5^2\varepsilon(\tau)=\frac{3 \pi ^4 \Lambda ^{8/3}}{2 }\frac{1}{\tau^{4/3}}\nonumber\\
    &+\frac{\pi^2\Lambda^2}{2^{7/4}3^{1/4}}\frac{3xh(x)-4f(x)}{ f(x)^{3/4}}\frac{1}{\tau^2}+\mathcal{O}(\tau^{-8/3}),
\label{eq:energyNS}
\end{align}
where $\Lambda$ is an energy scale related to the initial energy density of the plasma~\cite{Chesler:2009cy,Janik:2005zt}.\footnote{Notice that the dominant term for the Bjorken flow, Navier-Stokes hydrodynamic energy density~\eqref{eq:energyNS} in the $\tau\to\infty$ limit is: $\varepsilon(\tau\to\infty)\to 3\pi^4\Lambda^{8/3}/2\kappa_5^2\tau^{4/3}=3\pi^2 N_c^2\Lambda^{8/3}/8\tau^{4/3}=3\pi^2 N_c^2 T^4/8$, which does coincide with the equilibrium SYM result~\cite{Baier:2007ix,Finazzo:2014cna}; we used in this manipulation the fact that at leading order in the Bjorken flow, $T(\tau)=\Lambda^{2/3}/\tau^{1/3}$~\cite{Chesler:2009cy,Janik:2005zt}.} For any quantity $X$, we define here the notation $\hat{X}\equiv \kappa_5^2 X = 4\pi^2 X/N_c^2$.

%%%%%%%%%%%%%%%%%%%%%%%%%%%%%%%%%%%%%%%%%%%%%%%%

\subsection{Ansatze for the Bulk Fields and the Main Algorithm}
\label{sec:mainalgo}

In terms of five-dimensional generalized infalling Eddington-Finkelstein coordinates, the ansatze for the bulk EMD fields compatible with the Bjorken symmetry can be written as~\cite{Chesler:2009cy,Critelli:2018osu,Rougemont:2022piu},
\begin{subequations}
\label{eq:BFDansatze}
\begin{align}
ds^2 &= 2 d\tau [dr - A(\tau, r)d\tau] + \Sigma(\tau, r)^2\nonumber\\
&\times\left[e^{-2B(\tau, r)} d\xi^2 + e^{B(\tau, r)}(dx^2 + dy^2)\right], \\
A_\mu dx^\mu &= \Phi(\tau, r)d\tau, \\
\phi &= \phi(\tau, r),
\end{align}
\end{subequations}
with $r$ being the radial coordinate of the asymptotically AdS$_5$ spacetime, whose boundary is at $r\rightarrow\infty$, where $\tau$ becomes the usual proper time of the four-dimensional Bjorken flow for the boundary plasma.

By plugging~\eqref{eq:BFDansatze} into~\eqref{eq:EMDequations}, one obtains the following set of coupled partial differential equations (PDEs) in terms of five background functions to be determined, $\{A(\tau,r), \allowbreak \Sigma(\tau,r), \allowbreak B(\tau,r), \allowbreak \phi(\tau,r), \allowbreak \mathcal{E}(\tau,r)\equiv -\partial_r\Phi(\tau,r)\}$. These equations of motion are determined by five dynamical equations~\cite{Critelli:2018osu,Rougemont:2022piu},
\begin{subequations}
\label{eq:EOMs}
\begin{align}
\frac{3 \Sigma '}{\Sigma } + \frac{\partial_\phi f}{f}\phi ' + \frac{\mathcal{E}'}{\mathcal{E}} =0&,\label{eq:pdea}\\
4 \Sigma (d_+\phi)'+6 \phi'd_+\Sigma + 6 \Sigma'd_+\phi\hspace{1cm}&\nonumber\\
+\Sigma\mathcal{E}^2 \partial_\phi f - 2 \Sigma  \partial_\phi V =0&,\label{eq:pdeb}\\
(d_+\Sigma)' +\frac{2 \Sigma '}{\Sigma }d_+\Sigma+\frac{\Sigma}{12} \left(2V + f \mathcal{E}^2\right) =0&,\label{eq:pdec}\\
\Sigma \, (d_{+}B)'+\frac{3}{2}(B' d_+\Sigma + \Sigma'd_+ B) =0&,\label{eq:pded}\\
A'' + \frac{1}{12} \Big(18 B'd_{+}B-\frac{72 \Sigma'd_{+}\Sigma}{\Sigma^2}\hspace{1cm}& \nonumber\\
+ 6 \phi'd_{+}\phi-7 f \mathcal{E}^2-2 V\Big) =0&,\label{eq:pdee}
\end{align}
\end{subequations}
and three constraint equations~\cite{Critelli:2018osu,Rougemont:2022piu},
\begin{subequations}
\label{eq:constEOMs}
\begin{align}
\partial_v\mathcal{E}+A\mathcal{E}'+\left(3\frac{d_+\Sigma}{\Sigma}+\frac{\partial_\phi f}{f}d_+\phi\right)\mathcal{E} &=0,\label{eq:consta}\\
\Sigma'' + \frac{\Sigma}{6} \left(3 \left(B'\right)^2+\left(\phi '\right)^2\right) &=0,\label{eq:constb}\\
d_+(d_+\Sigma)+\frac{\Sigma}{2}(d_+B)^2-A'd_+\Sigma+\frac{\Sigma}{6}(d_+\phi)^2 &=0,\label{eq:constc}
\end{align}
\end{subequations}
where the notation $X'\equiv \partial_r X$ represent a directional derivative along infalling radial null geodesics, and $d_+X\equiv[\partial_\tau+A(\tau,r)\partial_r]X$ is the directional derivative along outgoing radial null geodesics.

The full set of Eqs.~\eqref{eq:EOMs} and \eqref{eq:constb} constitute a nested set of equations of motion which can be systematically integrated using numerical techniques, while the constraint Eqs. \eqref{eq:consta} and \eqref{eq:constc} can be used to monitor the accuracy of the numerical solutions so obtained. Before solving these PDEs, we first need to determine the boundary conditions required for the solutions to correctly describe the dual QFT dynamics at the boundary. This can be done by finding the near-boundary ultraviolet (UV) expansions of the bulk fields which are consistent with the boundary conditions associated with the Bjorken flow.

Since both the 1RCBH and 2RCBH background geometries are asymptotically AdS$_5$, in the generalized infalling Eddington-Finkelstein coordinates, the bulk metric must recover the following form close to the boundary,
\begin{equation}
\label{eq:limds2}
    \lim_{r\to \infty} ds^2 = ds_{\text{AdS}_5}^2=2d\tau dr + r^2 ds^2_{\text{Mink}_4},
\end{equation}
where $r^2$ is the global AdS$_5$ conformal factor. Therefore, by comparing the near-boundary form~\eqref{eq:limds2} of Eq.~\eqref{eq:BFDansatze} with the Minkowski metric~\eqref{eq:metricbjorken}, one obtains the following boundary conditions for the metric coefficients,
\begin{subequations}
\label{eq:metricbc}
\begin{align}
&A(\tau,r\to\infty)\sim\frac{r^2}{2},\\
&B(\tau,r\to\infty)\sim-\frac{2}{3}\ln(\tau), \\ &\Sigma(\tau,r\to\infty)\sim\tau^{1/3}r.
\end{align}    
\end{subequations}

According to the holographic dictionary, the mass $m$ of a scalar field in an asymptotically AdS$_5$ spacetime is related to the conformal dimension $\Delta$ of its dual QFT operator at the boundary by the formula, $m^2L^2=\Delta(\Delta-4)$. Since the expansion of the Dilaton potential from Eqs.~\eqref{eq:Vphi1R} and \eqref{eq:Vphi2R} in powers of $\phi$, $V(\phi)=-12/L^2-2\phi^2/L^2+\dots$, reveals a tachyonic mass squared $m^2=-4/L^2$, the resulting conformal dimension for both the 2RCBH and 1RCBH models can be identified as $\Delta=2$. Therefore, it follows that the leading contribution for the dilaton field near the boundary is given by,
\begin{align}
\phi(\tau,r\to\infty)\sim \frac{ \phi_{4-\Delta_\phi}(\tau)}{r^{4-\Delta_\phi}} = \frac{\phi_2(\tau)}{r^2}.
\label{eq:dilatonbc}
\end{align}
Finally, the nontrivial component of the Maxwell field is connected to the R-charge chemical potential of the quantum fluid at the boundary by the relation,
\begin{align}
\Phi(\tau\to\infty,r\to\infty)=\mu,
\label{eq:Maxbc}
\end{align}
where the $\tau\to\infty$ limit is also considered in the above identification of the equilibrium chemical potential $\mu$ of the boundary QFT, since only at asymptotic times the system relaxes to equilibrium. More precisely, in the case of the Bjorken flow, as it was clear, e.g., from the hydrodynamic expressions~\eqref{eq:rhoconser} and~\eqref{eq:energyNS}, which must be attained at late times in the dynamical evolution of the system, both the charge density and the energy density of the medium vanish in the infinite time limit. What we call here as ``equilibrium'' corresponds to the long time regime of the system where equilibrium thermodynamic relations like those in Eqs.~\eqref{eq:30} ---~\eqref{eq:32} become valid (within some small numerical tolerance). At early times, when the system is far-from-equilibrium, not even hydrodynamic relations are valid. Characteristic time scales when hydrodynamics becomes a good approximation to the dynamics of the system are called hydrodynamization times, which are generally different for different physical observables~\cite{Attems:2017zam,Rougemont:2021gjm}.

Considering Eqs.~\eqref{eq:metricbc},~\eqref{eq:dilatonbc} and~\eqref{eq:Maxbc} as the leading order terms for near-boundary UV expansions of the bulk fields, one can construct the following series expansion in inverse powers of the holographic radial coordinate $r$~\cite{Bianchi:2001kw,Critelli:2017euk,Critelli:2018osu},
\begin{subequations}
\label{eq:nearexpansion}
\begin{align}
A(\tau,r) &= \frac{1}{2}[r+\lambda(\tau)]^2 -\partial_\tau\lambda(\tau) +\sum_{n=1}^{\infty} \frac{a_n(\tau)}{r^n}, \label{eq:expA}\\
B(\tau,r) &= -\frac{2}{3}\ln(\tau) +\sum_{n=1}^{\infty} \frac{b_n(\tau)}{r^n}, \label{eq:expB}\\
\Sigma(\tau,r) &= \tau^{1/3}[r+\lambda(\tau)] +\sum_{n=0}^{\infty} \frac{s_n(\tau)}{r^n}, \label{eq:expS}\\
\phi(\tau,r) &= \sum_{n=2}^{\infty} \frac{\phi_n(\tau)}{r^n}, \label{eq:expDil}\\
\Phi(\tau,r) &= \Phi_0(\tau) + \sum_{n=2}^{\infty} \frac{\Phi_n(\tau)}{r^n}, \label{eq:expMax}
\end{align}
\end{subequations}
where $\lambda(\tau)$ is an arbitrary function of the proper time, and by which the radial coordinate $r$ can be additively shifted without changing the metric~\eqref{eq:BFDansatze}, which corresponds to a residual diffeomorphism invariance of the bulk theory~\cite{Chesler:2009cy,Chesler:2013lia}.

By substituting the above UV expansions into the EMD field equations, one obtains a set of algebraic equations that can be solved order by order near the boundary. By closely following the developments in Refs.~\cite{Critelli:2018osu,Rougemont:2022piu}, and working with expansions up to order $n=8$, one identifies four undetermined ultraviolet coefficients, namely $\{a_2(\tau),\phi_2(\tau),\Phi_0(\tau),\Phi_2(\tau)\}$. The values of $\{a_2(\tau),\phi_2(\tau)\}$ at the initial time slice $\tau_0$ constitute part of the initial data which must be chosen in the holographic Bjorken flow, as we are going to discuss in more detail in Section~\ref{sec:data_EC}. The remaining two coefficients are fixed by physical constraints: Eqs.~\eqref{eq:Maxbc} and~\eqref{eq:expMax} lead one to identify $\Phi_0(\tau)$ as the R-charge chemical potential $\mu$, while $\Phi_2(\tau)=-\rho_0/\tau$ sets the initial charge density via the choice of the parameter $\rho_0$ (which is part of the initial data of the holographic Bjorken flow). The latter relation can be deduced from the UV expansion of the bulk fields at order $n=6$, where one finds that $\Phi_2(\tau)=c/\tau$, with $c$ being a constant determined by comparing the expression obtained in Eq.~\eqref{eq:rhoconser}~for the R-charge density, $\hat{\rho}_c(\tau)\equiv\langle \hat{J}^{\tau}\rangle = \rho_0/\tau$, with the following formula fixed by the holographic renormalization procedure~\cite{Critelli:2017euk}, $\langle J^\tau\rangle=-\Phi_2(\tau)/\kappa_5^2$; these imply the aforementioned identification, $\Phi_2(\tau)=c/\tau=-\langle \hat{J}^\tau\rangle=-\rho_0/\tau$.

By integrating the Maxwell equation~\eqref{eq:pdea} using the Maxwell-dilaton coupling functions from Eqs.~\eqref{eq:fphi1R} and \eqref{eq:fphi2R}, one obtains $\mathcal{E}=-\Phi'=\Sigma^{-3}\,e^{\bar{c}}\,e^{C\phi}$, where $C=2\sqrt{2/3}$ for the 1RCBH model and $C=-\sqrt{2/3}$ for the 2RCBH model. The integration constant $\bar{c}$ is then determined by matching the solution's asymptotic behavior at the boundary $r\to\infty$. Comparing the near-boundary expansion, $\mathcal{E}(\tau,r\to\infty)\sim -\partial_r(\Phi_0+\Phi_2 r^{-2})=2\Phi_2 r^{-3}$, with the right-hand side's expansion, $e^{\bar{c}}\,\Sigma(\tau,r\to\infty)^{-3}\,e^{C\,\phi(\tau,r\to\infty)}\sim e^{\bar{c}}\,\tau^{-1}r^{-3}$, yields $e^{\bar{c}}=2\Phi_2(\tau)\tau=-2\rho_0$, therefore,
\begin{subequations}
\label{eq:MaxField2}
\begin{align}
&\text{1RCBH model:} &\nonumber\\
&\mathcal{E}(\tau,r)=-2\rho_0\,\Sigma(\tau,r)^{-3}\,e^{2\sqrt{2/3}\,\phi(\tau,r)},\\
&\nonumber\\
&\text{2RCBH model:}\nonumber\\
&\mathcal{E}(\tau,r)=-2\rho_0\,\Sigma(\tau,r)^{-3}\,e^{-\sqrt{2/3}\,\phi(\tau,r)}.
\end{align}
\end{subequations}
We see that the bulk `radial electric field' $\mathcal{E}=-\Phi'$ is determined once the metric coefficient $\Sigma$ and the dilaton field $\phi$ are known.

To numerically handle the holographic Bjorken flow of the 2RCBH and 1RCBH models, we employ here the so-called \textit{characteristic formulation of numerical general relativity}~\cite{Bondi:1962px,Sachs:1962wk}, as adequately adapted by Chesler and Yaffe to asymptotically AdS spacetimes in~\cite{Chesler:2008hg,Chesler:2009cy,Chesler:2013lia}. In the characteristic formulation, the bulk spacetime manifold is foliated in null hypersurfaces of constant $\tau$, which in the generalized EF coordinates of~\eqref{eq:BFDansatze} correspond to the paths of infalling radial null geodesics. In such a framework, one has: (i) a nested set of coupled PDEs which can be solved sequentially, (ii) dynamical equations which are first order in time derivatives.

Finally, we outline below the general steps followed in this work to solve the nested set of $1+1$ PDEs describing the Bjorken flow dynamics of both the 2RCBH and 1RCBH models. The numerical algorithm proceeds iteratively as follows~\cite{Rougemont:2022piu}:
\begin{enumerate}[i.]
    \item \textit{Setting the initial conditions:} At the initial time slice $\tau_0$, we choose the initial profiles for the metric anisotropy $B(\tau_0,r)$ and the dilaton field $\phi(\tau_0,r)$, besides the initial values for the charge density $\rho_0$ and the dynamical UV coefficient $a_2(\tau_0)$;\footnote{As detailed in Sec.~\ref{sec:apparent-hor}, the initial radial shift function $\lambda(\tau_0)$ must also be specified. Its subsequent time evolution is determined by requiring the apparent horizon to remain at a fixed radial coordinate \cite{Chesler:2013lia}.}

    \item \textit{Solving the constraints:} The Hamiltonian constraint, Eq.~\eqref{eq:constb}, is radially solved to find $\Sigma(\tau_0,r)$. This, in turn, determines $\mathcal{E}(\tau_0,r)$ via Eq.~\eqref{eq:MaxField2};

    \item \textit{Calculating the time derivatives:} Next, we radially solve in sequence the dynamical equations~\eqref{eq:pdec},~\eqref{eq:pded},~\eqref{eq:pdeb}, and~\eqref{eq:pdee}, which determine, respectively, the directional derivatives $d_+\Sigma(\tau_0,r)$, $d_+B(\tau_0,r)$, and $d_+\phi(\tau_0,r)$, as well as the metric function $A(\tau_0,r)$. Using the definition of the directional derivative along radial outgoing radial null geodesics, $d_+\equiv\partial_\tau+A(\tau,r)\partial_r$, we then compute the time derivatives $\partial_\tau B(\tau_0,r)$ and $\partial_\tau\phi(\tau_0,r)$. The derivative $\partial_\tau a_2(\tau_0)$ is also determined at this stage, as we shall discuss in Section~\ref{sec:apparent-hor};

    \item \textit{Evolving in time:} The complete set of initial data and their time derivatives allows us to evolve the system from $\tau_0$ to the next time slice, $\tau_0+\Delta\tau$. For this work, the radial integrations are performed using a pseudospectral method~\cite{boyd01}, while time integration is carried out with the fourth-order Adams-Bashforth method;

    \item \textit{Iterating:} The procedure between steps ii --- iv is repeated at each subsequent time slice, iteratively evolving the system until a desired final time $\tau_\textrm{end}$ is reached.
\end{enumerate}

\subsection{Subtracted Bulk Fields and Boundary Conditions}
\label{sec:2.3}
\allowdisplaybreaks
By substituting the near-boundary expansions~\eqref{eq:nearexpansion} into the equations of motion~\eqref{eq:EOMs}, and then algebraically solving for all possible UV coefficients in terms of the others, the UV asymptotic behavior of the EMD fields reads as follows~\cite{Rougemont:2022piu},
\begin{subequations}
\label{eq:UVasymptotic}
\begin{align}
    & A(\tau,r) = \frac{r^2}{2} + \lambda r + \frac{\lambda^2-2\partial_\tau\lambda}{2} + \frac{a_2}{r^2} \nonumber\\
    &\quad + \frac{-4\lambda a_2+\partial_\tau a_2}{2}\frac{1}{r^3} + \mathcal{O}\!\left(r^{-4}\right), \label{eq:expA2} \\
    & \Sigma(\tau,r) = \tau^{1/3}r + \frac{1+3\tau\lambda}{3\tau^{2/3}} - \frac{1}{9\tau^{5/3}r} + \frac{5+9\tau\lambda}{81\tau^{8/3}r^2} \nonumber\\
    &\quad - \frac{20+60\tau\lambda+54\tau^2\lambda^2+27\tau^4\phi_2^2}{486\tau^{11/3}r^3} + \mathcal{O}\!\left(r^{-4}\right), \label{eq:expS2} \\
    & B(\tau,r) = -\frac{2}{3}\ln(\tau) - \frac{2}{3\tau r} + \frac{1+2\tau\lambda}{3\tau^2r^2} - \frac{2+6\tau\lambda+6\tau^2\lambda^2}{9\tau^3r^3} \nonumber\\
    &\quad + \frac{1}{36\tau^4r^4}\Big[6+24\tau\lambda+36\tau^2\lambda^2-2\tau^4\phi_2^2+24\tau^3\lambda^3 \nonumber\\
    &\quad -36\tau^4a_2 -27\tau^5\partial_\tau a_2-3\tau^5\phi_2\partial_\tau\phi_2\Big] \nonumber\\
    &\quad -\frac{1}{180\tau^5r^5}\Big\{24+120\tau^2\lambda^2(2+2\tau\lambda+\tau^2\lambda^2) -48\tau^4a_2\nonumber\\
    &\quad +315\tau^5\partial_\tau a_2+35\tau^5\phi_2\partial_\tau\phi_2+15\tau^6 (\partial_\tau\phi_2)^2 \nonumber\\
    &\quad -20\tau\lambda\Big(-6+36\tau^4a_2+2\tau^4\phi_2^2+27\tau^5\partial_\tau a_2\nonumber\\
    &\quad +3\tau^5 \phi_2\partial_\tau\phi_2\Big)+135\tau^6\partial^2_\tau a_2 +15\tau^6\phi_2\partial^2_\tau\phi_2\Big\} \nonumber\\
    &\quad +\mathcal{O}\!\left(r^{-6}\right),\label{eq:expB2}\\
    & \phi(\tau,r) = \frac{\phi_2}{r^2} +\frac{-2\lambda\phi_2+\partial_\tau\phi_2}{r^3}+\frac{1}{12\tau r^4}\Big[36 \tau\lambda^2\phi_2\nonumber\\
    &\quad +\sqrt{6}\tau\phi_2^2+(3-36\tau\lambda)\partial_\tau\phi_2 +9\tau\partial_\tau^2\phi_2\Big] +\mathcal{O}\!\left(r^{-5}\right),\label{eq:expphi2}\\
    & [d_+B](\tau,r) = -\frac{1}{3\tau}+\frac{1}{3\tau^2r}-\frac{1+\tau\lambda}{3\tau^3r^2}+\frac{1}{18\tau^4r^3} \Big[6+12\tau\lambda \nonumber\\
    &\quad +6\tau^2\lambda^2+36\tau^4a_2+2\tau^4\phi_2^2 +27\tau^5\partial_\tau a_2\nonumber\\
    &\quad +3\tau^5\phi_2\partial_\tau\phi_2\Big]+\mathcal{O}\!\left(r^{-4}\right),\label{eq:expdB2}\\
    & [d_+\Sigma](\tau,r) = \frac{\tau^{1/3}r^2}{2}+\frac{(1+3\tau\lambda)r}{3\tau^{2/3}} +\frac{-1+2\tau\lambda+3\tau^2\lambda^2}{6\tau^{5/3}}\nonumber\\
    &\quad +\frac{10}{81\tau^{8/3}r} +\frac{-100-120\tau\lambda+972\tau^4a_2+81\tau^4\phi_2^2}{972\tau^{11/3}r^2} \nonumber\\
    &\quad +\mathcal{O}\!\left(r^{-3}\right),\label{eq:expdS2}\\
    & [d_+\phi](\tau,r) = -\frac{\phi_2}{r} +\mathcal{O}\!\left(r^{-2}\right),\label{eq:expdphi2}
\end{align}
\end{subequations}
where the above expansions are identical for both the 1RCBH and 2RCBH models up to the indicated orders, which provide all the necessary UV contributions for the calculations performed in this work.

To simplify the numerical integration of the equations of motion, we adopt a compact radial coordinate $u\equiv 1/r$, in terms of which the asymptotically AdS boundary is at $u=0$. Furthermore, the bulk fields, denoted here generically by $X(\tau,u)$, are conveniently rescaled and subtracted such that the redefined fields, $X_s(\tau,u)$, converge at the boundary to finite radial constants, allowing for a simpler extraction of the relevant UV coefficients. Specifically, the subtracted bulk fields are defined through the relation $u^p X_{\text{s}}(\tau,u)\equiv X(\tau,u)-X_{\text{UV}}(\tau,u)$, where $p$ is an integer, and $X_{\text{UV}}(\tau,u)$ is the UV expansion from \eqref{eq:UVasymptotic}, truncated at a given order.

Following the above considerations, the subtracted fields are specified here as follows~\cite{Rougemont:2022piu},
\begin{subequations}
\label{eq:subtracted}
\begin{align}
    & u^2A_{s}(\tau,u) \equiv A(\tau,u) - \frac{1}{2u^2} -\frac{\lambda}{u} -\frac{\lambda^2}{2}+\partial_\tau\lambda, \label{eq:subA}\\
    & u^3\Sigma_{s}(\tau,u) \equiv \Sigma(\tau,u) - \frac{\tau^{1/3}}{u} -\frac{1+3\tau\lambda}{3\tau^{2/3}} \nonumber\\
    &\quad +\frac{u}{9\tau^{5/3}} -\frac{(5+9\tau\lambda)u^2}{81\tau^{8/3}}, \label{eq:subS}\\
    & u^4B_{s}(\tau,u) \equiv B(\tau, u) +\frac{2\ln(\tau)}{3}+\frac{2u}{3\tau} \nonumber\\
    &\quad -\frac{(1+2\tau\lambda)u^2}{3\tau^2} +\frac{(2+6\tau\lambda+6\tau^2\lambda^2)u^3}{9\tau^3}, \label{eq:subB}\\
    & u^2\phi_{s}(\tau,u) \equiv \phi(\tau,u), \label{eq:subphi}\\
    & u^3[d_{+}B]_{s}(\tau,u) \equiv [d_{+}B](\tau,u) + \frac{1}{3\tau} \nonumber\\
    &\quad -\frac{u}{3\tau^2}+\frac{(1+\tau\lambda)u^2}{3\tau^3}, \label{eq:subdB}\\
    & u^2[d_{+}\Sigma]_{s}(\tau,u) \equiv [d_{+}\Sigma](\tau,u) -\frac{\tau^{1/3}}{2u^2} -\frac{1+3\tau\lambda}{3\tau^{2/3}u} \nonumber\\
    &\quad -\frac{-1+2\tau\lambda+3\tau^2\lambda^2}{6\tau^{5/3}}-\frac{10u}{81\tau^{8/3}}, \label{eq:subdS}\\
    & u[d_{+}\phi]_{s}(\tau,u) \equiv [d_{+}\phi](\tau,u), \label{eq:subdphi}\\
    & \mathcal{E}_s(\tau,u) = \mathcal{E}(\tau,u), \label{eq:subE}
\end{align}
\end{subequations}
whereas the corresponding boundary values can be expressed by the following radial constants~\cite{Critelli:2018osu,Rougemont:2022piu},
\begin{subequations}
\begin{align}
    & A_s(\tau,u=0) = a_2, \label{eq:bcAs}\\
    & \Sigma_s(\tau,u=0) = - \frac{20+60\tau\lambda+54\tau^2\lambda^2+27\tau^4\phi_2^2}{486\tau^{11/3}}, \label{eq:bcSs}\\
    & B_s(\tau,u=0) = \frac{1}{36\tau^4} \Big[6+24\tau\lambda+36\tau^2\lambda^2+24\tau^3\lambda^3 \nonumber\\
    &\quad -36\tau^4a_2 -2\tau^4\phi_2^2-27\tau^5\partial_\tau a_2-3\tau^5\phi_2\partial_\tau\phi_2\Big], \label{eq:bcBs}\\
    & \phi_s(\tau,u=0) = \phi_2, \label{eq:bcphis}\\
    & [d_+B]_s(\tau,u=0) = \frac{1}{18\tau^4} \Big[6+12\tau\lambda+6\tau^2\lambda^2+36\tau^4a_2 \nonumber\\
    &\quad +2\tau^4\phi_2^2 +27\tau^5\partial_\tau a_2+3\tau^5\phi_2\partial_\tau\phi_2\Big], \label{eq:bcdBs}\\
    & [d_+\Sigma]_s(\tau,u=0) = \frac{-100-120\tau\lambda+972\tau^4a_2+81\tau^4\phi_2^2}{972\tau^{11/3}},\label{eq:bcdSs}\\
    & [d_+\phi]_s(\tau,u=0) = -\phi_2. \label{eq:bcdphis}
\end{align}
\end{subequations}

By rewriting the original bulk fields in terms of the subtracted ones using Eqs.~\eqref{eq:subtracted}, and then substituting back into the original equations of motion~\eqref{eq:EOMs} and constraints~\eqref{eq:constEOMs}, one gets a nested set of $1+1$ PDEs for the subtracted fields written as functions of $(\tau,u)$. Numerical solutions for theses equations are obtained here by discretizing the radial and time directions, and employing the pseudospectral~\cite{boyd01} and fourth order Adams-Bashforth methods to integrate the equations in the radial and time directions, respectively. For the interested reader, we recommend Refs.~\cite{Critelli:2017euk,Rougemont:2021gjm} for more details about the general algorithm.

\subsection{Numerical Time Evolution}
\label{sec:apparent-hor}

Here we discuss the data required to implement the algorithm specified at the end of Section~\ref{sec:mainalgo}. The complete set of initial conditions to be specified at the initial time slice $\tau_0$ is $\{B_s(\tau_0,u),\phi_s(\tau_0,u),a_2(\tau_0),\rho_0;\lambda(\tau_0)\}$. As aforementioned, $\lambda(\tau)$ is a function related to the residual diffeomorphism invariance of the metric~\eqref{eq:BFDansatze}, whose details are going to be discussed in Subsection~\ref{sec:AH}. We set here $\lambda(\tau_0)=0$ and require its time derivative, $\partial_\tau\lambda(\tau)$, to satisfy Eq.~\eqref{eq:dlambda}. To evolve the system to subsequent time slices, we also need explicit expressions for the set of time derivatives $\{\partial_\tau B_s(\tau,u), \partial_\tau\phi_s(\tau,u), \partial_\tau a_2(\tau)\}$. The choice of the initial data set $\{B_s(\tau_0,u), \phi_s(\tau_0,u), a_2(\tau_0), \rho_0\}$, determining specific initial states of the holographic system, will be detailed in Section~\ref{sec:3}. Our task in the present section is to derive the aforementioned evolution equations.

To derive expressions for $\partial_\tau B_s$ and $\partial_\tau\phi_s$, one can simply rewrite expressions for $d_+B=\partial_\tau B+A\partial_r B$ and $d_+\phi=\partial_\tau \phi+A\partial_r \phi$ in terms of the subtracted fields and the radial coordinate $u\equiv 1/r$. After lengthy algebraic manipulations, one attains the following expressions~\cite{Rougemont:2022piu},
\begin{align}
\partial_\tau B_s(\tau,u) &= \frac{[d_+B]_s}{u} -\frac{2}{3\tau^4 u} - \frac{2A_s}{3\tau} + \frac{2uA_s}{3\tau^2} - \frac{2u^2A_s}{3\tau^3} \nonumber\\
&+ 4u^3A_sB_s + \frac{2B_s}{u} + \frac{B_s'}{2} + u^4A_sB_s' \nonumber\\
& +\, \left(4B_s-\frac{2}{u\tau^3}+\frac{4u A_s}{3\tau} - \frac{2u^2 A_s}{\tau^2}+uB_s'\right)\lambda\nonumber\\
&+\left(\! - \frac{1}{3\tau^3} \!-\! \frac{7}{3\tau^2 u} \!+\! 2uB_s \!-\! \frac{2u^2A_s}{\tau} +\frac{u^2B_s'}{2} \right)\lambda^2 \nonumber\\
& -\, \left( \frac{1}{\tau^2} + \frac{4}{3\tau u} \right)\lambda^3 - \frac{\lambda^4}{\tau} \nonumber\\
&+ \Big( \frac{2}{3\tau^3} - 4uB_s - u^2B_s' + \frac{2\lambda}{\tau^2} + \frac{2\lambda^2}{\tau} \Big)\partial_\tau\lambda,\label{eq:dtBs}\\
\partial_\tau\phi_s(\tau,u) & = \frac{[d_+\phi]_s}{u} +\frac{\phi_s'}{2}+u^4A_s\phi_s' +\frac{\phi_s}{u}+2u^3A_s\phi_s\nonumber\\
&+\left(u\phi_s'+2\phi_s\right)\lambda +\left(\frac{u^2\phi_s'}{2}+u\phi_s\right)\lambda^2\nonumber\\
&-\left(2u\phi_s+u^2\phi_s'\right)\partial_\tau\lambda.\label{eq:dtphis}
\end{align}
Here, the radial derivatives $X_s'(\tau,u)\equiv \partial_u X_s(\tau,u)$
are computed at each time slice by employing the pseudospectral differentiation matrix~\cite{boyd01} to the vector solution $X_s(\tau,u)$. The $N$ components of this vector are associated with the solution values at the $N$ collocation points of the Chebyshev-Gauss-Lobatto radial grid~\cite{Critelli:2017euk,Rougemont:2021gjm}.

The expression for $\partial_\tau \phi_2(\tau)$ is obtained from Eqs.~\eqref{eq:expphi2} and~\eqref{eq:subphi}~\cite{Rougemont:2022piu},
\begin{align}
    \phi_s(\tau,u)&\equiv\frac{1}{u^2}\phi(\tau,r=1/u) = \phi_2(\tau)+u\big(\partial_\tau \phi_2(\tau)\nonumber\\
    &\hspace{3cm}-2\lambda(\tau)\phi_2(\tau)\Big)+\mathcal{O}(u^4)\nonumber\\
    &\implies \phi_s'(\tau,u=0)\equiv\partial_u\phi_s(\tau,u)\Big|_{u=0}\nonumber\\
    &\hspace{2.75cm}=-2\lambda(\tau)\phi_2(\tau)+\partial_\tau\phi_2(\tau)\nonumber\\
    &\nonumber\\
&\implies\partial_\tau\phi_2(\tau) = \phi_s'(\tau,u=0)+2\lambda(\tau)\phi_2(\tau).
\label{eq:dtphi2}
\end{align}
By substituting Eq.~\eqref{eq:dtphi2} into Eq.~\eqref{eq:bcBs} and then solving for $\partial_\tau a_2(\tau)$, one obtains,
\begin{align}
\partial_\tau a_2(\tau) &= \frac{2}{9 \tau^5} - \frac{4 B_s(\tau, u=0)}{3 \tau} + \frac{8 \lambda(\tau)}{9 \tau^4} + \frac{4 \lambda(\tau)^2}{3 \tau^3} \nonumber\\
&+ \frac{8 \lambda(\tau)^3}{9 \tau^2} - \frac{4 a_2(\tau)}{3 \tau} - \frac{2 \phi_2(\tau)^2}{27 \tau} - \frac{1}{9} \phi_2(\tau) \partial_\tau \phi_2(\tau).
\end{align}

The radial integration algorithms used in \cite{Critelli:2017euk,Critelli:2018osu,Rougemont:2021qyk,Rougemont:2021gjm,Rougemont:2022piu,Rougemont:2024hpf,deOliveira:2025lhx} require the boundary values of the bulk fields to be calculated separately from the other bulk points. We implement this by using Eqs.~\eqref{eq:bcAs} --~\eqref{eq:bcdphis} together with the following results derived from the near-boundary expansions given in Eqs.~\eqref{eq:expA2} --~\eqref{eq:expdphi2}~\cite{Rougemont:2022piu},
\begin{align}
\partial_\tau B_s&(\tau,u=0) = -\frac{2}{3\tau^5}-\frac{2\lambda}{\tau^4}-\frac{2\lambda^2}{\tau^3}-\frac{2\lambda^3}{3\tau^2}+\frac{2\partial_\tau\lambda}{3\tau^3}\nonumber\\
&+\frac{2\lambda\partial_\tau\lambda}{\tau^2}+\frac{2\lambda^2\partial_\tau\lambda}{\tau}-\frac{7\partial_\tau a_2}{4}-\frac{7\phi_2 \partial_\tau\phi_2}{36}\nonumber\\
&-\frac{\tau(\partial_\tau\phi_2)^2}{12}-\,\frac{3\tau\partial^2_\tau a_2}{4}-\frac{\tau\phi_2\partial^2_\tau\phi_2}{12},\label{eq:dtBs0}\\
\partial_\tau\phi_s&(\tau,u=0) = \partial_\tau\phi_2,\label{eq:dtphis0}
\end{align}
where the following second order derivatives are derived, respectively, from Eqs.~\eqref{eq:expB2} and~\eqref{eq:expphi2}~\cite{Rougemont:2022piu},
\begin{align}
\partial^2_\tau &a_2(\tau) = -\frac{4 B'_s(\tau,u=0)}{3\tau}-\frac{8}{45\tau^6} -\frac{8\lambda}{9\tau^5}-\frac{16\lambda^2}{9\tau^4}-\frac{16\lambda^3}{9\tau^3} \nonumber\\
&-\frac{8\lambda^4}{9\tau^2}+\frac{16a_2}{45\tau^2}+\frac{16\lambda a_2}{3\tau}+\frac{8\lambda\phi_2^2}{27\tau}-\, \frac{7\partial_\tau a_2}{3\tau}+4\lambda\partial_\tau a_2 \nonumber\\
&-\frac{7\phi_2\partial_\tau\phi_2}{27\tau}+\frac{4\lambda\phi_2\partial_\tau\phi_2}{9} -\frac{(\partial_\tau\phi_2)^2}{9}-\frac{\phi_2\partial^2_\tau\phi_2}{9},\label{eq:dt2a2}\\
\partial^2_\tau &\phi_2(\tau) = \frac{2\phi''_s(\tau,u=0)}{3}-4\lambda^2\phi_2-\sqrt{\frac{2}{27}}\,\phi_2^2\nonumber\\
&\hspace{1cm}-\frac{\partial_\tau\phi_2}{3\tau} +4\lambda\partial_\tau\phi_2.\label{eq:dt2phi2}
\end{align}

%%%%%%%%%%%%%%%%%%%%%%%%%%%%%%%%%%%%%%%

\section{Boundary Physical Observables and Initial Conditions}
\label{sec:3}

Relevant information about the hydrodynamization properties of the 1RCBH and 2RCBH plasmas is encoded in the renormalized one-point functions of the dual QFT operators corresponding to the expectation values of the energy-momentum tensor $\langle T_{\mu\nu}\rangle$ (dual to the bulk metric field $g_{\mu\nu}$), the $U(1)$ vector current $\langle J_\mu\rangle$ (dual to the bulk Maxwell field $A_\mu$), and the scalar condensate $\langle O_\phi\rangle$ (dual to the bulk scalar field $\phi$). These expectation values can be related to the bulk gravity ultraviolet coefficients $\{a_2(\tau), \Phi_2(\tau) = -\rho_0/\tau, \phi_2(\tau)\}$ and their time derivative via holographic renormalization. Typically, the renormalization program for the 1-point functions is carried out using the Fefferman-Graham (FG) coordinates~\cite{Critelli:2017euk,Bianchi:2001kw}, where the bulk spacetime metric takes the form,
\begin{equation}
\label{eq:ansatzequil}
    ds^2_{\text{FG}}=\frac{d\rho^2}{4\rho^2}+\gamma_{\mu\nu}(\rho,x)\,dx^\mu dx^{\nu}.
\end{equation}
In this description, $\rho$ is the FG radial coordinate, which places the spacetime boundary at $\rho=0$. The Greek indices in Eq.~\eqref{eq:ansatzequil} run over the coordinates of the dual QFT. The term $\gamma_{\mu\nu}$ represents the induced metric on a constant $\rho$ hypersurface, and it relates to the bulk metric via $\gamma_{\mu\nu}(\rho,x)=g_{\mu\nu}(x)/\rho$.

In terms of the FG coordinates, the EMD field expansions near the boundary take the following form (where the coefficients of the logarithmic terms below vanish for the conformal 2RCBH and 1RCBH plasmas)~\cite{Critelli:2017euk,Bianchi:2001kw},
\begin{subequations}
\label{eq:NearBoundExpOPF}
    \begin{align}
    \gamma_{\mu\nu}(\rho,x)&=\frac{1}{\rho}\gamma_{(0)\mu\nu}(x)+\gamma_{(2)\mu\nu}(x)+\gamma_{(2,1)\mu\nu}(x)\ln\rho\nonumber\\
&+\rho\left[\gamma_{(4)\mu\nu}(x)+\gamma_{(4,1)\mu\nu}\ln \rho+\gamma_{(4,2)\mu\nu}\ln^2 \rho\right]\nonumber\\
&\hspace{4.5cm}+\mathcal{O}(\rho^2),\\
    A_\mu(\rho,x)&=A_{(0)\mu}(x)+\rho\left[A_{(2)\mu}(x)+A_{(2,1)\mu}(x)\ln\rho\right]\nonumber\\
&\hspace{4.5cm}+\mathcal{O}(\rho^2),\\
    \phi(\rho,x)&=\rho\left[\phi_{(0)}(x)+\phi_{(0,1)}(x)\ln\rho\right]+\mathcal{O}(\rho^2).
\end{align}
\end{subequations}

As discussed in detail in Appendix A of~\cite{Critelli:2017euk}, for the 1RCBH model, and also for the 2RCBH model, the aforementioned renormalized one-point functions can be expressed in terms of the UV coefficients of the FG near-boundary expansions~\eqref{eq:NearBoundExpOPF} as follows,
\begin{subequations}
\label{eq:opf}
    \begin{align}
        \langle T_{\mu\nu} \rangle &= \frac{1}{\kappa_5^2}\left(2\gamma_{(4)\mu\nu}+\gamma_{(0)\mu\nu}\frac{\phi_{(0)}^2}{6}\right),\label{eq:optmn}\\
        \langle J_{\mu} \rangle&=\frac{1}{\kappa_5^2}A_{(2)\mu},\label{eq:opJm}\\
         \langle \mathcal{O}_{\phi} \rangle&=-\frac{1}{\kappa_5^2}\phi_{(0)}.\label{eq:opphi}
    \end{align}
\end{subequations}
To relate the FG UV coefficients from Eqs.~\eqref{eq:opf} to their counterparts in EF coordinates, it is necessary to establish the coordinate transformation between the two charts, which we discuss in the following subsection.

\subsection{Equilibrium One-Point Functions}

In this section we calculate the equilibrium values of the one-point functions $\langle T_{\mu\nu}\rangle_{\text{eq}}$, $\langle J^\mu\rangle_{\text{eq}}$ and $\langle \mathcal{O}_\phi\rangle_{\text{eq}}$, as expressed by Eqs.~\eqref{eq:opf}. Since the expressions for the equilibrium solutions were written in Section~\ref{sec:sec1} in the modified EF coordinate, $\tilde{r}$, and the holographic renormalization formulas are expressed in the FG coordinate, $\rho$, we begin by rewriting the equilibrium line element given by Eq.~\eqref{eq:AnsatzEqs} in diagonal form,
\begin{align}
\label{eq:AnsatzEqs2}
ds^2=e^{2a(\tilde{r})}\left[-h(\tilde{r})dt^2 +d\mathbf{x}^2\right]+\frac{e^{2 b(\tilde{r})}}{h(\tilde{r})}d\tilde{r}^2.
\end{align}
Then, by imposing that the radial part of both line elements~\eqref{eq:AnsatzEqs2} and~\eqref{eq:ansatzequil} are equal, such that, $(g_{\tilde{r}\tilde{r}}\,d\tilde{r}^2)_\textrm{EF}=(g_{\rho\rho}\,d\rho^2)_\textrm{FG}$, one obtains the following integral expression~\cite{deOliveira:2025lhx},
\begin{equation}
    \int \frac{e^{b(\tilde{r})}}{\sqrt{h(\tilde{r})}}d\tilde{r}=- \int \frac{d\rho}{2\rho}=-\frac{1}{2}\ln\rho,
    \label{eq:integral-radialcoord}
\end{equation}
with the minus sign following from the observation that $\tilde{r}\to\infty\Rightarrow\rho\to 0$. By perturbatively solving Eq.~\eqref{eq:integral-radialcoord} with $\tilde{r}$ close to the boundary employing the equilibrium backgrounds, one obtains that~\cite{deOliveira:2025lhx},
\begin{subequations}
\begin{align}
&\text{1RCBH model:}\nonumber\\
&\tilde{r}(\rho)=\frac{1}{\sqrt{\rho }}-\frac{\sqrt{\rho } Q^2}{6}+\frac{1}{72} \rho ^{3/2} \left(9 M^2+Q^4\right)+\mathcal{O}(\rho^{5/2}),\\
&\text{2RCBH model:}\quad\nonumber\\
&\tilde{r}(\rho)=\frac{1}{\sqrt{\rho }}-\frac{\sqrt{\rho } Q^2}{3}+\frac{1}{72} \rho ^{3/2} \left(9 M^2-2Q^4\right)+\mathcal{O}(\rho^{5/2}).
\end{align}
\end{subequations}

The next step is to perform a series expansion of the equilibrium solutions, given by Eqs.~\eqref{eq:AnsatzAll} and \eqref{eq:AnsatzAll2R}, in the region close to the boundary. This expansion is most conveniently carried out using the diagonal coordinate system defined in Eq.~\eqref{eq:AnsatzEqs2}. After obtaining the series, we rewrite them in terms of the expression for $\tilde{r}(\rho)$ derived above. This allows for a direct comparison with the standard structure of the FG UV expansions shown in Eq.~\eqref{eq:ansatzequil}. Through this matching procedure, we find that~\cite{deOliveira:2025lhx},
\begin{subequations}
\label{eq:coeff}
\begin{align}
&\text{1RCBH model:}\nonumber\\
&\gamma_{\tau\tau}(\rho)=-\frac{1}{\rho }+\rho  \left(\frac{3 M^2}{4}+\frac{Q^4}{18}\right)\nonumber\\
&\hspace{1cm}+\rho ^2 \left(-\frac{7 M^2 Q^2}{24}-\frac{13 Q^6}{648}\right)+\mathcal{O}(\rho^3),\\
& \gamma_{xx}(\rho)=\frac{1}{\rho }+\rho  \left(\frac{M^2}{4}-\frac{Q^4}{18}\right)\nonumber\\
&\hspace{1cm}+\rho ^2 \left(-\frac{ M^2 Q^2}{24}+\frac{13 Q^6}{648}\right)+\mathcal{O}(\rho^3),\\
& \phi(\rho)=-\sqrt{\frac{2}{3}} \rho  Q^2+\frac{\rho ^2 Q^4}{3 \sqrt{6}}+\mathcal{O}(\rho^3),\\
&\Phi(\rho)=\frac{M Q}{Q^2+\tilde{r}_H^2}-M \rho  Q+\frac{2}{3} M \rho ^2 Q^3+\mathcal{O}(\rho^3),\\
&&&\nonumber\\
&\text{2RCBH model:}\nonumber\\
&\gamma_{\tau\tau}(\rho)=-\frac{1}{\rho }+\rho  \left(\frac{3 M^2}{4}+\frac{Q^4}{18}\right)\nonumber\\
&\hspace{1cm}+\rho ^2 \left(\frac{Q^6}{162}-\frac{7 M^2 Q^2}{12}\right)+\mathcal{O}(\rho^3),\\
&\gamma_{xx}(\rho)=\frac{1}{\rho }+\rho  \left(\frac{M^2}{4}-\frac{Q^4}{18}\right)\nonumber\\
&\hspace{1cm}+\rho ^2 \left(-\frac{M^2 Q^2}{12}-\frac{Q^6}{162}\right)+\mathcal{O}(\rho^3),\\
& \phi(\rho)=\sqrt{\frac{2}{3}} \rho  Q^2+\frac{\rho ^2 Q^4}{3 \sqrt{6}}+\mathcal{O}(\rho^3),\\
&\Phi(\rho)=\frac{\sqrt{2} M Q}{Q^2+r_H^2}-\sqrt{2} M \rho  Q+\frac{1}{3} \sqrt{2} M \rho ^2 Q^3+\mathcal{O}(\rho^3).
\end{align}
\end{subequations}
We then compare the coefficients of our asymptotic solutions in Eqs.~\eqref{eq:coeff} with the universal form of the FG UV expansions presented in Eqs.~\eqref{eq:NearBoundExpOPF}.\footnote{It should be noted that the logarithmic terms in the FG expansions vanish for both models. This is due to their conformal symmetry, the flat boundary, and the fact that the dual scalar operator $\mathcal{O}_\phi$ has a scaling dimension of $\Delta=2$, as discussed in~\cite{Critelli:2017euk}.} Upon identifying the equilibrium FG coefficients, we insert them into the expressions for the renormalized one-point functions from Eqs.~\eqref{eq:opf}, obtaining~\cite{deOliveira:2025lhx},
\begin{subequations}
\label{eq:equil-observables}
\begin{align}
&\text{1RCBH model:} \nonumber\\
&\varepsilon_{\text{eq}}\equiv\langle T_{\tau\tau}\rangle_{\text{eq}} = \frac{1}{\kappa_5^2}\frac{3M^2}{2} =\frac{N_c^2}{4\pi^2}\frac{3M^2}{2}, \label{eq:eeq}\\
&p_{\text{eq}}\equiv\langle T_x^x\rangle_{\text{eq}} = \frac{1}{\kappa_5^2}\frac{M^2}{2} =\frac{N_c^2}{4\pi^2}\frac{M^2}{2}, \label{eq:rho43eq}\\
&\rho_{c,{\text{eq}}}\equiv\langle J^t \rangle_{\text{eq}} = \frac{1}{\kappa_5^2}MQ =\frac{N_c^2}{4\pi^2}MQ, \label{eq:rhoeq}\\
&\langle \mathcal{O}_\phi \rangle_{\text{eq}} = \frac{1}{\kappa_5^2}\sqrt{\frac{2}{3}}Q^2 = \frac{N_c^2}{4\pi^2}\sqrt{\frac{2}{3}}Q^2, \label{eq:ScalCondEq}\\
&\nonumber\\
&\text{2RCBH model:}\nonumber\\
&\varepsilon_{\text{eq}}\equiv\langle T_{\tau\tau}\rangle_{\text{eq}}= \frac{1}{\kappa_5^2}\frac{3M^2}{2} =\frac{N_c^2}{4\pi^2}\frac{3M^2}{2},\\
& p_{\text{eq}}\equiv\langle T_x^x\rangle_{\text{eq}} = \frac{1}{\kappa_5^2}\frac{M^2}{2} =\frac{N_c^2}{4\pi^2}\frac{M^2}{2},\\
&\rho_{c,{\text{eq}}}\equiv\langle J^t \rangle_{\text{eq}} = \frac{1}{\kappa_5^2}\sqrt{2}MQ =\frac{N_c^2}{4\pi^2}\sqrt{2}MQ,\\
&\langle \mathcal{O}_\phi \rangle_{\text{eq}} = -\frac{1}{\kappa_5^2}\sqrt{\frac{2}{3}}Q^2 = -\frac{N_c^2}{4\pi^2}\sqrt{\frac{2}{3}}Q^2,
\end{align}
\end{subequations}
where the above thermodynamic observables, written in terms of the bulk black hole parameters $(M,Q)$, are the equilibrium values for the energy density $\varepsilon_{\text{eq}}$, the equilibrium isotropic pressure $p_{\text{eq}}$, the $U(1)$ R-charge density $\rho_{c,\text{eq}}$, and the scalar condensate $\langle \mathcal{O}_\phi \rangle_{\text{eq}}$.

Using the 1RCBH and 2RCBH expressions for the ratio $Q/\tilde{r}_H$, as given by Eqs.~\eqref{eq:Qoverr}, one can rewrite the formula for the dual boundary QFT parameters ($T,\mu$), given by Eqs.~\eqref{eq:Tmu}, in the following form,
\begin{subequations}
\label{eq:QTMT2}    
\begin{align}
&\text{1RCBH model:} &&\text{2RCBH model:}\nonumber\\
& \frac{Q}{T}=x\sqrt{y^2+1},&& \frac{Q}{T}=\frac{x}{\sqrt{2}},\\
& \frac{M}{T^2}=x^2\frac{(y^2+1)^{3/2}}{y^2},&&\frac{M}{T^2}=\pi^2\left(1+\frac{x^2}{2\pi^2}\right),
\end{align}
\end{subequations}
where $y\equiv Q/\tilde{r}_H=\sqrt{2}\left(1\pm\sqrt{1-\left(x/x_c\right)^2}\right)/(x/x_c)$, with $x\equiv \mu/T$ and $x_c\equiv \pi/\sqrt{2}$.

Plugging relations~\eqref{eq:QTMT2} into Eqs.~\eqref{eq:equil-observables}, one recovers the exact expressions for the dimensionless quantities $s_{\text{eq}}/N_c^2T^3$, $\rho_{c,\text{eq}}/N_c^2T^3$, and $p_{\text{eq}}/N_c^2T^4$, as given by Eqs.~\eqref{eq:seq}, \eqref{eq:rho12}, and 
\eqref{eq:p12}. On the other hand, the equilibrium value for the scalar condensate can be obtained directly from the renormalized one-point function~\eqref{eq:ScalCondEq}, which in terms of $x\equiv\mu/T$ read as follows~\cite{deOliveira:2025lhx},
\begin{subequations}
\label{eq:thermo-Os}
\begin{align}
&\text{1RCBH model:}\nonumber\\
& \frac{\langle \mathcal{O}_{\phi}\rangle_{\text{eq}}}{N_c^2 T^2}=\frac{1}{4\sqrt{6}}\left[3\pm \sqrt{1-\left({x}/{x_c}\right)^2}\right]\left[1\pm  \sqrt{1-\left({x}/{x_c}\right)^2}\right],
\label{eq:O12}\\
&\text{2RCBH model:}\nonumber\\
&\frac{\langle \mathcal{O}_{\phi}\rangle_{\text{eq}}}{N_c^2 T^2}=-\frac{x^2}{4\sqrt{6}\pi^2}.
\end{align}
\end{subequations}

Moreover, in the relativistic hydrodynamic description of the Bjorken flow dynamics for the 2RCBH and 1RCBH models, it will turn out to be convenient to express the above observables in terms of the dimensionless ratios $\hat{\rho}^{4/3}/\hat{\varepsilon}$, $\langle\hat{O}_{\phi} \rangle/\hat{\varepsilon}^{1/2}$ and $\hat{s}^{4/3}/\hat{\varepsilon}$. Their respective equilibrium expressions are written as follows,
\begin{subequations}
\label{eq:thermo-ratios}
\begin{align}
&\text{1RCBH model:}\nonumber\\
& \frac{\hat{\rho}^{4/3}_{c,\text{eq}}}{\hat{\varepsilon}_{\text{eq}}}= \frac{(4\pi^2)^{1/3}\left[ \frac{x}{16} \left( 3 \pm \sqrt{1-(x/x_c)^2} \right)^2 \right]^{4/3}}{\frac{3\pi^2}{128} \left( 3 \pm \sqrt{1-(x/x_c)^2} \right)^3 \left( 1 \mp \sqrt{1-(x/x_c)^2} \right)}, 
\label{eq:rho43eThermo}\\
&\frac{\langle \hat{O}_{\phi}\rangle_{\text{eq}}}{\hat{\varepsilon}^{1/2}_{\text{eq}}}=\frac{4 \left[ \frac{1 \pm \sqrt{1-(x/x_c)^2}}{x/x_c} \right]^2}{3 \sqrt{1 + 2 \left[ \frac{1 \pm \sqrt{1-(x/x_c)^2}}{x/x_c} \right]^2}}, \label{eq:Oe12Thermo}\\
&\frac{\hat{s}^{4/3}_{\text{eq}}}{\hat{\varepsilon}_{\text{eq}}}=(4\pi^2)^{1/3} \frac{\left[ \frac{\pi^2}{16} \left( 3 \pm \sqrt{1-(x/x_c)^2} \right)^2  \right]^{4/3}}{\frac{3\pi^2}{128} \left( 3 \pm \sqrt{1-(x/x_c)^2} \right)^3}\nonumber\\
&\hspace{2cm}\times\frac{\left( 1 \mp \sqrt{1-(x/x_c)^2} \right)^{4/3}}{ \left( 1 \mp \sqrt{1-(x/x_c)^2} \right)}, \label{eq:s43eThermo}\\
&\text{2RCBH model:}\nonumber\\
&\frac{\hat{\rho}^{4/3}_{c,\text{eq}}}{\hat{\varepsilon}_{\text{eq}}}=\frac{2^{5/3}}{3}x^{4/3} \frac{  \left(x^2+2 \pi^2\right)^{4/3}}{\left(x^2+2 \pi^2\right)^2},\\
&\frac{\langle \hat{O}_{\phi}\rangle_{\text{eq}}}{\hat{\varepsilon}^{1/2}_{\text{eq}}}=-\frac{2}{3} \frac{x^2}{ x^2+2 \pi^2},\\
&\frac{\hat{s}^{4/3}_{\text{eq}}}{\hat{\varepsilon}_{\text{eq}}}=\frac{8 }{3}\pi ^{8/3}\frac{ \left(x^2+2 \pi^2\right)^{4/3}}{ \left(x^2+2 \pi^2\right)^2}.
\end{align}
\end{subequations}

\begin{figure*}
\centering  
\subfigure[Equilibrium value for $\langle \mathcal{O}_{\phi}\rangle_{\text{eq}}/N_c^2 T^2$.]{\includegraphics[width=0.475\linewidth]{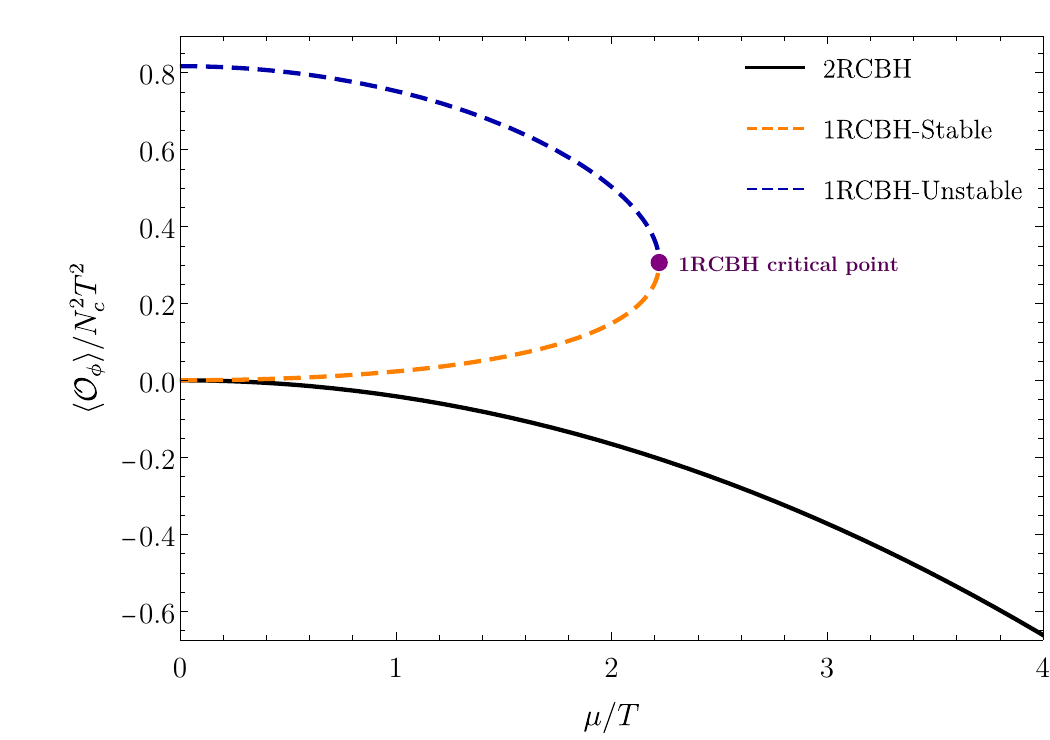}\label{fig:OThermo}}
\subfigure[Equilibrium value for $\langle\hat{O}_{\phi} \rangle/\hat{\varepsilon}^{1/2}$. ]
{\includegraphics[width=0.475\linewidth]{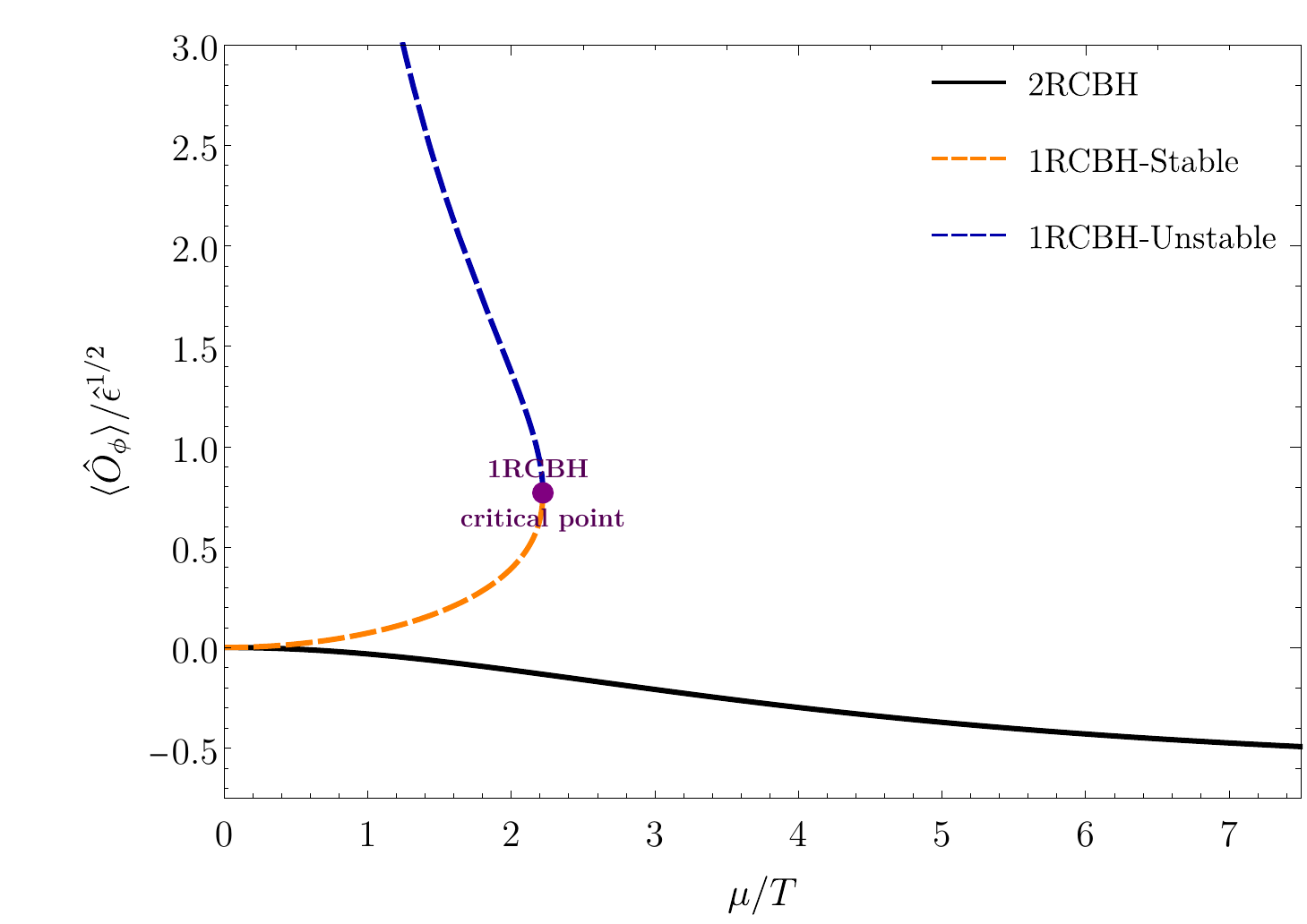}\label{fig:Oepsilon12Thermo}}
\subfigure[Equilibrium value for $\hat{\rho}^{4/3}/\hat{\varepsilon}$. ]
{\includegraphics[width=0.475\linewidth]{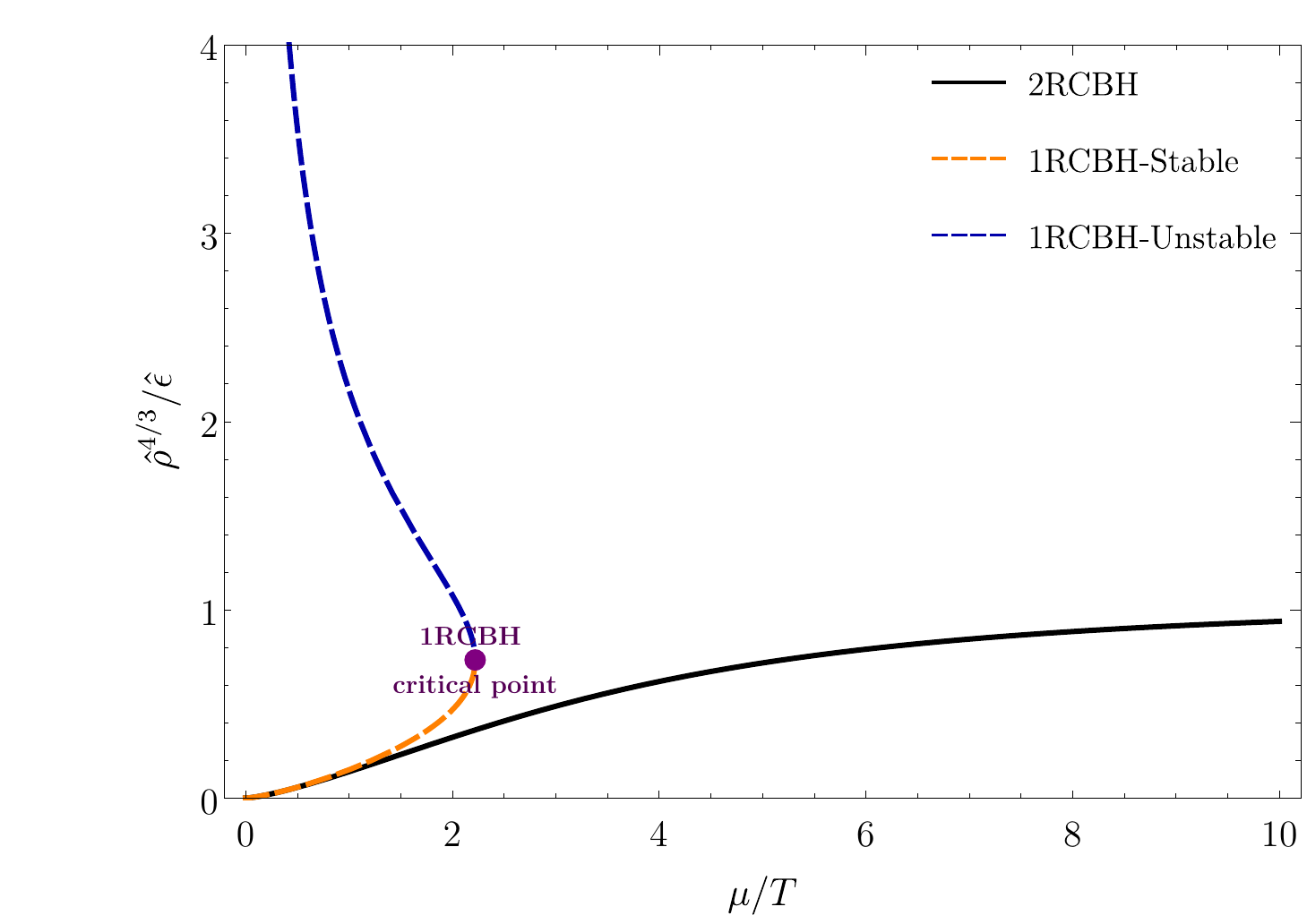}\label{fig:rho43epsilonThermo}}
\subfigure[Equilibrium value for $\hat{s}^{4/3}/\hat{\varepsilon}$. ]
{\includegraphics[width=0.475\linewidth]{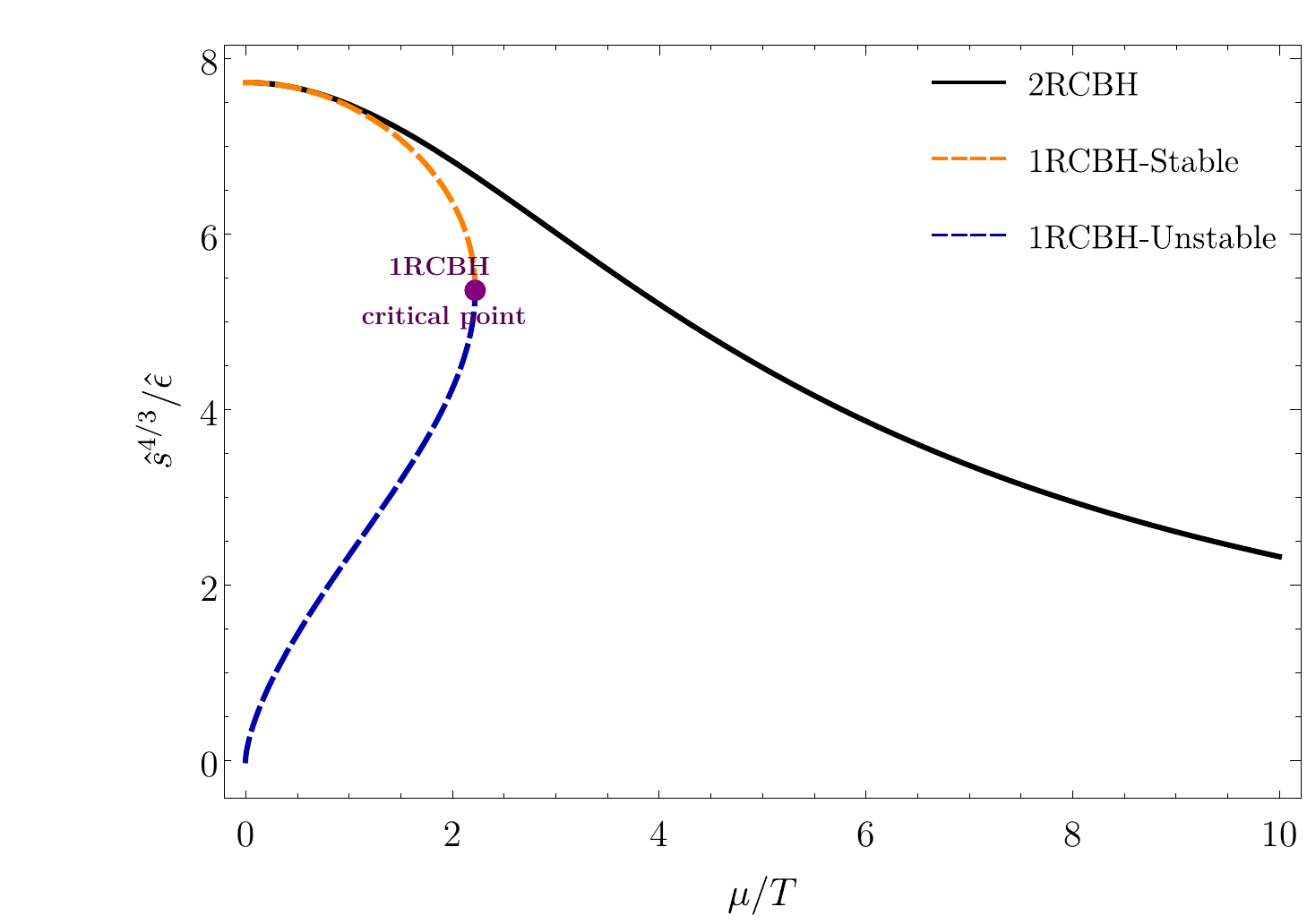}\label{fig:s43epsilonThermo}}
\caption{Plots for equilibrium dimensionless ratios of thermodynamic observables, both for the 2RCBH and 1RCBH models.}
\label{fig:Thermo2}
\end{figure*}

Finally, the plots for the dimensionless quantities expressed in Eqs.~\eqref{eq:thermo-Os} and~\eqref{eq:thermo-ratios} are displayed in Fig.~\ref{fig:Thermo2}.

\subsection{Dynamical Black Hole Horizons and Holographic Non-Equilibrium Entropy}

Before defining the non-equilibrium one-point functions in Subsection~\ref{Sec:NonEquilibrium-1PointFunctions}, we briefly review the calculation of the non-equilibrium entropy density in terms of the area of the event horizon (EH) and the apparent horizon (AH). To this end, we start by discussing the notions of black hole event and apparent horizons, and the determination of their radial location within the bulk.

\subsubsection{Event Horizon}

The event horizon of a black hole within an asymptotically $\text{AdS}_5$ bulk is defined as the surface formed by the family of congruent light rays that separates trajectories capable of reaching the conformal boundary from those trapped in the deep interior. This causal disconnection signifies that light rays contained within the horizon must remain unreachable to boundary observers. Consequently, to avoid losing physical information when numerically solving the bulk geometry, one needs to integrate the radial part of the bulk equations of motion at least from the EH (whenever one is present) to the conformal boundary.

In this framework, the radial location of the EH, $r_{EH}(\tau)$, is determined by solving the outgoing radial null geodesic equation, with the asymptotic equilibrium condition that at asymptotically large times the radial position of the event horizon is given by the largest zero of the metric coefficient $A(\tau,r)$,\footnote{Eq.~\eqref{eq:radialnullgeodesic} can be obtained by calculating the holographic radial distance from the black hole center where null rays cannot escape to the boundary. This can be derived directly from the spacetime ansatze Eq.~\eqref{eq:BFDansatze} by applying the following two physical constraints: (i) null condition, $ds^2=0$; and (ii) holographic radial condition, $d\xi=dx=dy=0$.}
\begin{equation}
    \frac{dr_{\text{EH}}(\tau)}{d\tau}=A(r_{\text{EH}}(\tau)), \qquad r_{\text{EH}}(\tau\to\infty)=r_{\text{EH}}^{\text{eq}},\label{eq:radialnullgeodesic}
\end{equation}
where $r_{\text{EH}}^{\text{eq}}$ is the largest root of equation $A(\tau\to\infty,r)=0$.

By substituting the inverse radial coordinate $u=1/r$ and the subtracted metric field $A_s(\tau,u)$ given by Eq.~\eqref{eq:subA} into Eq.~\eqref{eq:radialnullgeodesic}, the time evolution of the EH is governed by the following differential equation,
\begin{align}
    &\frac{du_{\text{EH}}(\tau)}{dt} = -u_{\text{EH}}^2(\tau) A(\tau, u_{\text{EH}}(\tau)) =- \frac{1}{2}- u_{\text{EH}}(\tau)\lambda(\tau)\nonumber\\
    & \hspace{0.5cm}-u_{\text{EH}}^4(\tau) A_s(\tau, u_{\text{EH}}(\tau))   - u_{\text{EH}}^2(\tau) \left( \frac{\lambda^2(\tau)}{2} - \partial_\tau \lambda(\tau) \right) \\
    &u_{\text{EH}}(\tau \to \infty) = u_{\text{EH}}^{\text{eq}}(\mu/T),\label{eq:uEH}
\end{align}
where $u_{\text{EH}}^{\text{eq}}(\mu/T)$ is the smallest simple root of the following equation,
\begin{align}
    A(\tau \to \infty, u) &= u^2 A_s(\tau \to \infty, u) + \frac{1}{2u^2} + \frac{\lambda(\tau \to \infty)}{u} \nonumber\\
    &+ \frac{\lambda^2(\tau \to \infty)}{2} - \partial_\tau \lambda \Big|_{\tau \to \infty} = 0.
    \label{eq:EHmetricsubtracted}
\end{align}

Operationally, the asymptotic time limit $\tau\to\infty$ in Eq.~\eqref{eq:EHmetricsubtracted} is approximated by its evaluation at $\tau=\tau_{\text{end}}$, provided the numerical simulation runs long enough for the geometry to reach a stationary state within some numerical tolerance. Moreover, notice that due to the asymptotic late-time equilibrium condition given by Eq.~\eqref{eq:uEH}, the EH is understood as a global-in-time structure of the bulk geometry. This means that its determination demands the specification of the entire time evolution of the metric coefficient $A_s(\tau,u)$. For non-equilibrium settings, where one needs to radially integrate the bulk equations of motion at each time slice, a more convenient solution to prevent the loss of physical information at the boundary is to consider a different causal surface, the apparent horizon, which is local in time, as we discuss below.

\subsubsection{Apparent Horizon}
\label{sec:AH}

In general relativity, a trapped null surface constitutes a compact surface traced out by light where both ingoing and outgoing light rays are converging (i.e., its surface area is decreasing). In a dynamic situation, multiple trapped surfaces can form. The outermost trapped null surface is then the boundary enclosing all other trapped surfaces where the expansion of the outgoing null geodesics is precisely zero. In this context, the apparent horizon is defined locally as the outermost trapped null surface. It delineates the spacetime region where outward-directed null geodesics begin to converge from the region where they can still expand outward. The radial location of the apparent horizon, $r_{\text{AH}}$, is identified by the condition where the expansion of outgoing null geodesics vanishes~\cite{Chesler:2013lia},
\begin{align}
[d_+\Sigma](\tau,r_{\textrm{AH}})=0.
\label{eq:AppHor}
\end{align}

In dynamic scenarios like the holographic Bjorken flow, the radial position of the apparent horizon can fluctuate dramatically over time. These large oscillations may eventually cause the apparent horizon to cross a fixed infrared cutoff for radial integration in the Gauss-Lobatto grid~\cite{Critelli:2017euk,Rougemont:2021gjm}, potentially compromising the reliability of the numerical simulation by allowing information causally connected to boundary observers to be missed. To avoid this issue, we exploit the residual diffeomorphism invariance related to the radial shift function $\lambda(\tau)$. Since $\lambda(\tau)$ is an arbitrary function of time, we follow~\cite{Chesler:2013lia} and impose that the apparent horizon remains at a fixed radial coordinate, $r_{AH}$, throughout the entire evolution of the system.

By requiring that $\partial_\tau r_{\textrm{AH}}(\tau)=0$ and that Eq.~\eqref{eq:AppHor} remains valid at all times, it follows that $\partial_\tau [d_+\Sigma](\tau,r_{\textrm{AH}})=0$, implying that $d_+[d_+\Sigma](\tau,r_{\textrm{AH}})\allowbreak =A(\tau,r_{\textrm{AH}})\partial_r [d_+\Sigma](\tau,r_{\textrm{AH}})$. Substituting this result into the constraint Eq.~\eqref{eq:constb}, and then combining it with other components of Einstein's equations, one obtains that,
\begin{align}
A(\tau,u_{\textrm{AH}}) = \frac{6([d_+B](\tau,u_{\textrm{AH}}))^2 + 2([d_+\phi](\tau,u_{\textrm{AH}}))^2}{2V+f\mathcal{E}^2},
\label{eq:Astar}
\end{align}
where we used Eq.~\eqref{eq:AppHor} together with the definition of the inverse radial coordinate $u=1/r$.

By using now Eq. \eqref{eq:subA} evaluated at the apparent horizon position $u=u_{\textrm{AH}}$, it follows that,
\begin{align}
\partial_\tau\lambda(\tau) &= u_{\textrm{AH}}^2 A_s(\tau,u_{\textrm{AH}})+\frac{1}{2u_{\textrm{AH}}^2}+\frac{\lambda(\tau)}{u_{\textrm{AH}}}\nonumber\\
&+\frac{\lambda^2(\tau)}{2} -A(\tau,u_{\textrm{AH}}).
\label{eq:dlambda}
\end{align}
{The temporal evolution of the function $\lambda(\tau)$ from a chosen initial value, $\lambda(\tau_0)$, is dictated by Eq.~\eqref{eq:dlambda}, a method employed to anchor the apparent horizon at a constant radial coordinate. Adopting the methodology of Refs.~\cite{Rougemont:2021qyk,Rougemont:2021gjm}, we impose the initial condition $\lambda(\tau_0)=0$. The initial horizon location is then computed by finding the root of Eq.~\eqref{eq:AppHor}, combined with Eq.~\eqref{eq:subS}, via the Newton-Raphson algorithm. This approach ensures the horizon's position is held static at this initial value for the duration of the evolution, within some numerical tolerance.}

\subsubsection{Non-Equilibrium Entropy}

Since the event horizon is non-local in time, while the apparent horizon is local, in dynamic holographic scenarios one usually considers the definition of the non-equilibrium entropy through the area of the apparent horizon~\cite{Figueras:2009iu,Chesler:2009cy,Heller:2011ju,Heller:2012je,vanderSchee:2014qwa,Jankowski:2014lna,Buchel:2016cbj,Grozdanov:2016zjj,Engelhardt:2017aux,Muller:2020ziz}. For dissipative systems, as in the case of the holographic Bjorken flow of the 1RCBH and 2RCBH models, the apparent horizon lies behind and asymptotically converges to the event horizon as the black hole relaxes towards a stationary state in the long time regime, guaranteeing that the non-equilibrium entropy defined through the area of the apparent horizon coincides with the usual Bekenstein-Hawking's entropy relation in thermodynamic equilibrium, which is defined through the area of the equilibrium event horizon~\cite{Bekenstein:1973ur,Hawking:1975vcx}.

In the present work, we shall obtain numerical results for both, the time evolution of the holographic non-equilibrium entropy defined through the area of the apparent horizon, and via the area of the event horizon. As we shall discuss, when transient plateaus are formed in the non-equilibrium entropy of the AH, they correlate with some patterns of the pressure anisotropy of the system, while such correlations cannot be seen in the non-equilibrium entropy of the EH. Consequently, the AH seems to holographically encode some far-from-equilibrium correlations with the energy momentum-tensor of the dual boundary QFT, which are not present in the EH.

In the presentation of the results for the time evolution of the holographic non-equilibrium entropy, $S_\text{H}(\tau)$, the subscript H denotes either the AH or the EH. The holographic non-equilibrium entropy is defined in terms of the non-stationary version of the Bekenstein-Hawking relation,
\begin{align}
S_{\textrm{H}}(\tau) = \frac{A_{\textrm{H}}(\tau)}{4G_5},
\label{eq:BHrel}
\end{align}
where $A_H(\tau)$ is the time-dependent hyperarea of either the AH or EH,
\begin{align}
A_{\textrm{H}}(\tau) &= \int d^3x \sqrt{-g}\,\biggr|_{u=u_{\textrm{H}}} = \int dx dy d\xi \sqrt{-g}\,\biggr|_{u=u_{\textrm{H}}} \nonumber\\
&= \sqrt{-g}\,\biggr|_{u=u_{\textrm{H}}} \mathcal{A} = |\Sigma(\tau,u_{\textrm{H}})|^3 \mathcal{A}.
\label{eq:AreaHor}
\end{align}
In Eq.~\eqref{eq:AreaHor}, $\mathcal{V}(\tau) = \tau \mathcal{A} = \tau \int dx dy d\xi$ is the expanding volume of the medium in the Bjorken flow. Therefore, the entropy density $s_{H}(\tau)$ scaled by the gravitational constant reads as follows,
\begin{align}
\hat{s}_{\textrm{H}}(\tau) \equiv \kappa_5^2\, s_{\textrm{H}}(\tau) = \kappa_5^2\, \frac{S_{\textrm{H}}(\tau)}{\mathcal{V}(\tau)} = \frac{2\pi|\Sigma(\tau,u_{\textrm{H}})|^3}{\tau},
\label{eq:hats}
\end{align}
where $\Sigma(\tau,u_{\textrm{H}})$ is determined from the numerical result for $\Sigma_s(\tau,u_{\textrm{H}})$ via the relation in Eq.~\eqref{eq:subS}, evaluated at the horizon's radial coordinate $u_H$.

The dimensionless ratio between the entropy and energy densities, $\hat{s}_\textrm{H}^{4/3}/\hat{\varepsilon}$, corresponds to an important physical observable to track during the time evolution of the system. This is mainly due to the equilibrium formula given by Eq.~\eqref{eq:s43eThermo}, which serves as a highly nontrivial consistency check of the late-time validity of the numerical simulations, associated to the late-time convergence of the dynamical bulk geometry to the corresponding equilibrium black hole solutions discussed before. Besides that, such a comparison also allows to extract characteristic equilibration times for the fluid~\cite{Rougemont:2021gjm}.

On the other hand, since the energy density is itself a nontrivial function of time, in order to understand specifically the time evolution of the entropy function $S_{\textrm{H}}(\tau)$, which is related to the apparent or event horizon area $A_\textrm{H}(\tau)$ by Eq.~\eqref{eq:BHrel}, we also consider the following alternative dimensionless normalization,
\begin{align}
\frac{\tau\hat{s}_\textrm{H}(\tau)}{\Lambda^2} = \frac{\hat{S}_\textrm{H}(\tau)}{\mathcal{A}\Lambda^2} = \frac{2\pi A_\textrm{H}(\tau)}{\mathcal{A}\Lambda^2} = \frac{2\pi|\Sigma(\tau,u_\textrm{H})|^3}{\Lambda^2}.
\label{eq:normS}
\end{align}
The parameter $\Lambda$ is a characteristic energy scale determined for each individual initial state being evolved in time. Its value is obtained by fitting the late-time behavior of our numerical energy density to the established analytical solution from Navier-Stokes hydrodynamics \cite{Critelli:2018osu,Rougemont:2021qyk,Rougemont:2021gjm,Rougemont:2022piu}.

The energy scale $\Lambda$ is also employed here to construct a dimensionless time measure, $\omega_\Lambda(\tau) \equiv \tau T_{\textrm{eff}}(\tau)$. Here, as also done in Refs.~\cite{Rougemont:2021qyk,Rougemont:2021gjm,Rougemont:2022piu}, we take the `effective non-equilibrium temperature', $T_{\textrm{eff}}(\tau)$, to be defined according to the following third order hydrodynamic expression derived for the purely thermal SYM plasma \cite{Booth:2009ct,Florkowski:2017olj}:
\begin{align}
    T_{\textrm{3rd}}^{\text{SYM}}(\tau) &= \frac{\Lambda}{(\Lambda\tau)^{1/3}} \Big[ 1 - \frac{1}{6\pi(\Lambda\tau)^{2/3}} + \frac{-1+\ln 2}{36\pi^2(\Lambda\tau)^{4/3}} \nonumber\\
&\hspace{1.5cm}+\frac{-21+2\pi^2+51\ln(2)-24\ln^2(2)}{1944\pi^3(\Lambda\tau)^2} \Big].
    \label{eq:time}
\end{align}

\subsection{Non-Equilibrium One-Point Functions \label{Sec:NonEquilibrium-1PointFunctions}}

Before presenting the non-equilibrium versions of the one-point functions, $\langle T_{\mu\nu}\rangle $, $\langle J^\mu\rangle$ and $\langle O_\phi\rangle$, we need to establish a map between the EF radial coordinate, $r$, in terms of which the non-equilibrium metric ansatz in \eqref{eq:BFDansatze} is expressed, and the FG radial coordinate, $\rho$, in terms of which the renormalized one-point functions are written.

This can be accomplished by comparing the diagonalized version of the EF metric \eqref{eq:BFDansatze}, with the FG metric \eqref{eq:ansatzequil}. By introducing the EF time coordinate, given by Eq.~\eqref{eq:EFtime}, into Eq.~\eqref{eq:BFDansatze}, together with the requirement that the resulting $rt$-component is zero, we obtain that, $g_{tt}=-2A$ and $g_{rr}=1/2A$, which diagonalizes Eq.~\eqref{eq:BFDansatze}. Therefore, by imposing the equality $(g_{rr}dr^2)_{\text{EF}}=(g_{\rho\rho}d\rho^2)_{\text{FG}}$, it follows that,
\begin{equation}
    \int \frac{dr}{\sqrt{2 A(\tau,r)}}=- \int \frac{d\rho}{2\rho}=-\frac{1}{2}\ln\rho, \label{eq:RelIntRo}
\end{equation}
where the minus signs is a direct consequence  of the condition $r\to\infty\Rightarrow\rho\to 0$. By expanding the integral~\eqref{eq:RelIntRo} as a perturbative series in terms of negative powers of $r$ with the help of Eq.~\eqref{eq:expA2}, we obtain the following relation between the radial coordinates $r$ and $\rho$~\cite{Rougemont:2021gjm},
\begin{align}
r(\rho)=\frac{1}{\sqrt{\rho}}-\frac{a_2(\tau)\rho^{3/2}}{4}-\frac{\partial_\tau a_2(\tau)\rho^2}{10} + \mathcal{O}(\rho^{5/2}).
\label{eq:rrho}
\end{align}
Eq.~\eqref{eq:rrho} is the same for both the 1RCBH and 2RCBH models. Matching the FG metric coefficients from Eq.~\eqref{eq:ansatzequil} with the EF ansatze in Eq.~\eqref{eq:BFDansatze}, and substituting the $r(\rho)$-expansion from Eq.~\eqref{eq:UVasymptotic} into the resulting expression yields,
\begin{subequations}
\begin{align}
\gamma_{\tau\tau}(\rho)&=-\frac{1}{\rho }-\frac{3}{2} \rho  a_2(\tau)+\mathcal{O}(\rho^{3/2}),\\
\gamma_{xx}(\rho)&=\frac{1}{\rho }+\frac{1}{12} \rho  \Big(-9 \tau \partial_\tau a_2(\tau)-18 a_2(\tau)\nonumber\\
&-\tau \phi _2(\tau) \partial_\tau\phi_2(\tau)-2 \phi _2(\tau){}^2\Big)+\mathcal{O}(\rho^{3/2}),\\
\gamma_{\xi\xi}(\rho)&=1+\frac{\tau^2}{\rho }+\frac{2 \tau}{\sqrt{\rho }}+\frac{\tau^2}{6}\rho   \Big(9 \tau \partial_\tau a_2(\tau)+9 a_2(\tau)\nonumber\\
&+\tau \phi _2(\tau) \partial_\tau\phi_2(\tau)\Big)+\mathcal{O}(\rho^{3/2}),\\
\phi(\rho)&=\rho\phi_2(\tau)+\mathcal{O}(\rho^{3/2}),\\
\Phi(\rho)&=\Phi_0(\tau)+\rho\Phi_2(\tau)+\mathcal{O}(\rho^{3/2}).
\end{align}
\end{subequations}

Inserting Eq.~\eqref{eq:rrho} into the near-boundary expansions given by~\eqref{eq:expA2} ---~\eqref{eq:expdphi2} allows us to extract the UV coefficients in FG coordinates. These coefficients are required for the holographic renormalization of the dual QFT one-point functions~\cite{Critelli:2017euk}. Consequently, the renormalized one-point functions for both the 1RCBH and 2RCBH plasmas subjected to the Bjorken flow are given by~\cite{Critelli:2018osu},
\begin{subequations}
\begin{align}
&\hat{\varepsilon}(\tau) \equiv \kappa_{5}^{2}\langle T_{\tau\tau} \rangle(\tau) = -3a_2(\tau)-\frac{1}{6}\phi_{2}(\tau)^{2},\label{eq:hatE}\\
&\hat{p}_T(\tau) \equiv \kappa_{5}^{2}\langle T_x^x \rangle(\tau) =\kappa_{5}^{2}\langle T_y^y \rangle(\tau) = -3a_2(\tau)\nonumber\\
&\qquad-\frac{1}{6}\phi_{2}(\tau)^{2}-\frac{3}{2}\tau \partial_\tau a_2(\tau) -\frac{1}{6}\tau\phi_{2}(\tau)\partial_\tau \phi_2(\tau), \label{eq:hatpT}\\
&\hat{p}_L(\tau) \equiv \kappa_{5}^{2}\langle T_\xi^\xi \rangle(\tau) =  3a_2(\tau)+\frac{1}{6}\phi_{2}(\tau)^{2}+3\tau \partial_\tau a_2(\tau) \nonumber\\
&\qquad+\frac{1}{3}\tau\phi_{2}(\tau)\partial_\tau \phi_2(\tau), \label{eq:hatpL}\\
&\hat{\rho}_c(\tau) \equiv \kappa_{5}^{2}\langle J^{\tau} \rangle(\tau) = -\Phi_2(\tau) =  \frac{\rho_0}{\tau}, \label{eq:hatrho}\\
&\langle\hat{O}_\phi\rangle(\tau) \equiv \kappa_{5}^{2}\langle \mathcal{O}_{\phi} \rangle(\tau) = -\phi_2(\tau). \label{eq:hatOphi}
\end{align}
\end{subequations}
From the above holographic formulas for the physical observables in the 2RCBH and 1RCBH plasmas, it is straightforward to verify that the boundary system is conformal, possessing a vanishing trace anomaly: $\langle T_\mu^\mu\rangle = -\varepsilon+p_L+2p_T=0$. Here, $p_L$ and $p_T$ denote the longitudinal and transverse pressures, respectively, while $\Delta p\equiv p_T - p_L$ represents the pressure anisotropy. Furthermore, expressing the bulk electric field from Eq.~\eqref{eq:MaxField2} in terms of the R-charge density $\rho_c(\tau)=\kappa_5^{-2} \rho_0/\tau$, we find,
\begin{subequations}
\begin{align}
&\text{1RCBH model:}\nonumber\\
&\mathcal{E}(\tau,r)= -\frac{8\pi^2\tau\rho_c(\tau)}{N_c^2}\,\Sigma^{-3}(\tau,r)\,e^{2\sqrt{\frac{2}{3}}\phi(\tau,r)},\\
&\nonumber\\
&\text{2RCBH model:}\nonumber\\
&\mathcal{E}(\tau,r) = -\frac{8\pi^2\tau\rho_c(\tau)}{N_c^2}\,\Sigma^{-3}(\tau,r)\,e^{-\sqrt{\frac{2}{3}}\phi(\tau,r)}.
\label{eq:Efield2R2}
\end{align}
\end{subequations}

Finally, to determine the value of the dimensionless ratio of R-charge chemical potential over temperature, $\mu/T$, we evolve our numerical simulation until it reaches its late time regime, and compare the asymptotic value for the ratio $[\hat{\rho}_c^{4/3}/\hat{\varepsilon}](\tau)$ with its corresponding stable equilibrium solutions in Eqs.~\eqref{eq:rho43eThermo}. We adopt this approach because the ratio $\mu/T$ is not an initial data for the holographic Bjorken flow, but an emergent result extracted from the late time asymptotic behavior of the system. For accurate estimations of the values of $\mu/T$ analyzed in the present work, we choose to perform the above comparison at a large late time $\tau_\text{end}=60$, where the physical observables are expected to have converged to their asymptotic values within some numerical tolerance.\footnote{For increasing values of the initial charge density parameter $\rho_0$, such a convergence is slower, and since $\mu/T$ generally increases with $\rho_0$, larger values of $\mu/T$ require larger values of $\tau_\text{end}$ in order to be accurately estimated. This is particularly important for the 2RCBH model which has $\mu/T\in[0,\infty)$.}

By substituting the value of $\mu/T$, as extracted from the aforementioned analysis, into the known analytical expressions for the stable equilibrium observables $[\langle\hat{O}_\phi\rangle/\hat{\varepsilon}^{1/2}]_\textrm{eq}$ and $[\hat{s}_\textrm{H}^{4/3}/\hat{\varepsilon}]_\textrm{eq}$, given by Eqs.~\eqref{eq:thermo-ratios}, we obtain the equilibrium asymptotes that the corresponding dynamical quantities are expected to approach at late times. Since both the equilibrium and near-equilibrium hydrodynamic values for these quantities are independent of the specific far-from-equilibrium initial conditions, depending only on the ratio $\mu/T$, these asymptotes serve as a robust analytical validation of our numerical results.

Analogously, the pressure anisotropy, $[\Delta \hat{p}/\hat{\varepsilon}](\tau)$, must converge at late times to the analytical result derived from Navier-Stokes (NS) hydrodynamics in Eq.~\eqref{eq:pressure-anisotropy-EOM}, with $\hat{\varepsilon}_\textrm{NS}(\tau,\mu/T)$ given by Eq.~\eqref{eq:energytau} evaluated at the same value of $\mu/T$ obtained from the late time analysis of $ [\hat{\rho}_c^{4/3}/\hat{\varepsilon}](\tau)$.  Indeed, for all initial conditions, we consistently observe the convergence of these different physical observables to their corresponding analytical hydrodynamic or thermodynamic limits, fixed by the same value of $\mu/T$. These outcomes constitute independent and highly nontrivial consistency checks of our numerical simulations against analytical results in the late time regime of the system.

\subsection{Initial Data and Energy Conditions\label{sec:data_EC}}

As mentioned in Section~\ref{sec:apparent-hor}, the numerical simulation of the fluid's expansion requires the specification of the initial data $\{B_s(\tau_0,u),\phi_s(\tau_0,u),a_2(\tau_0),\rho_0;\lambda(\tau_0)\}$. In this work, we set $\lambda(\tau_0)=0$ and consider the following general formulas for the initial profile of the subtracted dilaton field $\phi_s(\tau_0,u)$, and metric anisotropy coefficient $B_s(\tau_0,u)$,
\begin{subequations}
\label{eq:ICprofileformula}
    \begin{align}
\phi_s(\tau_0,u) &= \sum_{i=0}^{3}c_i u^i + \zeta\, e^{-u^2/\sigma} %+ \alpha_1 \cos(\omega_1 u) + \alpha_2 \sin(\omega_2 u),
\label{eq:phis0}\\
B_s(\tau_0,u) &= \Omega \sin(\gamma u) + \sum_{i=0}^{5}\beta_i u^i  + \frac{\alpha}{u^4} \Big[-\frac{2}{3} \ln\left(1+ \frac{u}{\tau _0}\right) \nonumber\\
&\hspace{2cm}+ \frac{2 u^3}{9 \tau_0^3} - \frac{u^2}{3 \tau _0^2}+\frac{2 u}{3 \tau _0}\Big].\label{eq:Bs0}
\end{align}
\end{subequations}
We remark that formulas~\eqref{eq:ICprofileformula} are particular cases of more general functional forms employed in Ref.~\cite{Rougemont:2022piu}.

Throughout this work, we employed a set of four different combinations of initial conditions (ICs) for the profiles of the subtracted functions $\phi_s(\tau_0,u)$ and $B_s(\tau_0,u)$. These varied profiles enable the investigation of a diverse range of dynamical scenarios for the 1RCBH and 2RCBH plasmas undergoing Bjorken flow. The particular set of values for their parameters is enumerated in Table~\ref{tab:parameters}.

We remark that we tested several other ICs besides those specified by the particular values presented in Table~\ref{tab:parameters}. Based on the features we selected to analyze (which, as we are going to discuss, feature ICs that do not develop transient violations of energy conditions, as well ICs that present such transient violations with different structures for the extrema of the pressure anisotropy; ICs without transient plateaus for the entropy, as well ICs with such plateaus), at least from the larger set of ICs that we effectively simulated, the 4 ICs displayed in Table~\ref{tab:parameters} are a representative set of typical situations, in the sense that the other ICs that we have not plotted behave in qualitatively similar ways to those plotted, at least with regard to the selected features we analyzed. A more systematic approach, scanning for other possible general features and correlations that we may have eventually missed in our analysis, would likely require evolving in time a very large set of ICs (which is a computationally demanding task) and then using some neural network to identify other possible patterns that we eventually did not notice.

\begin{table*}[h!]
\centering
% O resizebox ajusta a tabela para a largura do texto
\begin{tabular}{|c||c|c|c|c|c|c|c|c|c||c|c|c|c|c|c|}
\hline
\multicolumn{1}{|c||}{IC} & \multicolumn{9}{c||}{$B_s$ parameters}  & \multicolumn{6}{c|}{$\phi_s$ parameters} \\
\hline
\textbf{\#} & $\Omega$ & $\gamma$ & $\beta_0$ & $\beta_1$ & $\beta_2$ & $\beta_3$ & $\beta_4$ & $\beta_5$ & $\alpha$ & $c_0$ & $c_1$ & $c_2$ & $c_3$ & $\zeta$ & $\sigma$ \\
\hline
\hline
01 & 0.0 & 0.0 & 0.5 & -0.5 & 0.4 & 0.2 & -0.3 & 0.1 & 1.0 & 0.7 & -0.9 & 0.0 & 0.4 & 0.0 & 1.0 \\
\hline
02 & 0.0 & 0.0 & -0.2 & -0.5 & 0.0 & 0.0 & 0.0 & 0.0 & 1.0 & 0.0 & 0.0 & 0.0 & 0.0 & 0.0 & 1.0 \\
\hline
03 & 0.0 & 0.0 & -0.3 & -0.5 & 0.0 & 0.0 & 0.0 & 0.0 & 1.0 &-0.8 & 1.1 & 0.6 & 0.0 & 0.0 & 1.0 \\
\hline
04 & 1.0 & 8.0 & 0.0 & 0.0 & 0.0 & 0.0 & 0.0 & 0.0 & 1.0 & 0.0 & 0.0 & 0.0 & 0.0 & 0.5 & 2.0 \\
\hline
\end{tabular}%
\caption{Set of parameters for the initial profile of the subtracted dilaton field \eqref{eq:phis0} and the metric anisotropy \eqref{eq:Bs0} analyzed in this work.}
\label{tab:parameters}
\end{table*}

As established in Section~\ref{sec:mainalgo}, the initial charge density of the medium, $\hat{\rho}_c(\tau_0)=\rho_0/\tau_0$, is determined by the parameter $\rho_0$, whereas the initial energy density, $\hat{\varepsilon}(\tau_0)$, is in turn specified by the combination of $a_2(\tau_0)$ and $\phi_2(\tau_0)=\phi_s(\tau_0,u=0)$ via Eq.~\eqref{eq:hatE}. In the subsequent section, we are going to explore the effects of varying the initial charge density for fixed configurations of the subtracted function profiles and the initial value of $a_2(\tau)$. In particular, to observe the effective thermalization of the scalar condensate at late times, we evolve the numerical simulations from an initial proper time $\tau_0=0.2$ to the final time $\tau_\text{end}=60$. This time duration is approximately two times longer than that used in Ref.~\cite{Rougemont:2022piu}.

To complete our analysis, we explore if the weak energy condition (WEC) and the dominant energy condition (DEC) are valid for the strongly interacting conformal fluids which we are studying under the Bjorken flow expansion. The latter are conditions, usually enforced on classical matter sources in general relativity, to guarantee, for example, energy positivity and the requirement of causality. It is important to remark, that although being a criterion for reasonability of possible classical matter configurations, there exists certain quantum phenomena, such as the Casimir effect, that are known to violate such energy conditions~\cite{Visser:1999de,Costa:2021hpu}. 

\begin{figure*}
\centering  
\subfigure[]{\includegraphics[width=0.475\linewidth]{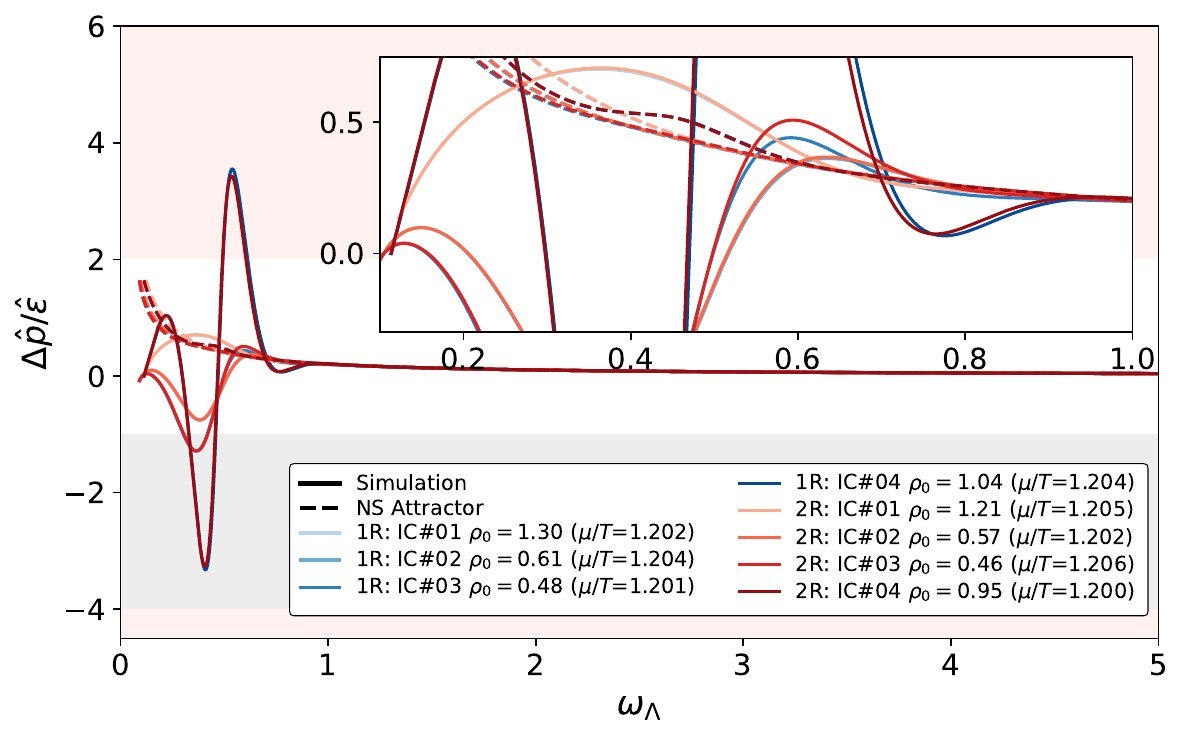}\label{fig:Comparisona}}
\subfigure[]{\includegraphics[width=0.475\linewidth]{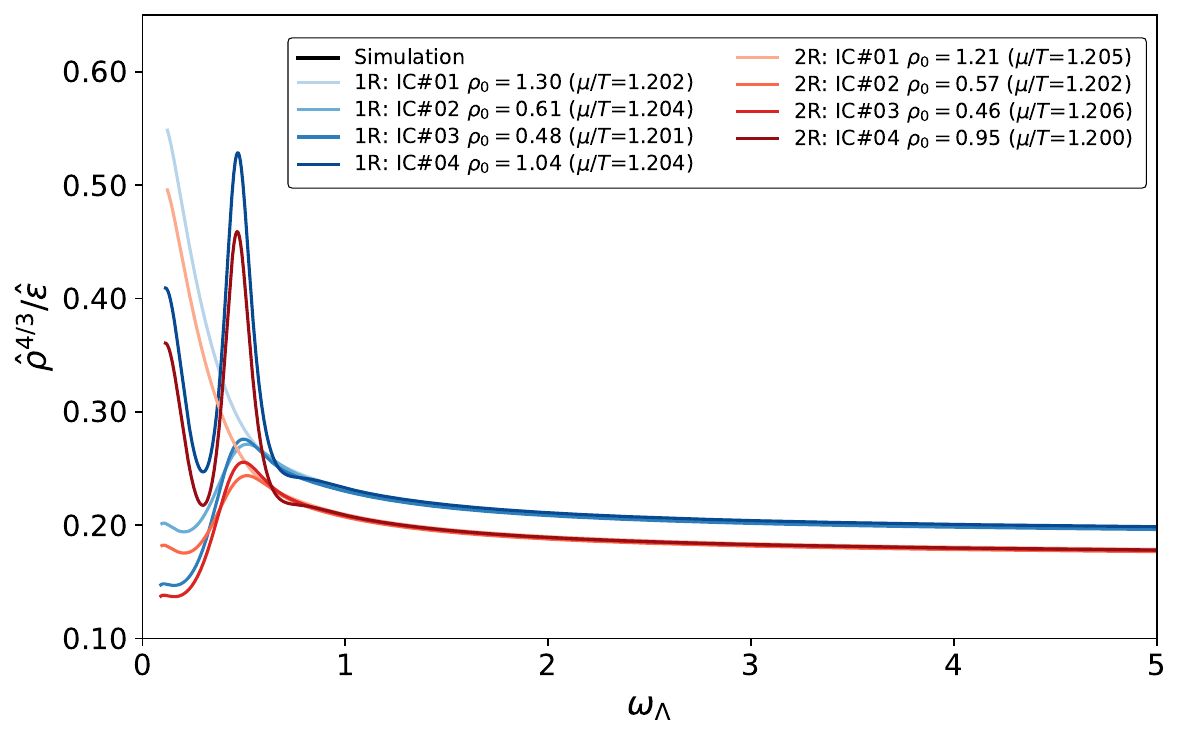}\label{fig:Comparisonb}}
\subfigure[]{\includegraphics[width=0.475\linewidth]{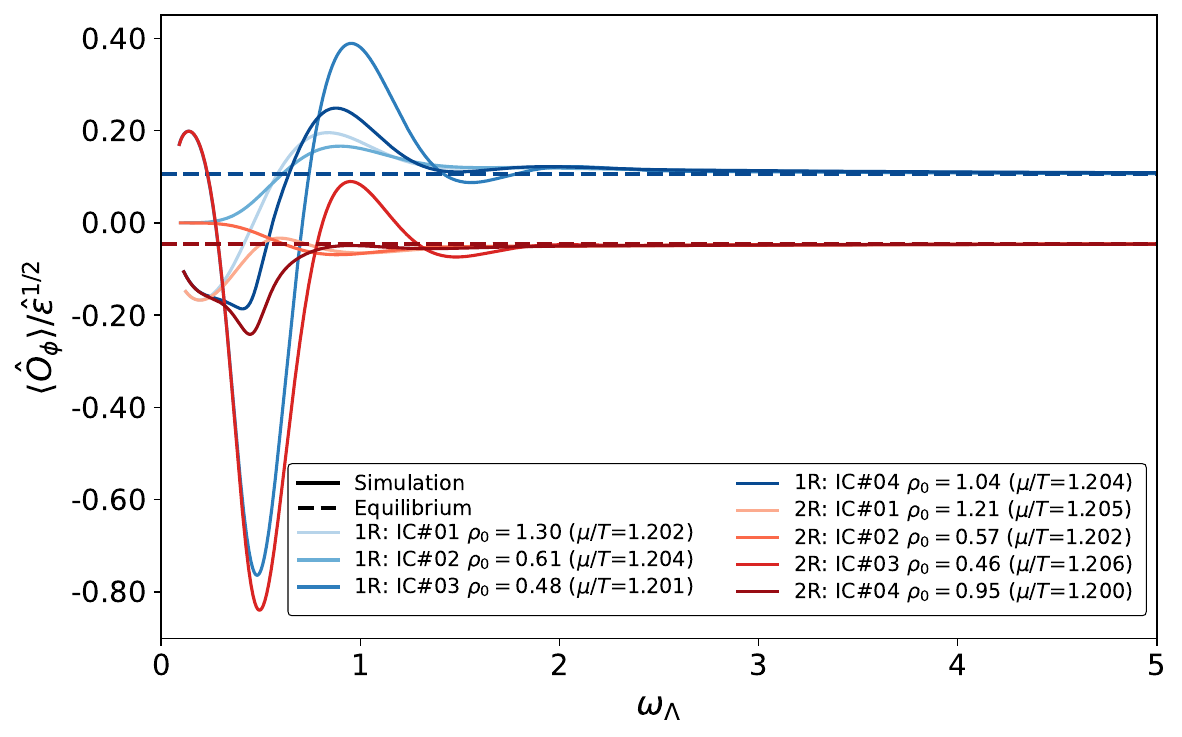}\label{fig:Comparisonc}}
\subfigure[]{\includegraphics[width=0.475\linewidth]{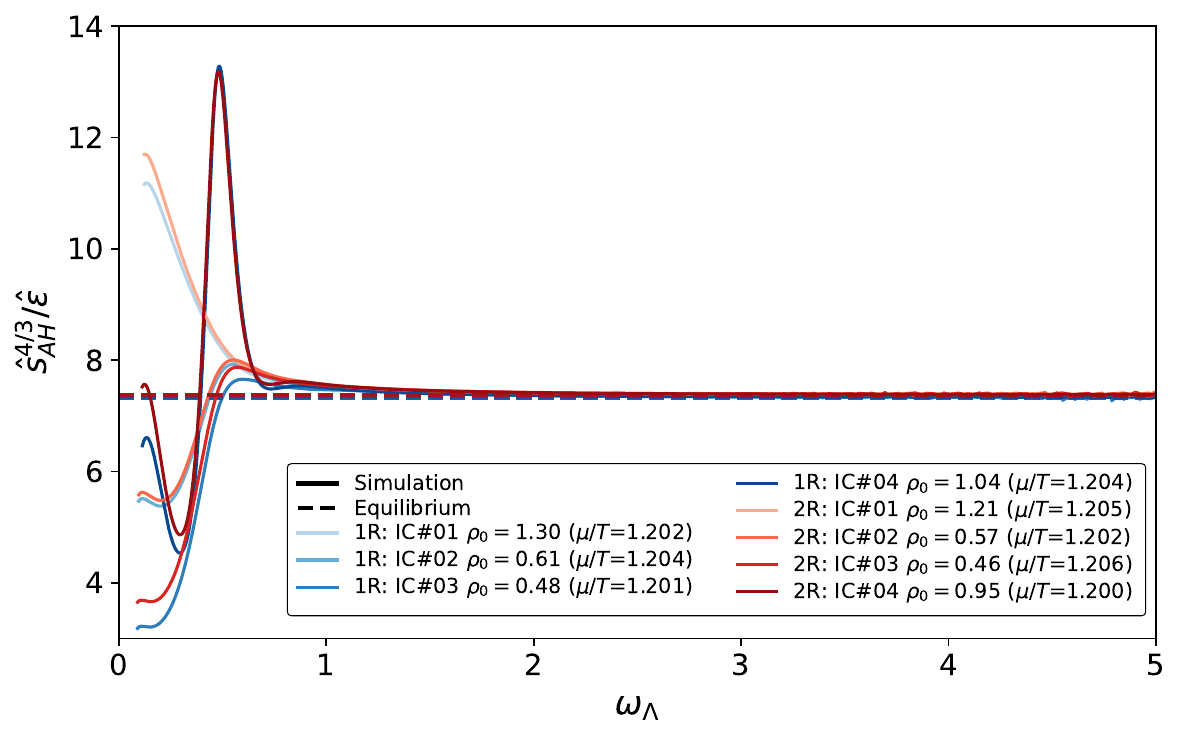}\label{fig:Comparisond}}
\subfigure[]{\includegraphics[width=0.475\linewidth]{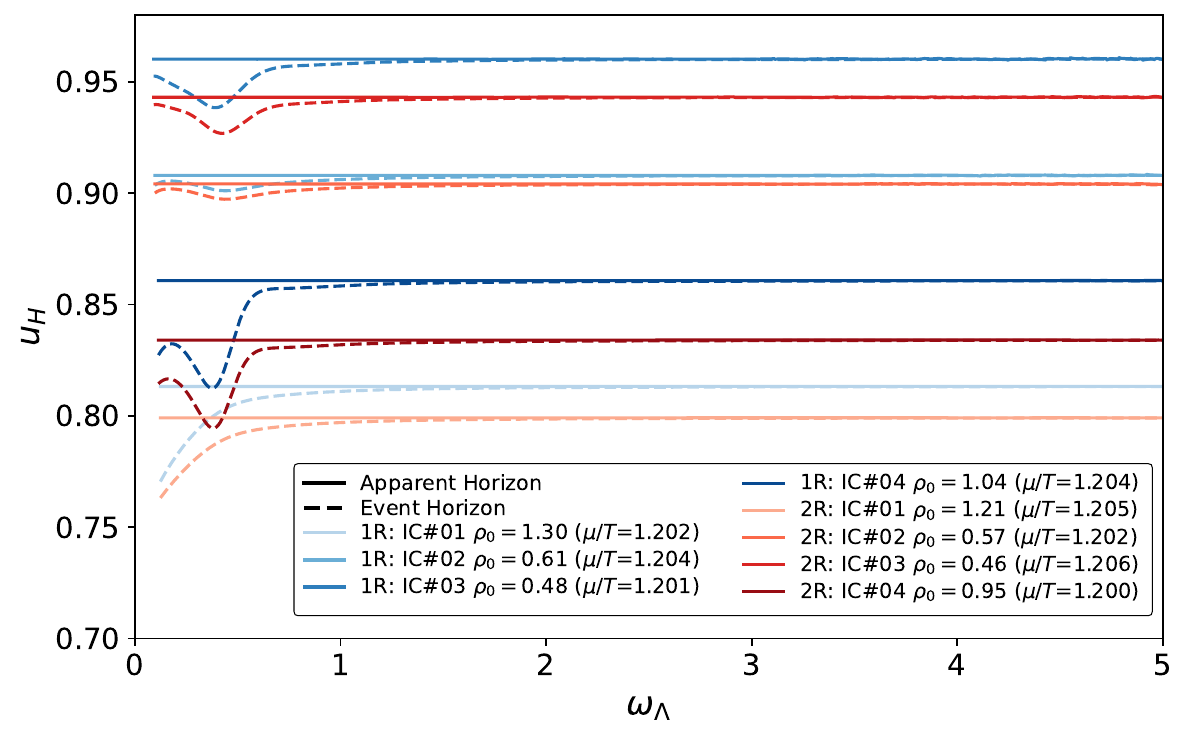}\label{fig:Comparisone}}
\subfigure[]{\includegraphics[width=0.475\linewidth]{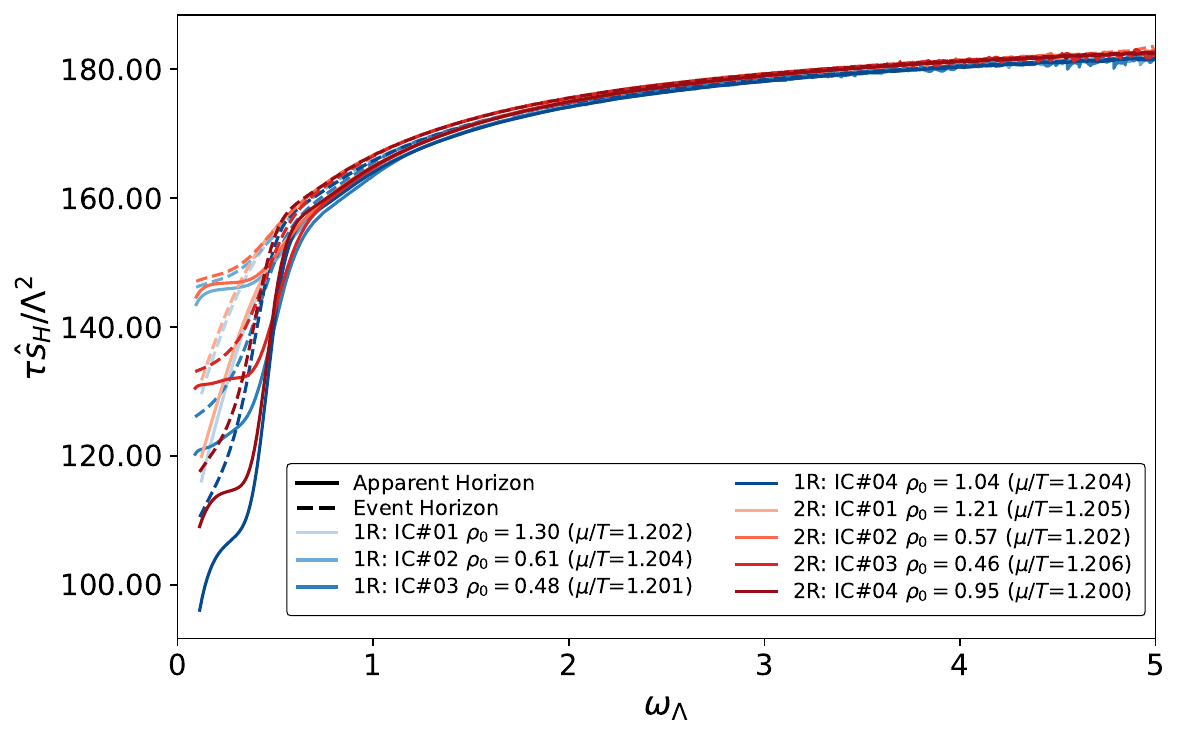}\label{fig:Comparisonf}}
\caption{(a) Normalized pressure anisotropy (solid lines) and the corresponding hydrodynamic Navier-Stokes attractor given by Eqs.~\eqref{eq:pressure-anisotropy-EOM} and~\eqref{eq:energyNS} (dashed lines) --- the gray and red zones indicate violations of the DEC and WEC, respectively; (b) normalized charge density; (c) normalized scalar condensate (solid lines) and the corresponding thermodynamic stable equilibrium result (dashed lines); (d) normalized non-equilibrium entropy density associated to the apparent horizon $\hat{s}_\text{AH}^{4/3}/\hat{\varepsilon}$ (solid lines) and the corresponding thermodynamic stable equilibrium result (dashed lines); (e) radial location of the apparent horizon (solid lines) and event horizon (dashed lines); and (f) non-equilibrium entropy $\hat{S}_H/\mathcal{A}\Lambda^2=\tau \hat{s}_H/\Lambda^2$ calculated from the area of the apparent horizon (solid line) and from the area of the event horizon (dashed lines). The plots make use of each of the four ICs in Table~\ref{tab:parameters} to compare the time evolution of different initial states of the 1RCBH and 2RCBH plasmas. The results were obtained by calibrating the initial charge density parameter $\rho_0$ such as to produce evolutions equilibrating towards the same approximate value of $\mu/T\approx 1.2$. For all the curves, the initial energy density was fixed using $a_2(\tau_0)=-7.40$.}
\label{fig:comparison}
\end{figure*}

Formally, the WEC asserts that, for any timelike vector $t^{\mu}$, the condition $\langle \hat{T}_{\mu\nu}\rangle t^\mu t^\nu\ge 0$ is true. Crudely, this means that the energy detected by an arbitrary observer in spacetime must always be non-negative. In the case of the Bjorken expansion of a conformal fluid, this condition entails the fulfillment of the following constraints \cite{Janik:2005zt,Rougemont:2021qyk,Rougemont:2021gjm},
\begin{align}
\hat{\varepsilon}(\tau)\ge 0\,\,\,\,\,\text{and}\,\,\,\,\, -4 \le \left[\frac{\Delta\hat{p}}{\hat{\varepsilon}}\right](\tau)\le 2.
\label{eq:WEC}
\end{align}
Consequently, a WEC violation must originate either from a negative energy density or through a normalized pressure anisotropy going outside the above interval --- in this work, as well as in~\cite{Rougemont:2021qyk,Rougemont:2021gjm,Rougemont:2022piu,deOliveira:2025lhx}, the observed WEC violations originated purely from the latter case. Such violations occurred only in the transient far-from-equilibrium regime of the time evolution of some initial states, while the energy density remained positive throughout the entire time evolution. Notably, for conformal systems, the strong energy condition (SEC), defined by $\langle\hat{T}_{\mu\nu}\rangle t^\mu t^\nu\ge -\langle \hat{T}_\mu^\mu\rangle /2$, reduces to the WEC as a consequence of the null trace of the energy-momentum tensor. The DEC, on the other hand, establishes that the vector $X^\mu \equiv -\langle\hat{T}^{\mu\nu}\rangle t_\nu$ needs to be a future-directed timelike or null vector for any timelike vector $t^{\mu}$ who is also future-directed.\footnote{By future-directed we mean a vector whose time component is positive. It indicates that it points towards the future (increasing time) and not towards the past.} In simple language, it can be understood as a sufficient condition, although not necessary, for matter to propagate in a causal manner~\cite{Wald:1984rg}. As demonstrated in~\cite{Rougemont:2021qyk,Rougemont:2021gjm}, in regard to conformal fluids, the DEC enforces the following bounds,
\begin{align}
\hat{\varepsilon}(\tau) \ge 0\,\,\,\,\,\text{and}\,\,\,\,\,  -1 \le\left[\frac{\Delta\hat{p}}{\hat{\varepsilon}}\right](\tau)\le 2.
\label{eq:DEC}
\end{align}

%%%%%%%%%%%%%%%%%%%%%%%%%%%%%%%%%
%%%%%%%%%%%%%%%%%%%%%%%%%%%%%%%%%

\section{Numerical Results and Physical Analysis}
\label{sec:results}

\begin{figure*}
\centering  
\subfigure[]{\includegraphics[width=0.475\linewidth]{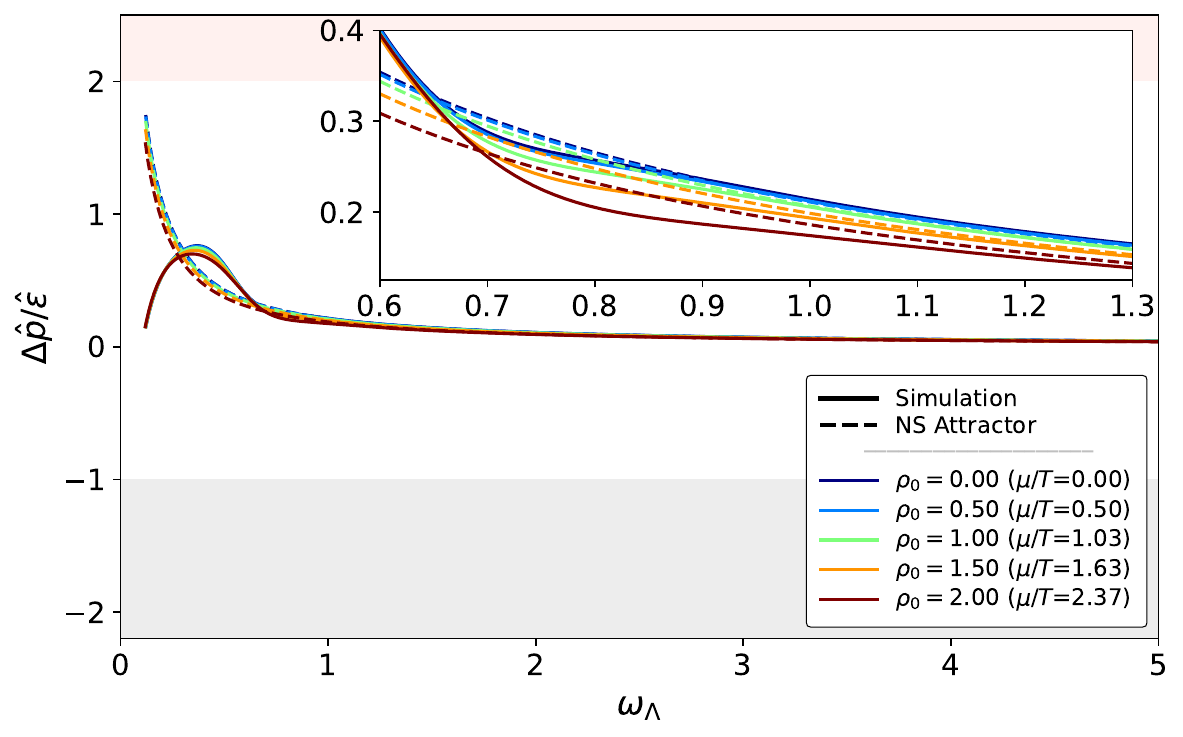}\label{fig:01a}}
\subfigure[]{\includegraphics[width=0.475\linewidth]{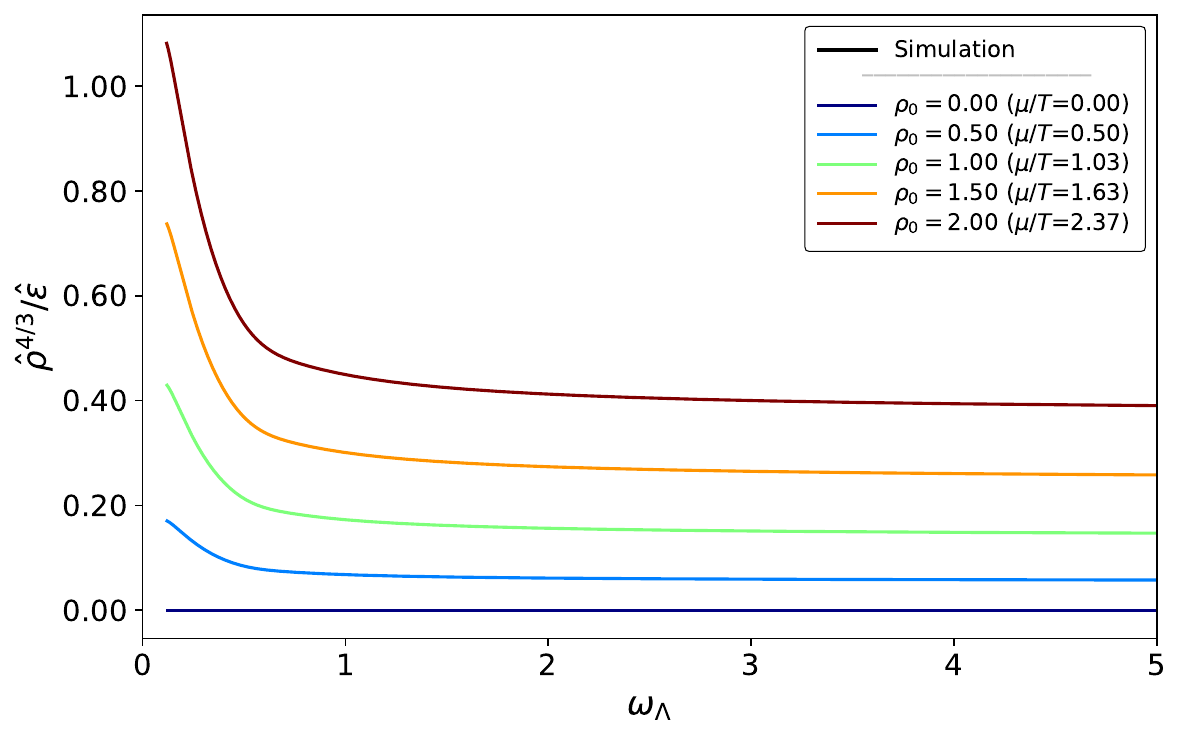}\label{fig:01b}}
\subfigure[]{\includegraphics[width=0.475\linewidth]{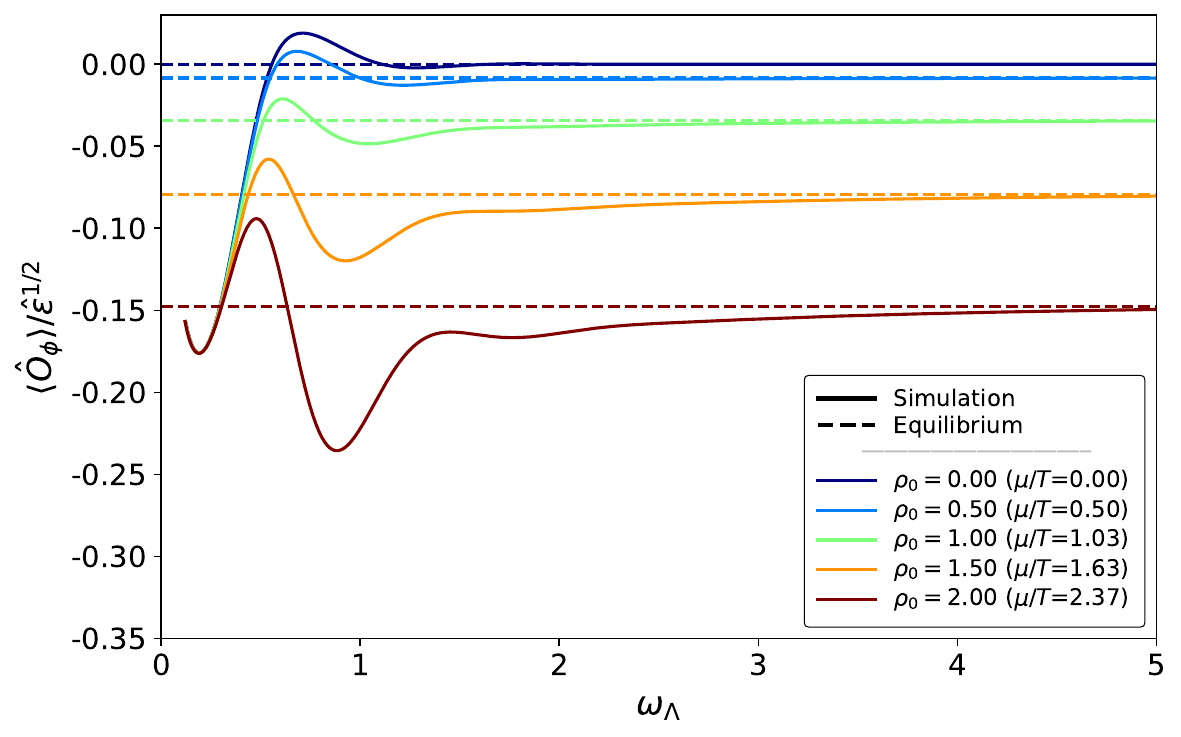}\label{fig:01c}}
\subfigure[]{\includegraphics[width=0.475\linewidth]{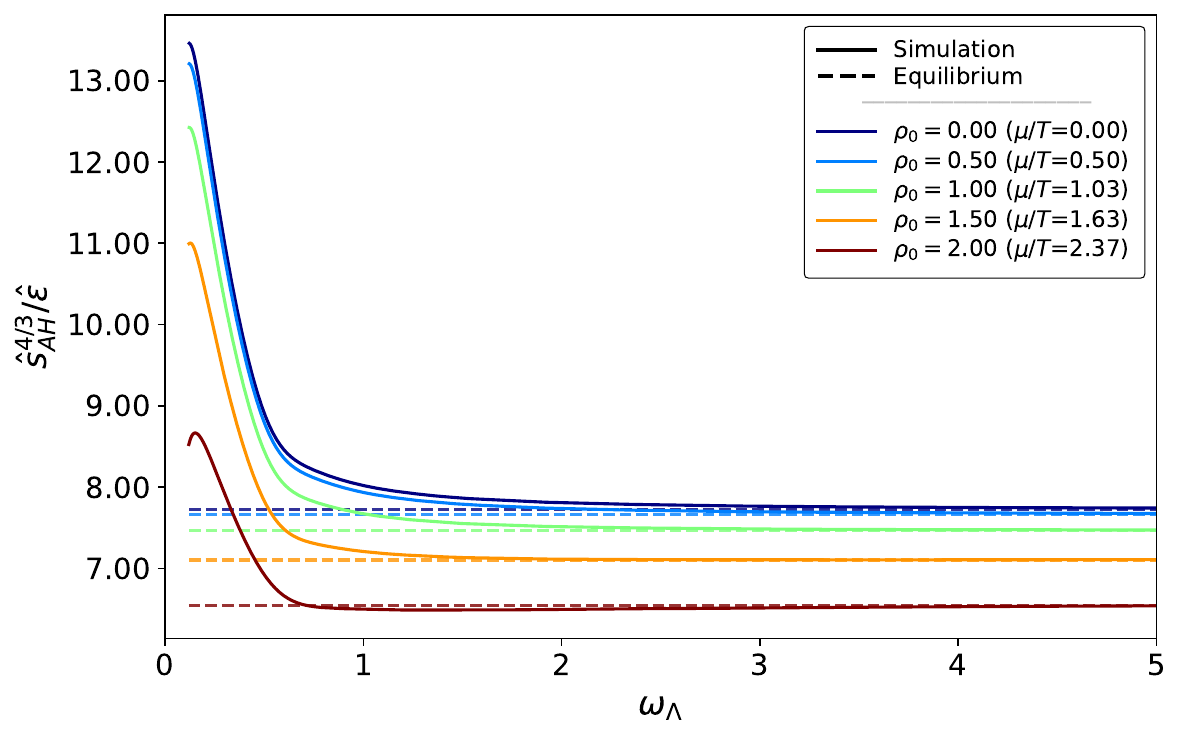}\label{fig:01d}}
\subfigure[]{\includegraphics[width=0.475\linewidth]{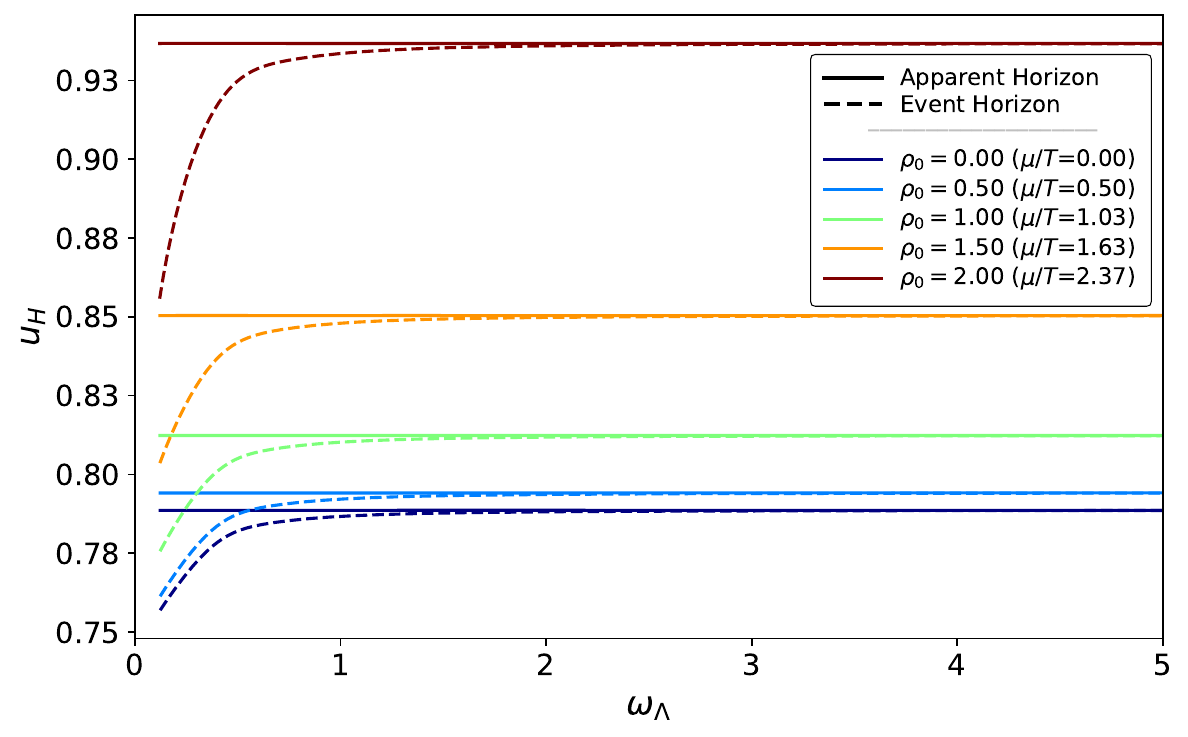}\label{fig:01e}}
\subfigure[]{\includegraphics[width=0.475\linewidth]{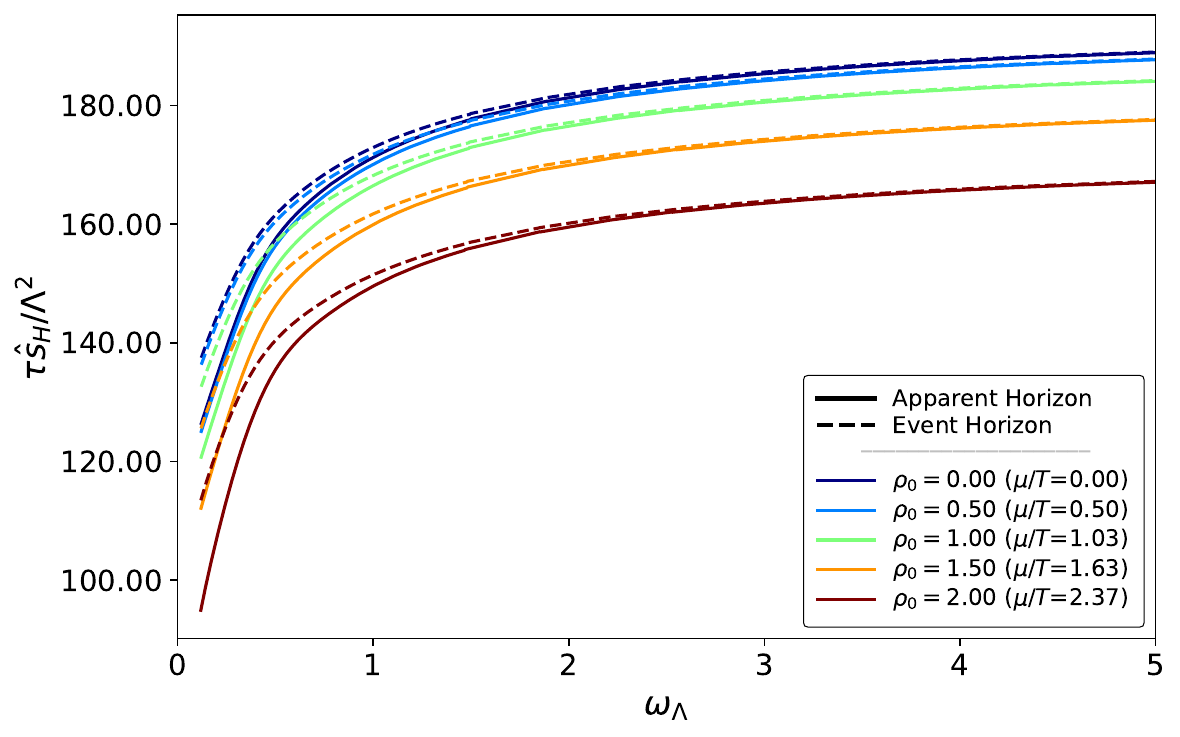}\label{fig:01f}}
\caption{(a) Normalized pressure anisotropy (solid lines) and the corresponding hydrodynamic Navier-Stokes attractor given by Eqs.~\eqref{eq:pressure-anisotropy-EOM} and~\eqref{eq:energyNS} (dashed lines) --- the gray and red zones indicate violations of the DEC and WEC, respectively; (b) normalized charge density; (c) normalized scalar condensate (solid lines) and the corresponding thermodynamic stable equilibrium result (dashed lines); (d) normalized non-equilibrium entropy density associated to the apparent horizon $\hat{s}_\text{AH}^{4/3}/\hat{\varepsilon}$ (solid lines) and the corresponding thermodynamic stable equilibrium result (dashed lines); (e) radial location of the apparent horizon (solid lines) and event horizon (dashed lines); and (f) non-equilibrium entropy $\hat{S}_H/\mathcal{A}\Lambda^2=\tau \hat{s}_H/\Lambda^2$ calculated from the area of the apparent horizon (solid line) and from the area of the event horizon (dashed lines). The results were obtained by varying the initial charge density parameter $\rho_0$, while keeping $B_s$ and $\phi_s$ fixed according to the IC\#01 in Table \ref{tab:parameters}, with $a_2(\tau_0)=-6.6667$.}
\label{fig:01}
\end{figure*}

\begin{figure*}
\centering  
\subfigure[]{\includegraphics[width=0.475\linewidth]{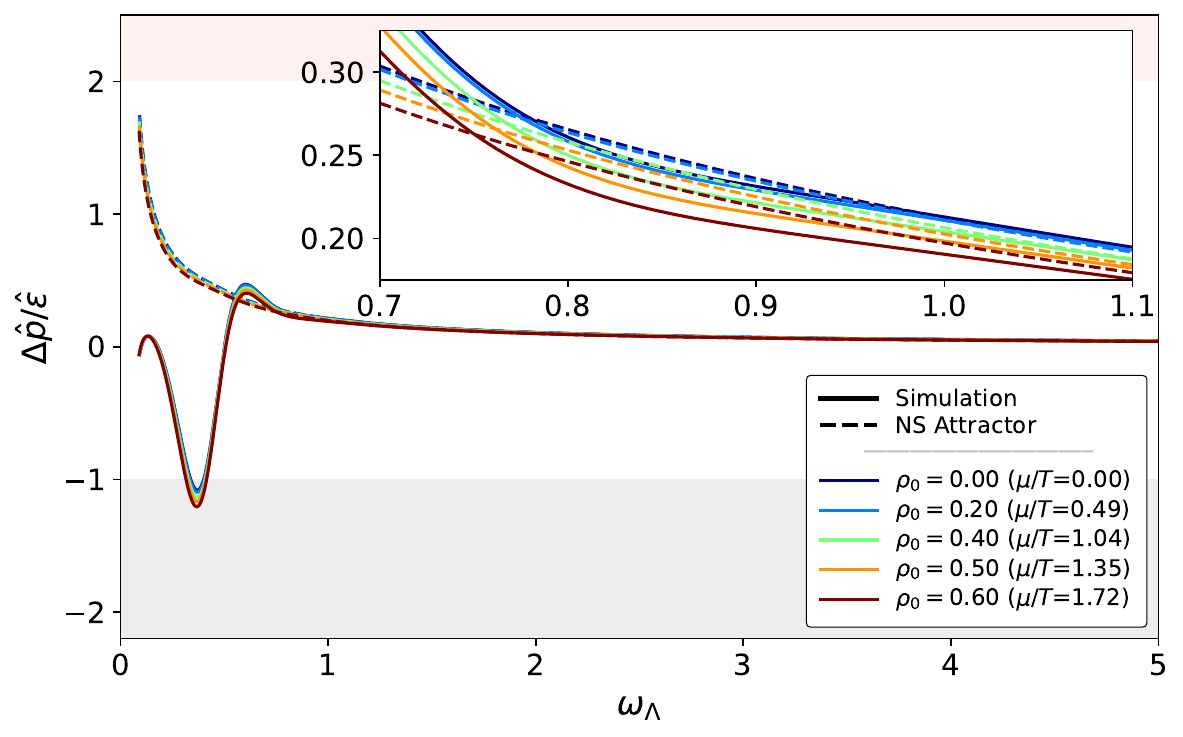}\label{fig:02a}}
\subfigure[]{\includegraphics[width=0.475\linewidth]{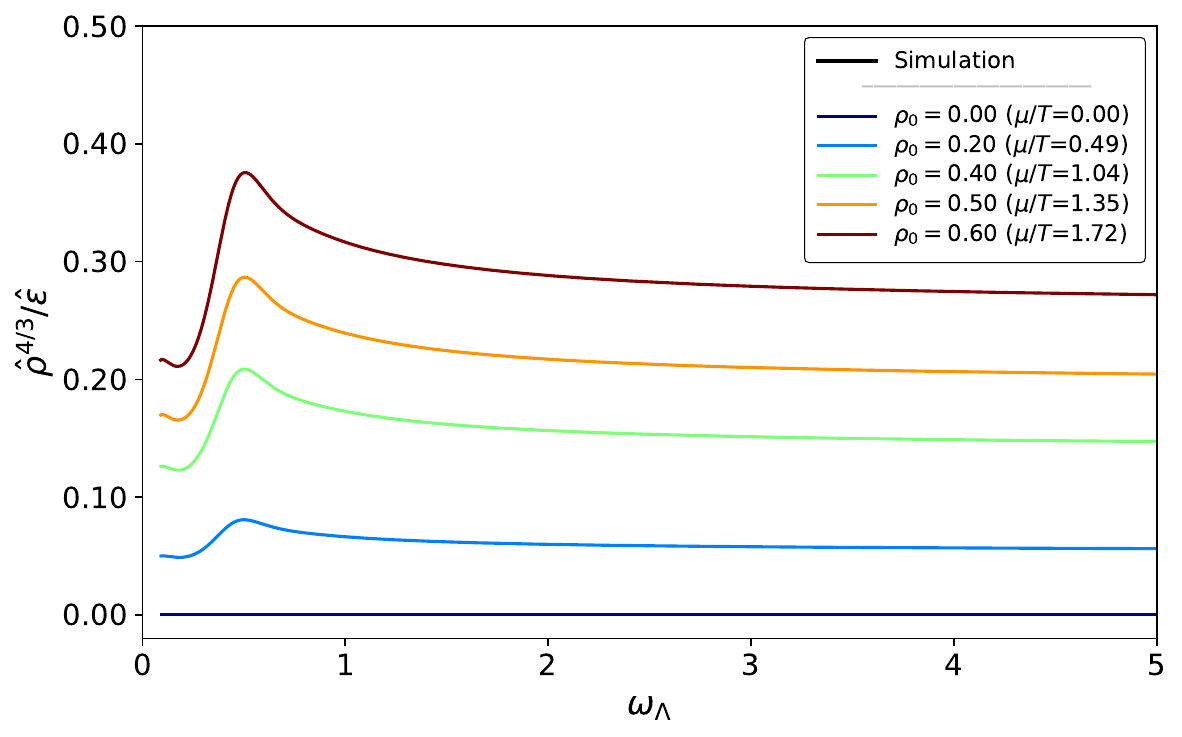}\label{fig:02b}}
\subfigure[]{\includegraphics[width=0.475\linewidth]{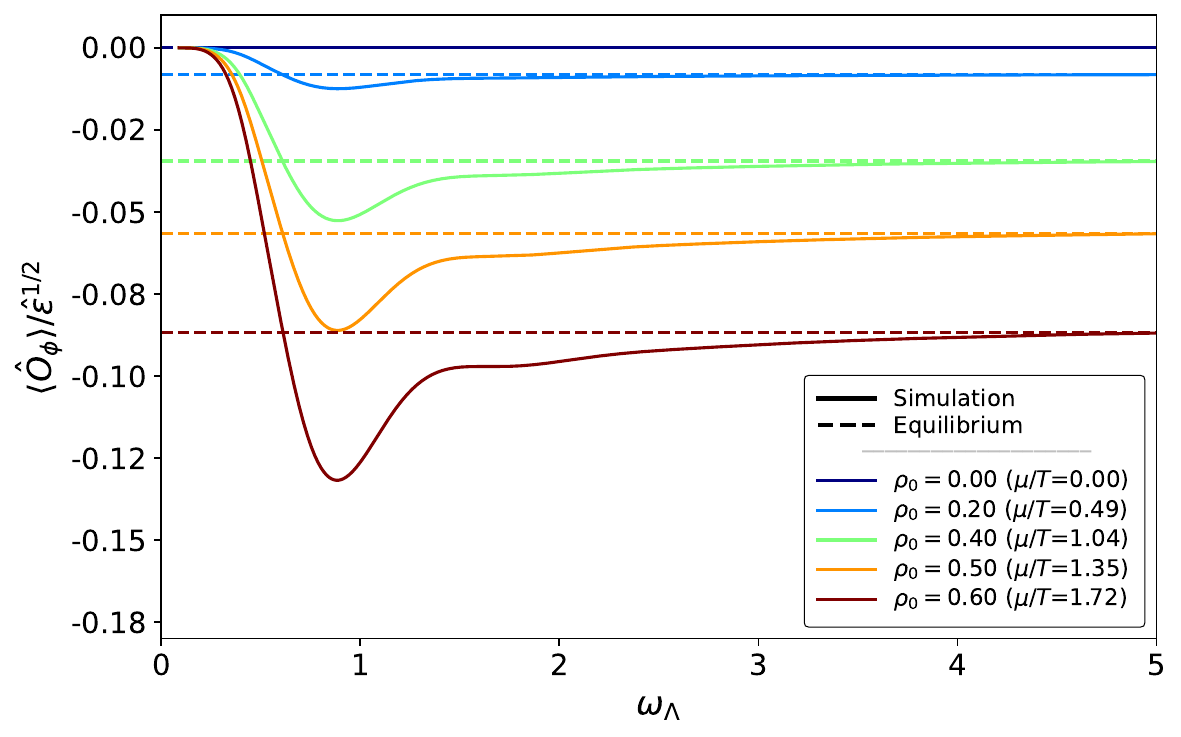}\label{fig:02c}}
\subfigure[]{\includegraphics[width=0.475\linewidth]{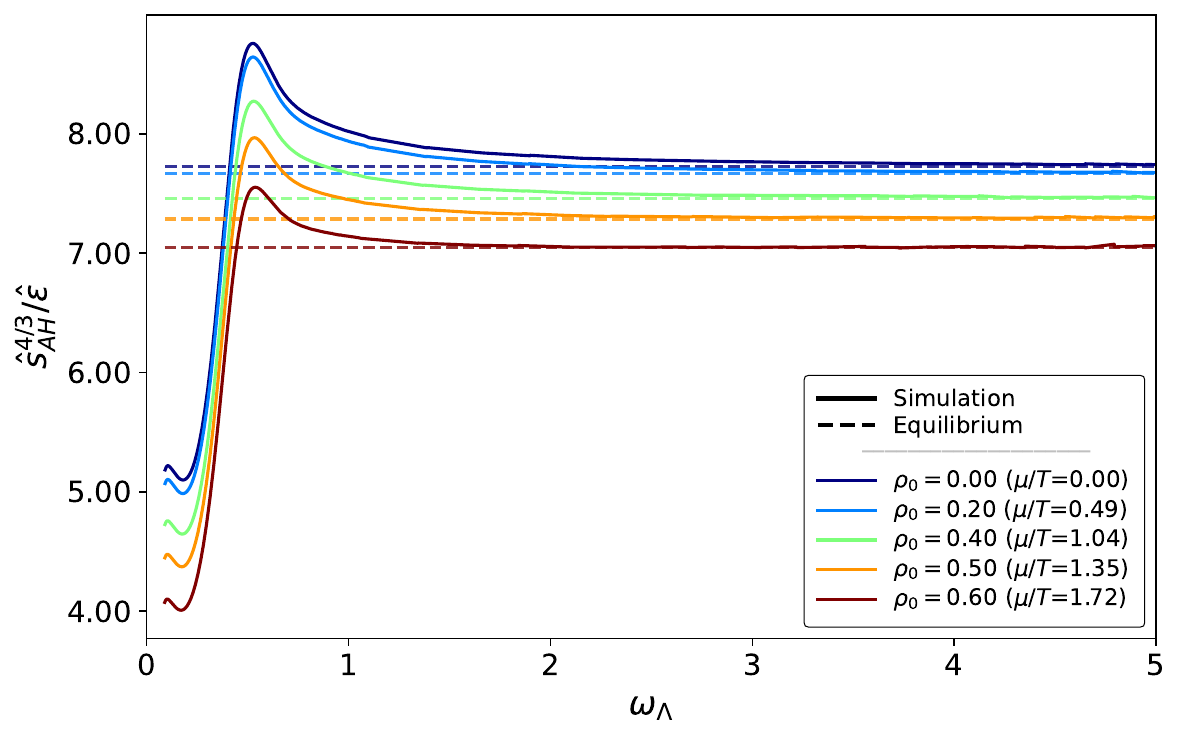}\label{fig:02d}}
\subfigure[]{\includegraphics[width=0.475\linewidth]{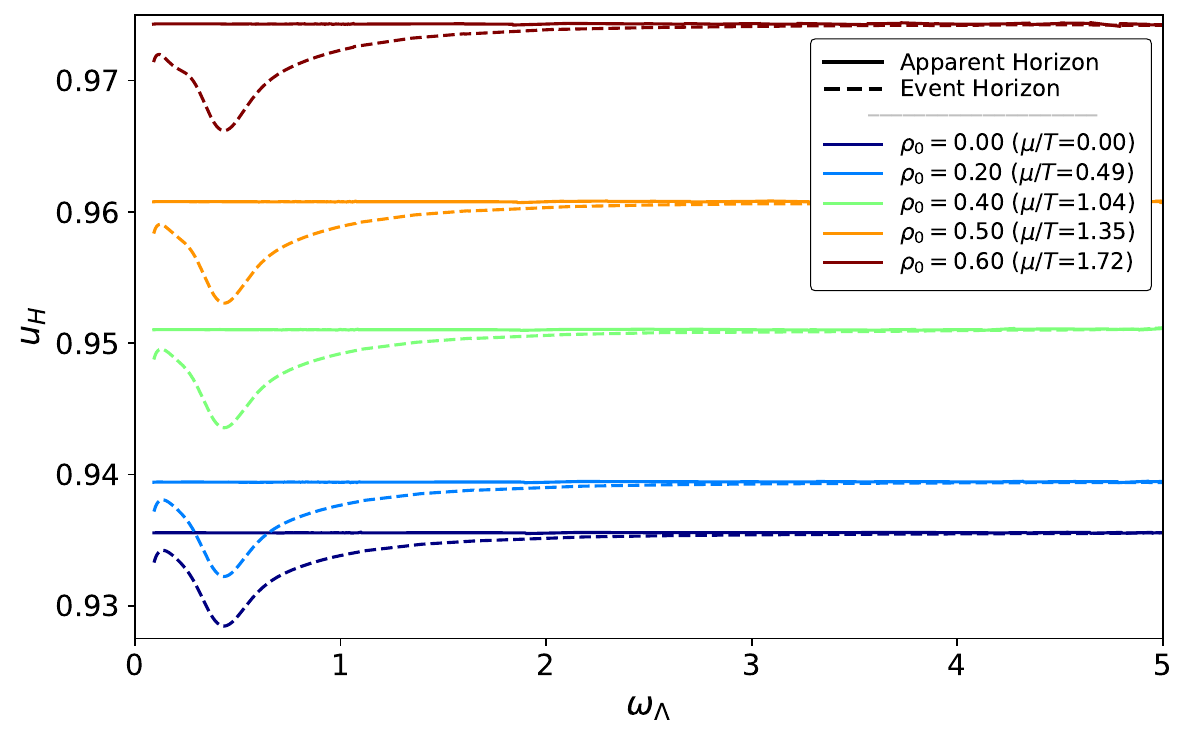}\label{fig:02e}}
\subfigure[]{\includegraphics[width=0.475\linewidth]{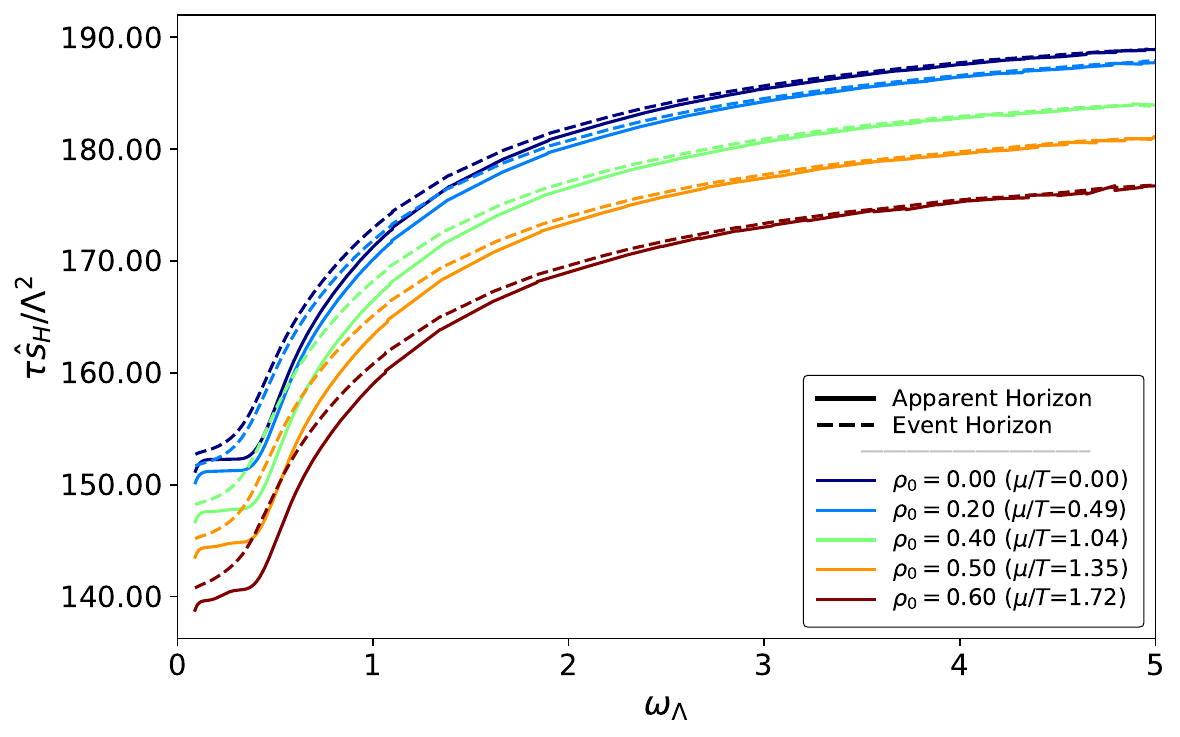}\label{fig:02f}}
\caption{(a) Normalized pressure anisotropy (solid lines) and the corresponding hydrodynamic Navier-Stokes attractor given by Eqs.~\eqref{eq:pressure-anisotropy-EOM} and~\eqref{eq:energyNS} (dashed lines) --- the gray and red zones indicate violations of the DEC and WEC, respectively; (b) normalized charge density; (c) normalized scalar condensate (solid lines) and the corresponding thermodynamic stable equilibrium result (dashed lines); (d) normalized non-equilibrium entropy density associated to the apparent horizon $\hat{s}_\text{AH}^{4/3}/\hat{\varepsilon}$ (solid lines) and the corresponding thermodynamic stable equilibrium result (dashed lines); (e) radial location of the apparent horizon (solid lines) and event horizon (dashed lines); and (f) non-equilibrium entropy $\hat{S}_H/\mathcal{A}\Lambda^2=\tau \hat{s}_H/\Lambda^2$ calculated from the area of the apparent horizon (solid line) and from the area of the event horizon (dashed lines). The results were obtained by varying the initial charge density parameter $\rho_0$, while keeping $B_s$ and $\phi_s$ fixed according to the IC\#02 in Table \ref{tab:parameters}, with $a_2(\tau_0)=-6.6667$.}
\label{fig:02}
\end{figure*}

In this section we present our main results. Figs.~\ref{fig:comparison} ---~\ref{fig:04} comprise six plots each, displaying results for the following quantities: $[\Delta \hat{p}/\hat{\varepsilon}](\omega_{\Lambda})$, $[\hat{\rho}_c^{4/3}/\hat{\varepsilon}](\omega_{\Lambda})$, $[\langle  \hat{O}_{\phi}\rangle/\hat{\varepsilon}^{1/2}](\omega_{\Lambda})$, $[\hat{s}_H^{4/3}/\hat{\varepsilon}](\omega_{\Lambda})$, $u_H(\omega_\Lambda)$ and $[\tau \hat{s}_H/\Lambda^2](\omega_{\Lambda})$. Fig.~\ref{fig:comparison} shows a comparison between the 1RCBH and the 2RCBH models, while Figs.~\ref{fig:01} ---~\ref{fig:04} focus on different time evolutions for the 2RCBH model.

\subsection{Comparison between the 1RCBH and 2RCBH Models\label{Sec:ComparisonModels}}

In Fig.~\ref{fig:comparison}, we compare the 1RCBH and 2RCBH models using the four ICs given in Table~\ref{tab:parameters}, with $a_2(\tau_0)=-7.40$. For each of these initial data, we calibrated the initial charge density parameter $\rho_0$ to ensure that all evolution trajectories converge in the late time regime to approximately the same equilibrium value of $\mu/T \approx 1.2$. The results for the 1RCBH and 2RCBH models are depicted in varying shades of blue and red curves, respectively.

Fig.~\ref{fig:Comparisona} displays a comparison between the evolution of the normalized pressure anisotropy for the different ICs (solid lines) and their corresponding Navier-Stokes hydrodynamic attractors (dashed lines). At early times the system exhibits a strong dependence on the initial conditions, displaying a diverse range of transient behaviors. On the other hand, at late times this dependence fades away as the different curves converge to the corresponding hydrodynamic Navier-Stokes attractor, and the systems are said to have hydrodynamized. For most of the evolution, curves for the pressure anisotropy with the same IC from the two different plasmas reveals no clear separation for the considered value of $\mu/T \approx 1.2$. It is only within a short transient window of $0.5 \lesssim \omega_\Lambda \lesssim 0.9$ that curves from the 1RCBH and 2RCBH models are more easily distinguished. Some of the time evolutions transiently violate the DEC (gray zone) or even the WEC (red zone) when the system is far-from-equilibrium, even though all the initial states were chosen to satisfy such energy conditions at the initial time slice. In the late time hydrodynamic regime, no violations of energy conditions happen.

Fig.~\ref{fig:Comparisonb} shows the evolution of the normalized R-charge density for both holographic models. Although the systems are initialized with different charge densities, each set of colored curves relax towards constant values at late times, as expected in the equilibrium regime. The observation that the numerical results for the 1RCBH (blue) and 2RCBH (red) models converge into stable and distinct equilibrium values, provides a qualitative check on the consistency of the numerical simulations. The corresponding value of $\mu/T$ is estimated by employing the Newton-Raphson algorithm to numerically find the value of $\mu/T$ for which $[\hat{\rho}_{c,\text{eq}}^{4/3}/\hat{\varepsilon}_\text{eq}](\mu/T)$ in the equilibrium formula~\eqref{eq:rho43eThermo} matches the numerical value $[\hat{\rho}_c^{4/3}/\hat{\varepsilon}](\tau_\text{end})$ evaluated at the end time of the simulation, when the system effectively equilibrated.

Figs.~\ref{fig:Comparisonc} and~\ref{fig:Comparisond} present the numerical results for the normalized scalar condensate and entropy density (calculated from the area of the apparent horizon), respectively, alongside their thermodynamic equilibrium values derived from the analytical formulas~\eqref{eq:thermo-ratios}. Both plots reveal a clear separation between curves corresponding to distinct initial conditions and holographic models in the far-from-equilibrium regime. Importantly, all simulated trajectories converge to the correct equilibrium asymptotes calculated with the same value of $\mu/T$ extracted from the aforementioned analysis of the normalized charge density, providing a highly nontrivial quantitative consistency check of our numerical results. Notably, each set of four colored curves coalesces to a different equilibrium value determined by the corresponding holographic dual. While the scalar condensate asymptotes exhibit a significant separation, the entropy density curves converge to nearly identical equilibrium values (though not exactly the same).

Consistent with previous studies on the Bjorken flow of the 1RCBH model and also the purely thermal SYM plasma~\cite{Critelli:2018osu,Rougemont:2021qyk,Rougemont:2021gjm,Rougemont:2022piu}, we also see that for the 2RCBH model the effective thermalization of the scalar condensate occurs on a time scale generally longer than the hydrodynamization time of the pressure anisotropy. This delay appears slightly more pronounced for the 1RCBH trajectories compared to the 2RCBH case. Another important characteristic of the scalar condensate is immediately apparent in the asymptotic limit: the 1RCBH plasma (blue curves) relaxes towards a positive equilibrium value, whereas the 2RCBH plasma (red curves) relaxes towards a negative value. This behavior is consistent with the analytical thermodynamic expressions obtained for each model in Eqs.~\eqref{eq:thermo-ratios}.

Fig.~\ref{fig:Comparisone} illustrates the time evolution of the radial position of the apparent horizon (solid lines) and the event horizon (dashed lines). The inverse radial position $u_H=1/r_H$ of the AH is kept constant (within some numerical tolerance) due to the way the residual diffeomorphism radial shift function $\lambda(\tau)$ was fixed, while the EH evolves dynamically. The convergence of the EH curves to the AH curves at late times signals the thermalization of the black hole horizon. Finally, as expected, at early times, when the system is out of equilibrium, the AH is always behind the EH (remember that in the radial $u=1/r$ coordinate, the boundary lies at $u=0$).

Fig.~\ref{fig:Comparisonf} tracks the time evolution of the dimensionless non-equilibrium entropy, $\hat{S}_H/\mathcal{A}\Lambda^2=\tau \hat{s}_H/\Lambda^2$, calculated from the area of the AH (solid lines) and from the area of the EH (dashed lines). In several solid lines, we observe complex non-monotonic (albeit non-decreasing) behavior at early times before the system settles into the monotonic growth characteristic of the hydrodynamic regime. Specifically, we identify the emergence of transient plateaus and quasi-plateaus (periods of near-zero entropy production) far-from-equilibrium, a phenomenon consistent with previous analyses on the 1RCBH model and the purely thermal SYM plasma undergoing Bjorken flow~\cite{Rougemont:2021qyk, Rougemont:2021gjm,Rougemont:2022piu}. These earlier studies qualitatively spotted a correlation between the formation of plateaus in the far-from-equilibrium regime of the entropy associated to the apparent horizon and the subsequent production of local minima in the normalized pressure anisotropy (which may lead to DEC violations from below). A discussion about such correlations in the context of the 2RCBH model and the specific initial conditions considered here is provided below in Sec.~\ref{sec:comprehensive_analysis}. Finally, the agreement between the AH and EH entropy calculations at late times confirms the consistency of our numerical setup.

\subsection{Comprehensive Analysis of the Bjorken Flow for the 2RCBH Model\label{sec:comprehensive_analysis}}

In this subsection, we discuss in detail the time evolution of physical observables of the 2RCBH plasma for different values of the system's initial charge density parameter, $\rho_0$. The results are presented in four panels, each containing six plots corresponding to the initial conditions IC\#01 (Fig.~\ref{fig:01}), IC\#02 (Fig.~\ref{fig:02}), IC\#03 (Fig.~\ref{fig:03}), and IC\#04 (Fig.~\ref{fig:04}) specified in Table~\ref{tab:parameters}. Within each figure, five distinct colored curves represent different initial charge densities for the same IC. As in the previous subsection, solid curves denote the numerical simulation data, while dashed lines serve as analytical standards for the late time behavior of the system, either of hydrodynamic or thermodynamic origin.

\subsubsection{Normalized Pressure Anisotropy: \texorpdfstring{$[\Delta \hat{p}/\hat{\varepsilon}](\omega_{\Lambda})$}{Delta p/epsilon}}

\begin{figure*}
\centering  
\subfigure[]{\includegraphics[width=0.475\linewidth]{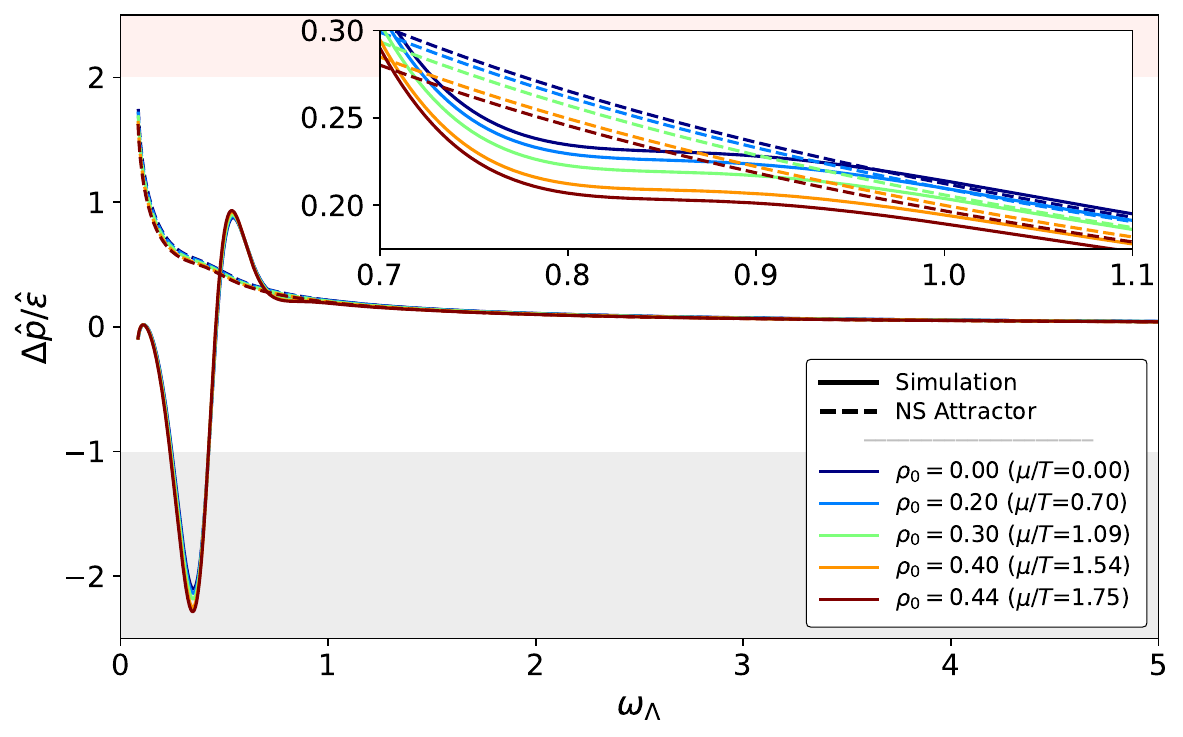}\label{fig:03a}}
\subfigure[]{\includegraphics[width=0.475\linewidth]{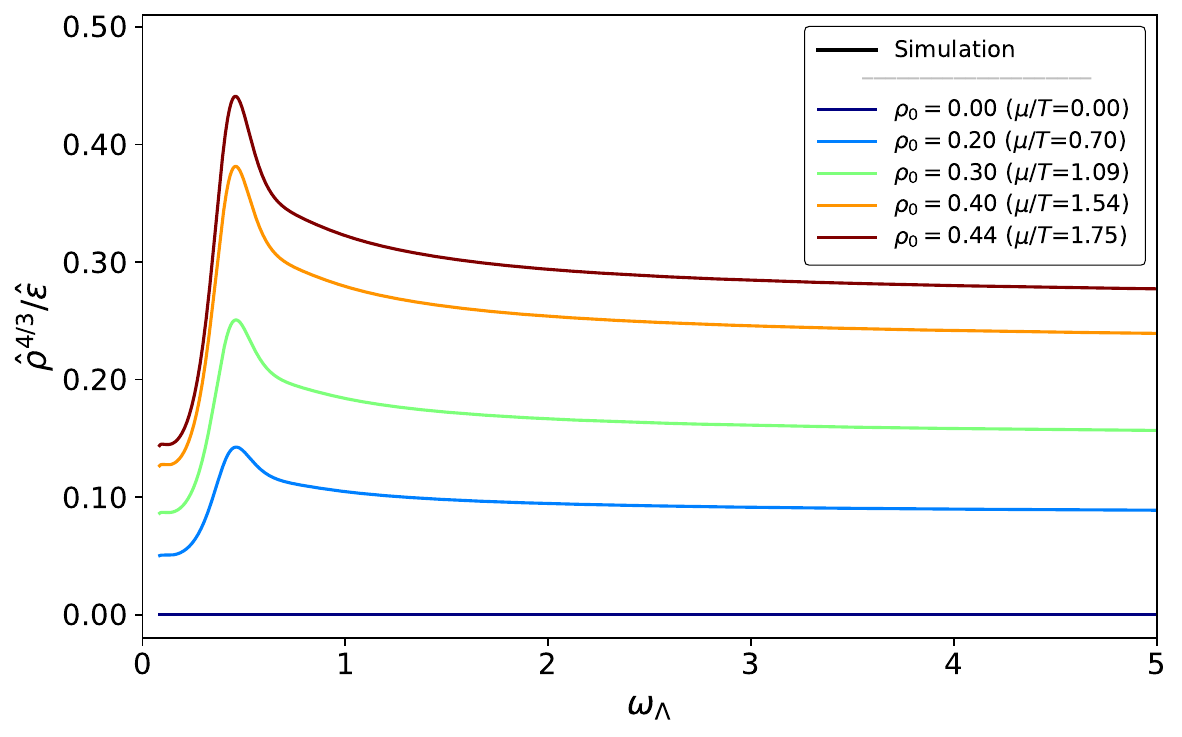}\label{fig:03b}}
\subfigure[]{\includegraphics[width=0.475\linewidth]{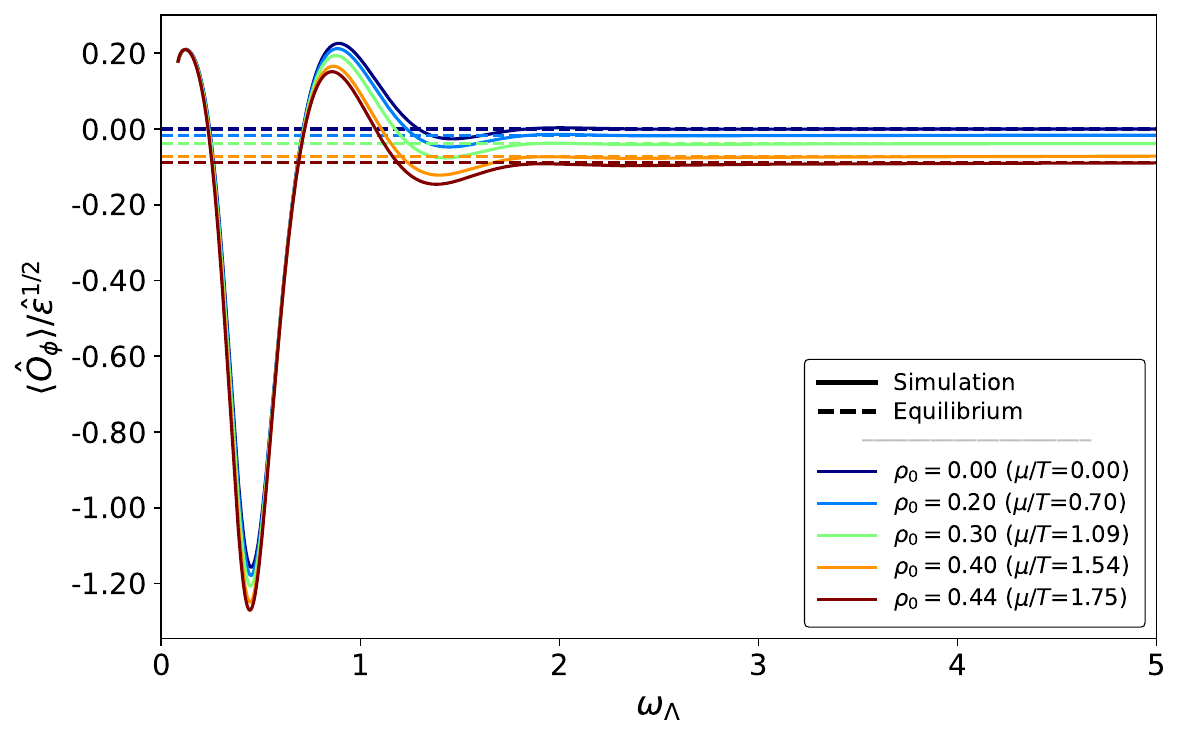}\label{fig:03c}}
\subfigure[]{\includegraphics[width=0.475\linewidth]{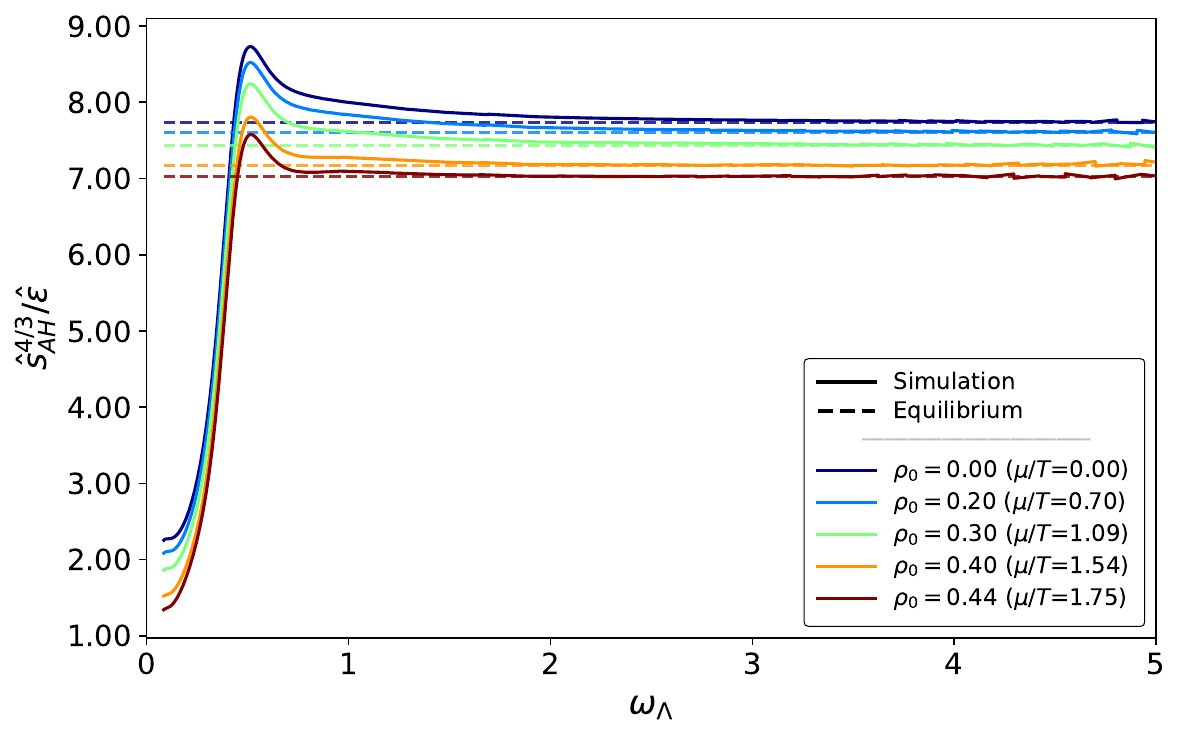}\label{fig:03d}}
\subfigure[]{\includegraphics[width=0.475\linewidth]{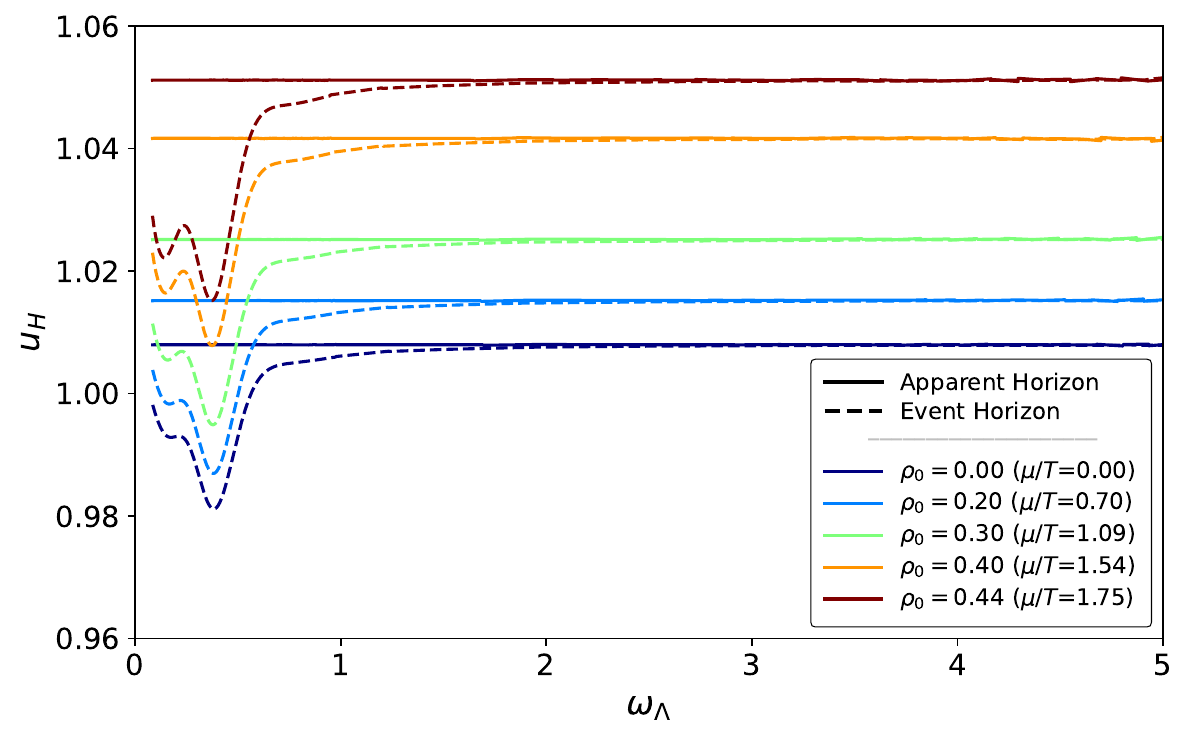}\label{fig:03e}}
\subfigure[]{\includegraphics[width=0.475\linewidth]{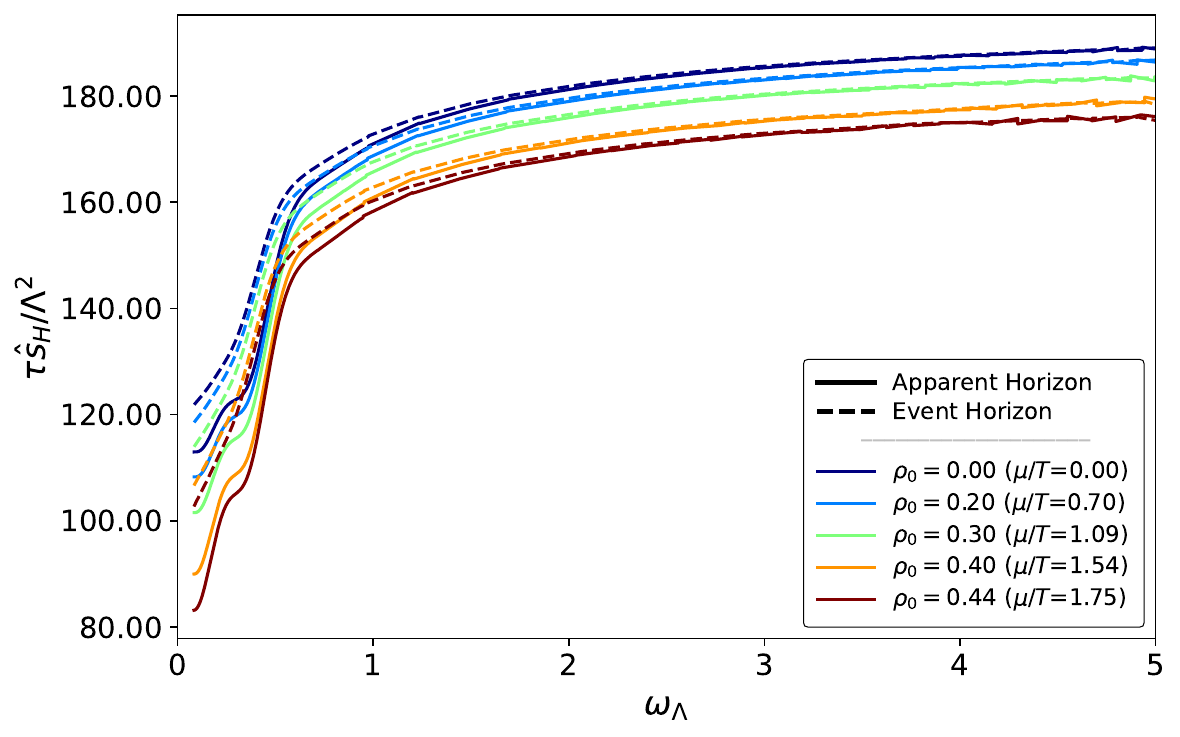}\label{fig:03f}}
\caption{(a) Normalized pressure anisotropy (solid lines) and the corresponding hydrodynamic Navier-Stokes attractor given by Eqs.~\eqref{eq:pressure-anisotropy-EOM} and~\eqref{eq:energyNS} (dashed lines) --- the gray and red zones indicate violations of the DEC and WEC, respectively; (b) normalized charge density; (c) normalized scalar condensate (solid lines) and the corresponding thermodynamic stable equilibrium result (dashed lines); (d) normalized non-equilibrium entropy density associated to the apparent horizon $\hat{s}_\text{AH}^{4/3}/\hat{\varepsilon}$ (solid lines) and the corresponding thermodynamic stable equilibrium result (dashed lines); (e) radial location of the apparent horizon (solid lines) and event horizon (dashed lines); and (f) non-equilibrium entropy $\hat{S}_H/\mathcal{A}\Lambda^2=\tau \hat{s}_H/\Lambda^2$ calculated from the area of the apparent horizon (solid line) and from the area of the event horizon (dashed lines). The results were obtained by varying the initial charge density parameter $\rho_0$, while keeping $B_s$ and $\phi_s$ fixed according to the IC\#03 in Table \ref{tab:parameters}, with $a_2(\tau_0)=-6.6667$.}
\label{fig:03}
\end{figure*}

\begin{figure*}
\centering  
\subfigure[]{\includegraphics[width=0.475\linewidth]{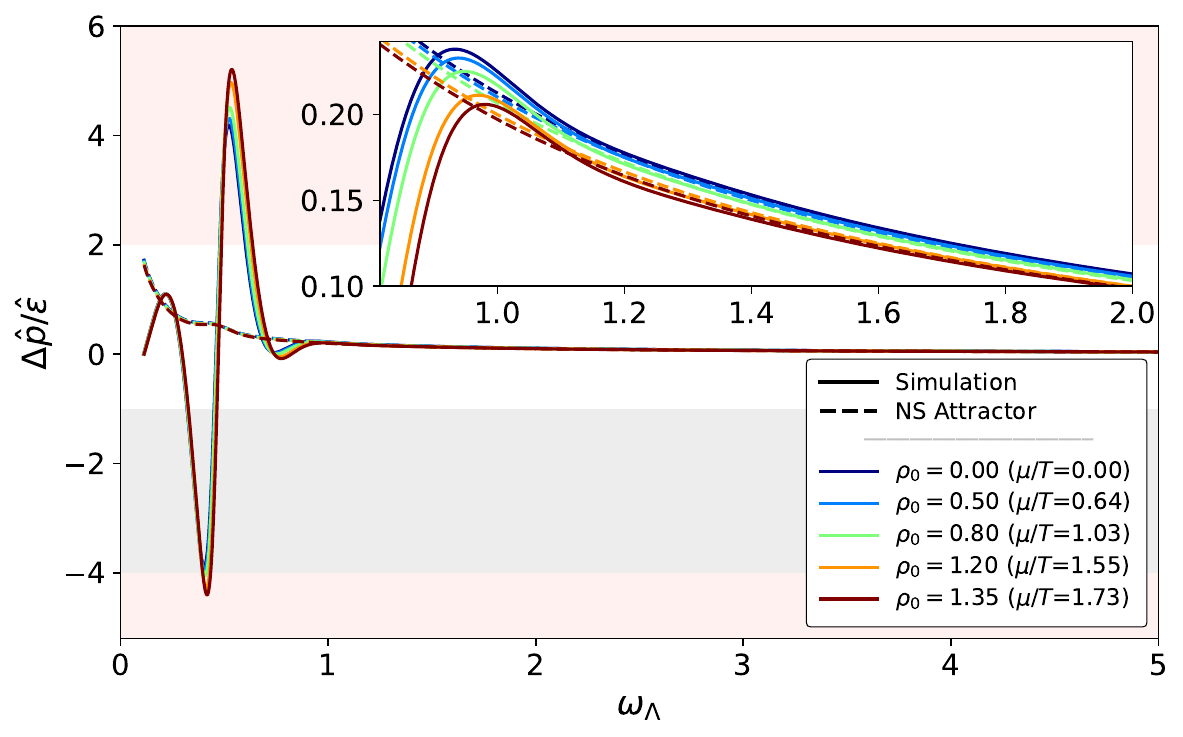}\label{fig:04a}}
\subfigure[]{\includegraphics[width=0.475\linewidth]{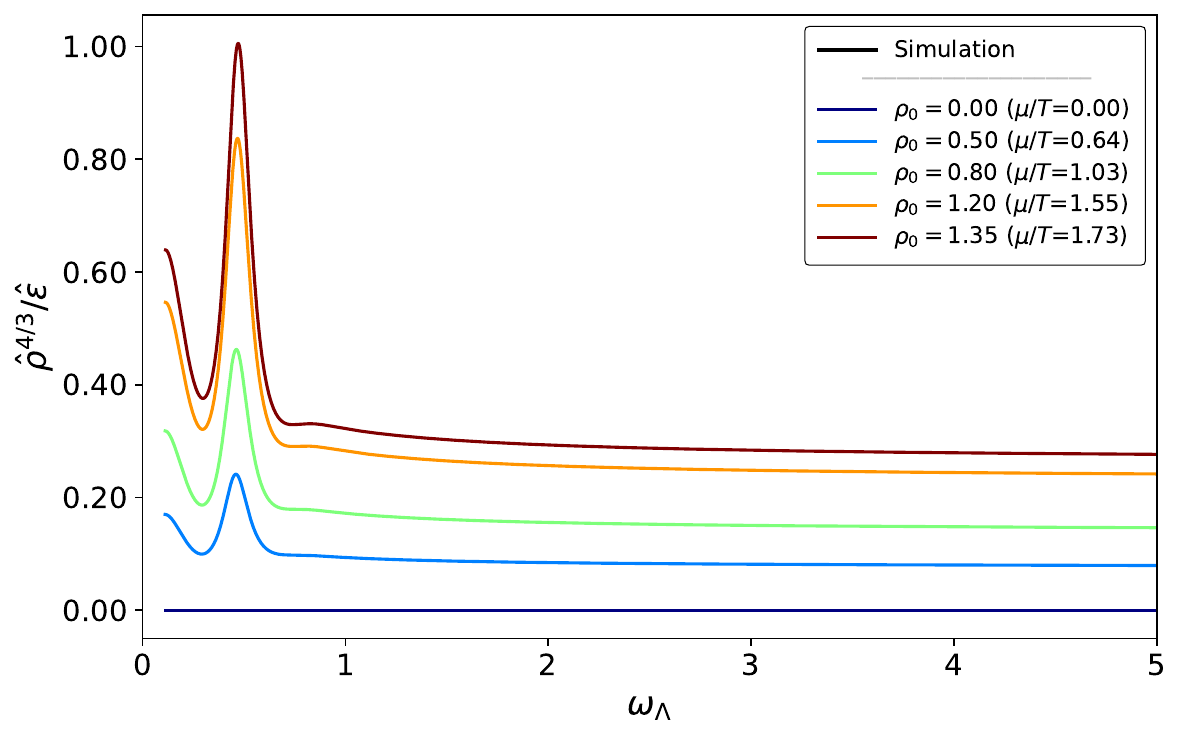}\label{fig:04b}}
\subfigure[]{\includegraphics[width=0.475\linewidth]{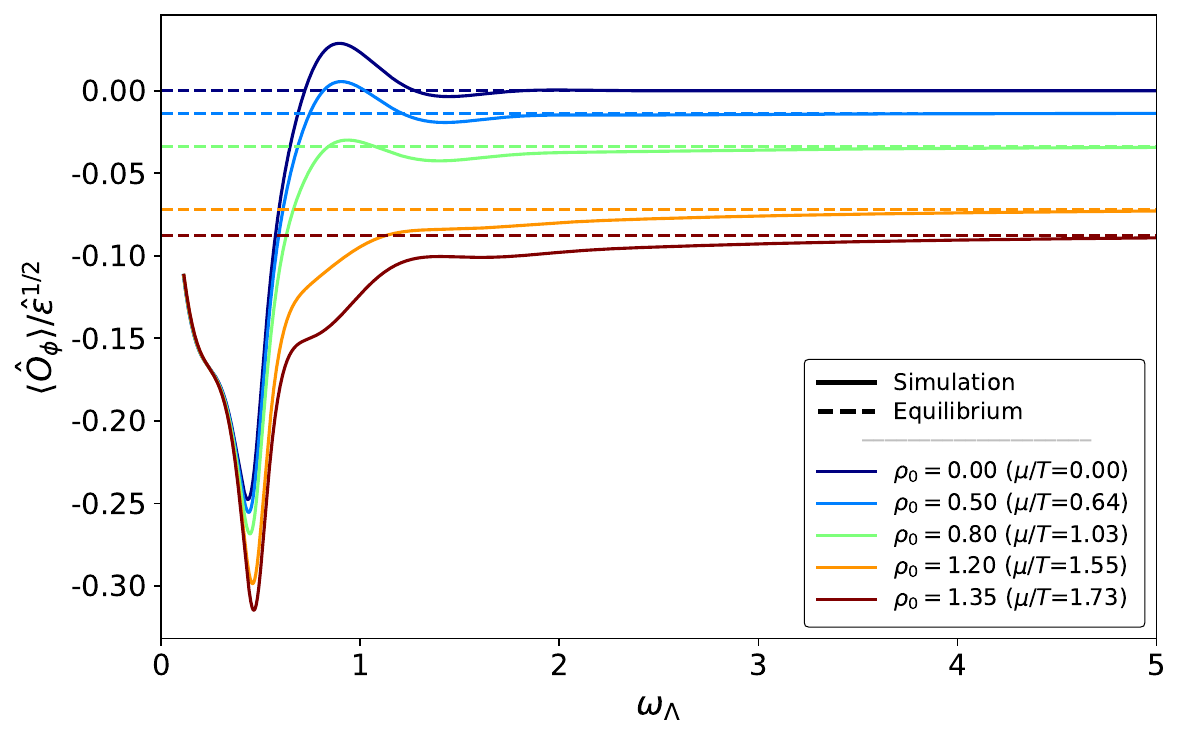}\label{fig:04c}}
\subfigure[]{\includegraphics[width=0.475\linewidth]{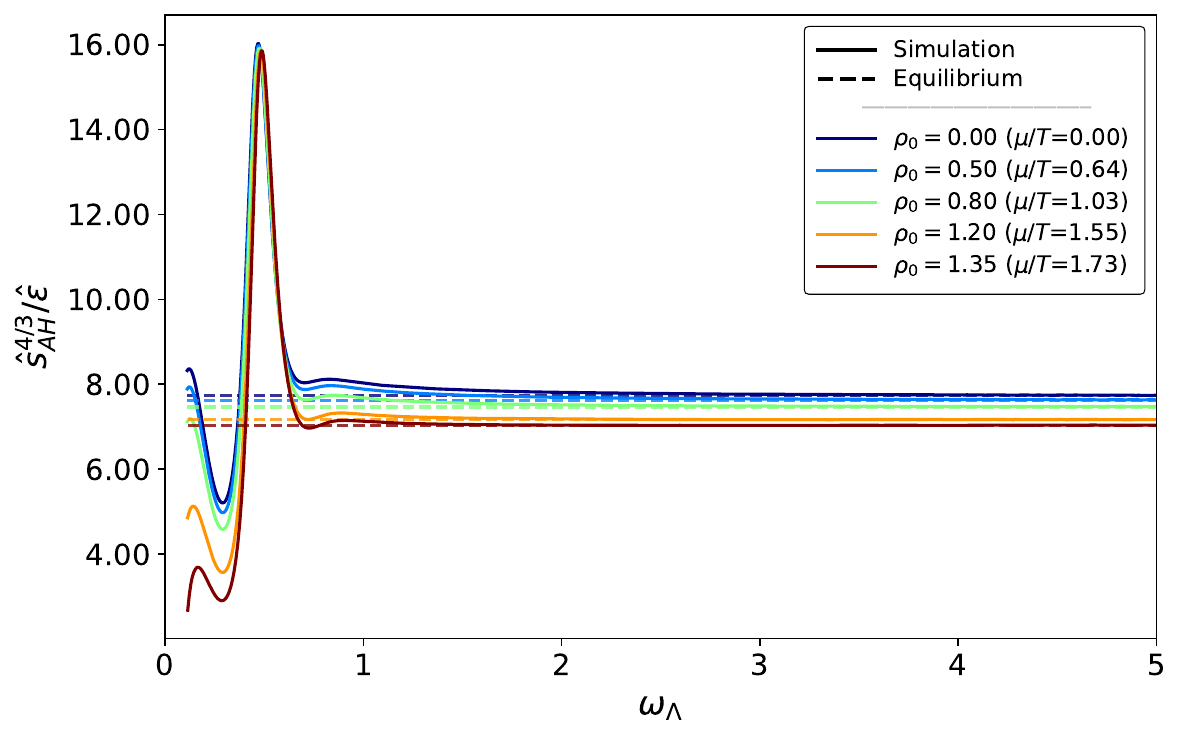}\label{fig:04d}}
\subfigure[]{\includegraphics[width=0.475\linewidth]{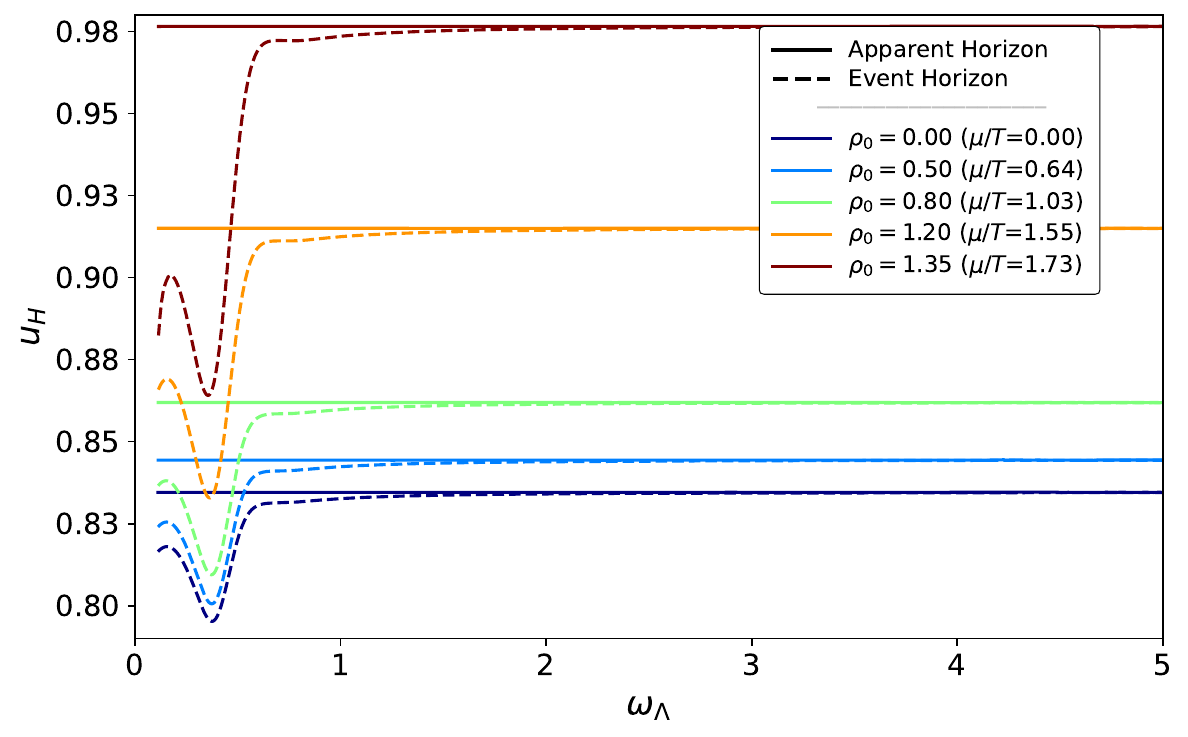}\label{fig:04e}}
\subfigure[]{\includegraphics[width=0.475\linewidth]{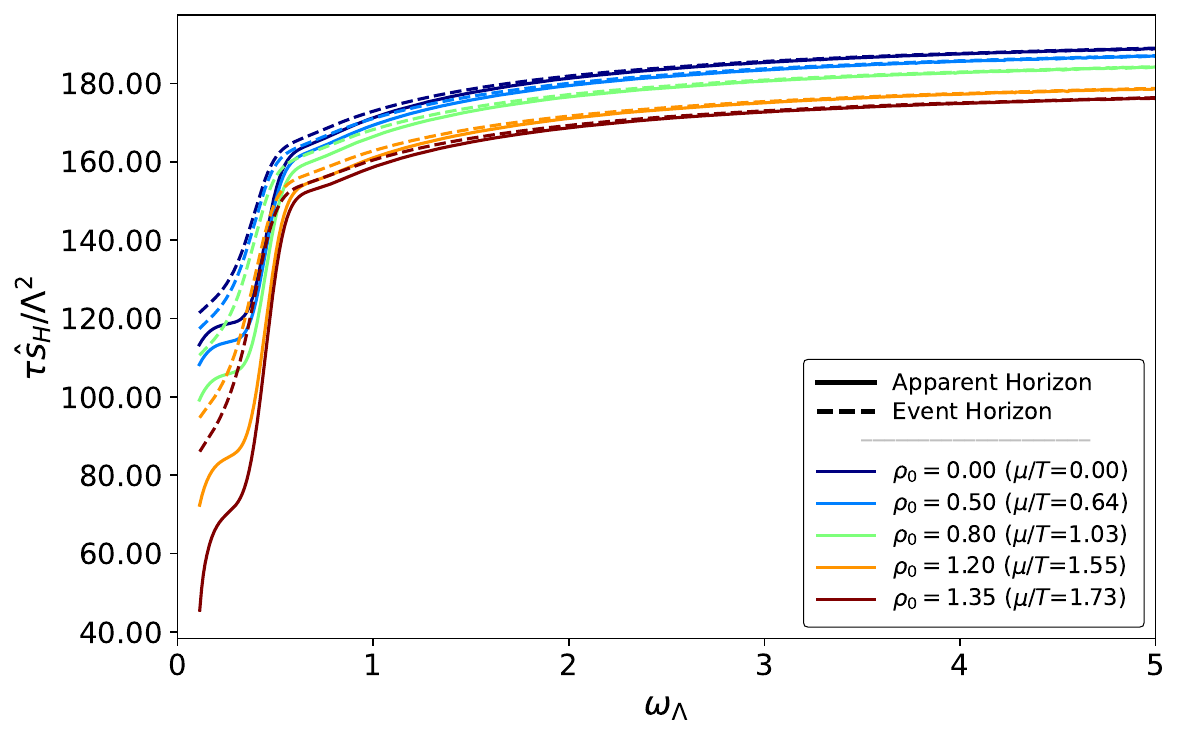}\label{fig:04f}}
\caption{(a) Normalized pressure anisotropy (solid lines) and the corresponding hydrodynamic Navier-Stokes attractor given by Eqs.~\eqref{eq:pressure-anisotropy-EOM} and~\eqref{eq:energyNS} (dashed lines) --- the gray and red zones indicate violations of the DEC and WEC, respectively; (b) normalized charge density; (c) normalized scalar condensate (solid lines) and the corresponding thermodynamic stable equilibrium result (dashed lines); (d) normalized non-equilibrium entropy density associated to the apparent horizon $\hat{s}_\text{AH}^{4/3}/\hat{\varepsilon}$ (solid lines) and the corresponding thermodynamic stable equilibrium result (dashed lines); (e) radial location of the apparent horizon (solid lines) and event horizon (dashed lines); and (f) non-equilibrium entropy $\hat{S}_H/\mathcal{A}\Lambda^2=\tau \hat{s}_H/\Lambda^2$ calculated from the area of the apparent horizon (solid line) and from the area of the event horizon (dashed lines). The results were obtained by varying the initial charge density parameter $\rho_0$, while keeping $B_s$ and $\phi_s$ fixed according to the IC\#04 in Table \ref{tab:parameters}, with $a_2(\tau_0)=-6.6667$.}
\label{fig:04}
\end{figure*}

Figs.~\ref{fig:01a} ---~\ref{fig:04a} display the time evolution of the normalized pressure anisotropy, with the gray and red zones describing, respectively, regions where the DEC~\eqref{eq:DEC} or even the WEC~\eqref{eq:WEC} are violated (from Eqs.~\eqref{eq:DEC} and~\eqref{eq:WEC}, one notes that any WEC violation also implies in DEC violation). In particular, the initial conditions used throughout this work were purposefully selected to illustrate a variety of scenarios, including cases that satisfy these energy conditions during the entire time evolution of the system, and cases which transiently violate these energy conditions, when the system is far-from-equilibrium (see Sec.~\ref{sec:data_EC}), although all the initial data considered here satisfy the DEC and WEC at the initial time $\tau_0$.

The evolution of the pressure anisotropy in the 2RCBH plasma reveals a rich variety of transient behaviors, heavily depending on the initial metric anisotropy profile. Each curve converges to its corresponding NS attractor at late times, which depends only on the value of $\mu/T$, confirming the late time hydrodynamization of the 2RCBH fluid undergoing Bjorken flow. For IC\#01 in Fig.~\ref{fig:01a}, the curves exhibit a single maximum and remain entirely within the region where all energy conditions are satisfied. In contrast, the remaining ICs in Table~\ref{tab:parameters} display different extrema, some of them accompanied by energy condition violations: IC\#02 and IC\#03 show violations of the DEC from below ($\Delta \hat{p}/\hat{\varepsilon} <-1$), while IC\#04 violates both the DEC and WEC, from above ($\Delta \hat{p}/\hat{\varepsilon} >2$) and also from below ($\Delta \hat{p}/\hat{\varepsilon} <-4$). Interestingly, whenever the pressure anisotropy presents a minimum, its magnitude increases with increasing $\mu/T$. This observation aligns with previous findings for the 1RCBH plasma~\cite{Rougemont:2022piu}, which also established that increasing $\mu/T$ delays the hydrodynamization time toward the NS attractor and intensifies energy condition violations (DEC or WEC) whenever they occur.

\subsubsection{Normalized Charge Density: \texorpdfstring{$[\hat{\rho}_c^{4/3}/\hat{\varepsilon}](\omega_{\Lambda})$}{rho43/epsilon}}

In Figs.~\ref{fig:01b} ---~\ref{fig:04b}, the normalized charge density is plotted as a function of the effective dimensionless time measure $\omega_{\Lambda}$. The general behavior observed across the tested ICs for the 2RCBH model is similar to the findings of Refs.~\cite{Critelli:2018osu,Rougemont:2022piu} for the 1RCBH model, where the magnitude of the normalized charge density consistently increases with $\mu/T$. However, the early time qualitative behavior of $\hat{\rho}_c^{4/3}/\hat{\varepsilon}$ varies significantly depending on the initial data considered.

IC\#01 follows a strictly monotonic, non-oscillatory decay from an initial maximum, devoid of any transient growth. In contrast, IC\#02 and IC\#03 exhibit a rapid transient rise toward a global maximum before relaxing towards equilibrium following a monotonic decay. Finally, IC\#04 distinguishes itself by exhibiting a pronounced local minimum followed by an even more pronounced maximum in the far-from-equilibrium regime, whose magnitudes increase with the increase of $\rho_0$ before settling into its final equilibrium state. In the late time regime, we observe that a higher initial $\rho_0$ leads to a higher equilibrium $\mu/T$. We also note that, at early times, for all the ICs analyzed in Table~\ref{tab:parameters}, the oscillatory patterns observed for $\hat{\rho}_c^{4/3}/\hat{\varepsilon}$ and $\hat{s}_\text{AH}^{4/3}/\hat{\varepsilon}$ are clearly correlated. Since the time evolution of the R-charge density is rather simple in the holographic Bjorken flow, $\rho_c(\tau)=\rho_0/\tau$ according to Eq~\eqref{eq:hatrho}, and the entropy is itself a non-decreasing function of time, such correlations appear related to the behavior of the energy density used in the normalizations.

\subsubsection{Normalized Scalar Condensate: \texorpdfstring{$[\langle \hat{O}_{\phi}\rangle/\hat{\varepsilon}^{1/2}](\omega_{\Lambda})$}{<O>phi/epsilon12}}

Figs.~\ref{fig:01c} ---~\ref{fig:04c} illustrate the graphical results for the normalized scalar condensate. All the numerical curves saturate at late times to the precise values predicted by the analytical thermodynamic formulas (represented by dashed lines), providing a further consistency check of the validity of our numerical data.

Similar to what was found in Ref.~\cite{Rougemont:2022piu} for the 1RCBH model, we also observe in the 2RCBH model that increasing $\mu/T$ delays the thermalization of the scalar condensate. This process is noticeably much slower than the hydrodynamization of the pressure anisotropy, highlighting the distinct relaxation timescales governing these observables. Also, the magnitude of the 2RCBH plasma scalar condensate always increases as $\mu/T$ is increased.

A prominent feature of the 2RCBH model, confirmed here numerically and consistent with the analytical formula~\eqref{eq:Oe12Thermo}, is the relaxation of the scalar condensate towards a negative equilibrium for nonzero R-charge densities, in direct contrast to the positive values found in the 1RCBH model.

\subsubsection{Normalized Entropy Density: \texorpdfstring{$[\hat{s}_\text{AH}^{4/3}/\hat{\varepsilon}](\omega_{\Lambda})$}{s43/epsilon} and \texorpdfstring{$[\tau \hat{s}_H/\Lambda^2](\omega_{\Lambda})$}{tau sH/Lambda}}

Figs.~\ref{fig:01d} ---~\ref{fig:04d} show the time evolution of the non-equilibrium entropy density over energy density ratio, $[\hat{s}_\text{AH}^{4/3}/\hat{\varepsilon}](\omega_{\Lambda})$. All numerical curves asymptotically converge to the constant values predicted by the thermodynamic equation of state. This agreement between the late time numerical results and the analytical equilibrium formulas reinforces once more the physical consistency of the numerical simulations, confirming that the system correctly relaxes towards the expected stable thermodynamic configurations.

In Figs.~\ref{fig:01e} ---~\ref{fig:04e}, we plot a comparison between the inverse radial location of the AH, $u_\text{AH}=1/r_\text{AH}$ (solid lines), and of the EH, $u_\text{EH}=1/r_\text{EH}$ (dashed lines). Consistently with the entropy results, a higher initial charge density parameter $\rho_0$ implies a larger final equilibrium value for $u_\text{H}=1/r_\text{H}$ across all initial conditions (notice that the larger the value of the inverse radial location of the horizon, the smaller is its area, and consequently the associated entropy).

Figs.~\ref{fig:01f} ---~\ref{fig:04f} display a comparison between the time evolution of the dimensionless ratio $\hat{S}_H/\mathcal{A}\Lambda^2=\tau \hat{s}_{AH}/\Lambda^2$ calculated both in terms of the apparent and the event horizon radial position. This normalization is interesting because it makes no use of the energy density to produce a dimensionless ratio, allowing to directly track the time evolution of the area of the horizon, which is non-decreasing along the entire time evolution of the system, in compatibility with the second law of (black hole) thermodynamics (see e.g.~\cite{Rougemont:2021gjm}). Similarly to the results for the 1RCBH model undergoing Bjorken expansion~\cite{Rougemont:2022piu}, we also observe that across every tested initial condition for the 2RCBH model, an increase in the equilibrium value of $\mu/T$ is always associated with a decrease in the magnitude of the entropy.\footnote{Our numerical results for the 2RCBH model show that $\tau \hat{s}_H/\Lambda^2$ decreases as the initial charge density (and, consequently, the late time equilibrium value of $\mu/T$) increases. At first sight, this behavior might appear counterintuitive given that, in thermodynamic equilibrium, the dimensionless ratio $s_{\text{eq}}/N_c^2T^3$ in Fig.~\ref{fig:Thermo} grows monotonically with $\mu/T$. The resolution lies on the fact that for a fixed black hole mass $M$, increasing the black hole charge $Q$ reduces the horizon radius $r_H$, see Eq.~\eqref{eq:bbrp}. Since the entropy is proportional to the horizon area, it must also decrease. The reason why the dimensionless ratio $s_{\text{eq}}/N_c^2T^3$ instead increases, is because besides the entropy density itself, also the temperature decreases with increasing $Q$ at fixed $M$, see Eq.~\eqref{eq:TThermDef1R} and \eqref{eq:TThermDef2R}.}

Furthermore, we observe a universal feature across all ICs: an inverse hierarchy regarding the behavior of the different physical observables as $\mu/T$ is increased: while $[\hat{\rho}_c^{4/3}/\hat{\varepsilon}](\omega_{\Lambda})$ increases, $[\langle \hat{O}_{\phi}\rangle/\hat{\varepsilon}^{1/2}](\omega_{\Lambda})$, $[\hat{s}_\text{AH}^{4/3}/\hat{\varepsilon}](\omega_{\Lambda})$, and $\tau \hat{s}_\text{H}(\omega_{\Lambda})/\Lambda^2$ decrease.

Another key finding shared with Refs.~\cite{Rougemont:2021qyk, Rougemont:2021gjm,Rougemont:2022piu} regarding the purely thermal SYM and the 1RCBH plasmas, is that also in the 2RCBH model different initial data lead to distinct entropy evolution scenarios, characterized by the presence or absence of transient (quasi-)plateaus structures for the far-from-equilibrium entropy associated specifically to the apparent horizon. In the present work, we also analyzed the behavior of the entropy calculated through the event horizon and we did not find any transient plateaus associated to it.

More specifically, in the cases of IC\#03 in Fig.~\ref{fig:03f} and IC\#04 in Fig.~\ref{fig:04f}, we found transient double and single quasi-plateaus, respectively, in the time evolution of the normalized non-equilibrium entropy of the apparent horizon, $\hat{S}_\text{AH}/\mathcal{A}\Lambda^2=\tau \hat{s}_\text{AH}/\Lambda^2$. These reported quasi-plateau structures for the non-equilibrium entropy were followed by local minima in the normalized pressure anisotropy which led to DEC violations from below with $\Delta \hat{p}/\hat{\varepsilon}<-1$ (in the case of IC\#04, some of the local minima also violated WEC from below with $\Delta \hat{p}/\hat{\varepsilon}<-4$); moreover, IC\#02 in Fig.~\ref{fig:02f}, displayed for the non-equilibrium entropy of the apparent horizon a sequential formation of exact double plateaus (for $\rho_0=0$ and $\rho_0=0.2$),\footnote{In these cases, the double plateaus are difficult to visually distinguish from a single plateau because they are very close to each other, but they can be distinguished by numerically evaluating the time derivative of the entropy --- see Appendix~\ref{sec:apA}.} which were progressively deformed into double quasi-plateaus (for $\rho_0=0.4$, $\rho_0=0.5$, and $\rho_0=0.6$), being followed by local minima in the pressure anisotropy with progressively increasing magnitudes leading to DEC violations from below with $\Delta \hat{p}/\hat{\varepsilon}<-1$. On the other hand, in the case of IC\#01 in Fig.~\ref{fig:01f}, our simulations showed no transient (quasi-)plateau structures for the non-equilibrium entropy of the apparent horizon and also no local minima for the pressure anisotropy and no energy conditions violations.

The overall picture found for the holographic Bjorken flow in Refs.~\cite{Rougemont:2021qyk,Rougemont:2021gjm}, concerning the purely thermal SYM plasma at zero R-charge density; in Ref.~\cite{Rougemont:2022piu}, regarding the 1RCBH plasma at finite R-charge density; and now in the present paper, regarding the 2RCBH plasma at finite R-charge density, is as follows. Transient and exact single or double plateaus in the apparent horizon entropy are correlated to a later formation of local minima for the pressure anisotropy with, respectively, $\Delta \hat{p}/\hat{\varepsilon}\sim-1$ or $\Delta \hat{p}/\hat{\varepsilon}<-1$ (thus, associated to DEC violations from below in the latter case).\footnote{Such correlations do not hold in reverse order, i.e. there are time evolutions violating DEC from below with no previous formation of plateaus in the entropy.} In face of the different systems analyzed, such far-from-equilibrium correlations between the entropy of the apparent horizon and the pressure anisotropy normalized by the energy density seem to be of a very general nature, at least for strongly interacting and conformal fluids, be them neutral or charged.

\section{Conclusions}
\label{sec:conc}

In the present work, we numerically investigated the far-from-equilibrium Bjorken flow dynamics of the holographic 2RCBH plasma, corresponding to a strongly interacting and conformal quantum fluid defined at finite temperature and R-charge density. We found that the hydrodynamization and the later thermalization of the fluid is delayed by increasing its initial R-charge density, which later results into higher values of $\mu/T$ in equilibrium. We also found that the normalized scalar condensate, $\langle \hat{O}_{\phi}\rangle/\hat{\varepsilon}^{1/2}$, and the non-equilibrium entropy associated either to the area of the apparent or the event horizon, are both reduced with increasing values of $\mu/T$. The normalized pressure anisotropy, $\Delta \hat{p}/\hat{\varepsilon}$, can display varied far-from-equilibrium behaviors depending on the initial states considered for the fluid, including some time evolutions transiently violating the dominant energy condition or even the weak energy condition, even though all the initial states considered here satisfy all the energy conditions. Such transient violations of energy conditions only happen when the system is far-from-equilibrium, ceasing before the fluid relaxes to the hydrodynamic regime. Also in the far-from-equilibrium regime, some time evolutions reveal the formation of plateau structures in the non-equilibrium entropy associated to the apparent horizon, although such plateaus are never seen in the non-equilibrium entropy associated to the event horizon.

Indeed, taken together with previous results in the literature for the purely thermal SYM plasma~\cite{Rougemont:2021qyk,Rougemont:2021gjm} and for the 1RCBH model~\cite{Rougemont:2022piu}, the present results for the 2RCBH model undergoing Bjorken flow provide strong qualitative evidence for a putative generality of the observed far-from-equilibrium correlations between the non-equilibrium entropy of the apparent horizon of dynamical bulk black branes and the expectation value of the energy-momentum tensor of the dual boundary QFT. Namely, in all the aforementioned strongly interacting conformal quantum fluids, exact single or double plateaus were observed to be correlated with subsequent developments of local minima for the normalized pressure anisotropy with, respectively, $\Delta \hat{p}/\hat{\varepsilon}\sim-1$ or $\Delta \hat{p}/\hat{\varepsilon}<-1$ (therefore, leading to DEC violations from below in the latter case). Such far-from-equilibrium correlations cannot be seen if one chooses to define the holographic non-equilibrium entropy through the area of the dynamical event horizon, due to the absence of transient plateau structures in its time evolution, which is a clear indication that the apparent horizon encodes information not available in the event horizon when the system is far from thermodynamic equilibrium. Moreover, such information is effectively erased when the system later enters into the hydrodynamic regime, by losing memory of the initial conditions which directly affects its pre-hydrodynamic stages.

The aforementioned far-from-equilibrium correlations were observed in the pre-hydrodynamic stages of holographic strongly interacting conformal fluids, both neutral and charged. It is important to investigate in the future whether similar correlations are also observed in the pre-hydrodynamic stages of non-conformal holographic models, such as in phenomenological EMD models designed to holographically emulate the quark-gluon plasma formed in relativistic heavy ion collisions~\cite{Rougemont:2023gfz}.

%%%%%%%%%%%%%%%%%%%%%%%%%%%%%%%%%
\begin{acknowledgement}
G.O. acknowledges financial support from the Coordination for the Improvement of Higher Education Personnel (CAPES) under grant number 88887.956852/2024-00. W.B. and R.R. acknowledge financial support by National Council for Scientific and Technological Development (CNPq) under grant number 407162/2023-2. R.R. also acknowledges financial support by CNPq under grant 305466/2024-0.
\end{acknowledgement}

%%%%%%%%%%%%%%%%%%%%%%%%%%%%%%%%%%
\appendix

\section{Smoothing of Numerical Noise}
\label{sec:apA}

Throughout the time evolution of the system implemented through a numerical code written in F90, the radial location of the apparent horizon typically displays a type of structured noise which largely increases at late times, affecting the plots for the corresponding non-equilibrium entropy. In the cases where it was not enough to simply select an appropriate number of radial collocation points in order to diminish the numerical noise increasing with the passage of time, we proceeded to smooth out the resulting data noise by using a dynamic smoothing procedure adjusted to distinct time windows.

The noise treatment procedure uses a multi-stage algorithm to separate the physical signals from spurious numerical artifacts. First, the data is downsampled and divided into variable time windows: we use shorter intervals during the early transient stages to resolve rapid changes, and longer time windows for the late time evolution. We then fit these segments using cubic splines. This method approximates the data using piecewise polynomials, which ensures that the filtered curves remain smooth and differentiable --- a property that is essential for our analysis. Furthermore, we apply adaptive smoothing strengths adjusted to the specific noise level of each quantity being analyzed: a stronger smoothing factor is used to filter the entropy, while a weaker factor is sufficient to treat the cleaner data of the horizon and energy density. This post-processing approach, implemented in Python using the Scipy Library, was essential to obtain stable results for the time derivative of the entropy, allowing us to distinguish plateaus and quasi-plateaus.

Our main F90 code and our initial Python post-processing filtering code were written solely by humans, but our final version of the Python filtering code was improved with the AI assistance of Gemini 3 Pro.

%
% BibTeX users please use
%%%%%%%%%%%%%%%%%%%%%%%%%%%%%%%%%%
\bibliographystyle{sn-mathphys-num} 
\bibliography{bibliography,extrabiblio}
% %
% % Non-BibTeX users please use
% \begin{thebibliography}{}
% %
% % and use \bibitem to create references.
% % Format for Journal Reference
% Author, Journal \textbf{Volume}, (year) page numbers.
% Author, \textit{Book title} (Publisher, place year) page numbers
% \end{thebibliography}

\end{document}